\documentclass[aps,10pt,prd,nofootinbib,superscriptaddress,twocolumn]{revtex4-2}

\usepackage[utf8]{inputenc} 			
\usepackage[ngerman, english]{babel}	

\newcommand{\papertitle}{Phase structure of quark matter and in-medium properties of mesons\\ from Callan-Symanzik flows}

\pdfoutput=1

\usepackage[fixamsmath,disallowspaces]{mathtools} 
\usepackage{amssymb}
\usepackage[all,warning]{onlyamsmath}
\usepackage{aligned-overset}
\usepackage{bbm}
\usepackage{slashed}
\usepackage{braket}

\usepackage{graphicx}
\setkeys{Gin}{height=5.6cm}
\usepackage[dvipsnames]{xcolor}
\usepackage{array}

\usepackage{xspace}
\usepackage{siunitx}
\usepackage{xfrac}
\usepackage{hyperref}
\usepackage[nameinlink]{cleveref}
\usepackage{appendix}

\usepackage{tikz}
\usetikzlibrary{shapes, arrows}
\tikzstyle{roundbox} = [rectangle, draw, text centered, rounded corners, 
minimum height=2em, minimum width=2em, draw=black, fill=blue!10]
\tikzstyle{process} = [rectangle, draw, minimum height=1em, 
minimum width=3em, text centered, draw=black, fill=green!10]
\tikzstyle{integration} = [ellipse, draw, text centered, minimum height=1em, 
minimum width=3em, draw=black, fill=red!10]

\usepackage{xifthen}
\usepackage{xcolor}
\hypersetup{
	colorlinks,
	linkcolor={red!75!black},
	citecolor={blue!75!black},
	urlcolor={blue!75!black}
}

\usepackage{array}

\graphicspath{
	{./}
}

\def\CO{{\mathcal O}}
\def\CS{{\mathcal S}}

\def\CM{{\mathcal M}}
\def\CI{{\mathcal I}}

\newcommand{\Tr}{\mathrm{Tr}}
\newcommand{\STr}{\mathrm{STr}}
\newcommand{\nn}{\nonumber}

\newcommand{\Nc}{N_{\rm c}}
\newcommand{\Nf}{N_{\rm f}}
\newcommand{\muc}{\mu_{\text{c}}}
\newcommand{\phifl}{\phi_{\text{fl}}}
\newcommand{\supsc}{(\text{sym})}
\newcommand{\mext}{m_{\text{ext}}}
\newcommand{\mextvac}{m_{\text{ext}}^{(0)}}
\newcommand{\tc}{T_{\text{c}}}
\newcommand{\tpc}{T_{\text{pc}}}
\renewcommand{\muc}{\mu_{\text{c}}}
\newcommand{\mupc}{\mu_{\text{pc}}}
\newcommand{\muSB}{\mu_{\text{SB}}}

\newcommand{\pF}{p_{\text{F}}}

\newcommand{\Gammat}{\tilde{\Gamma}}

\newcommand{\overbar}[1]{\mkern 1.8mu\overline{\mkern-1.8mu#1\mkern-1.8mu}\mkern 1.8mu}
\newcommand{\hb}{\overbar{h}}

\newcommand{\mb}{\overbar{m}}
\newcommand{\psib}{\bar{\psi}}

\newcommand*{\Hb}{\overbar{H}}
\newcommand*{\varphib}{\overbar{\varphi}}
\newcommand*{\sigmab}{\overbar{\sigma}}
\newcommand*{\pib}{\overbar{\pi}}

\newcommand{\dmom}[1]{(2 \pi)^4\ \delta^{(4)}\left(#1\right)}

\newcommand{\iu}{\mathrm{i}\mkern1mu}
\newcommand*{\e}{{\mathrm{e}\mkern1mu}}

\newcommand*{\pt}{\tilde{p}}
\renewcommand*\d{\mathop{}\!\mathrm{d}}
\newcommand{\dslash}{\slashed{\partial}}
\newcommand{\dt}{\partial_t}
\newcommand{\ct}{\text{CT}}
\newcommand*{\MeV}{\mskip3mu \text{MeV}}

\renewcommand*{\Im}[1]{\mathfrak{Im}\left\lbrace #1\right\rbrace}

\newcommand*{\uM}[1]{\mathbbm{1}_{\scriptscriptstyle #1 \times #1}}
\DeclareMathOperator{\arsinh}{arsinh}
\DeclareMathOperator{\artanh}{artanh}
\newcommand{\irchi}[2]{\raisebox{\depth}{$#1\chi$}} 
\DeclareRobustCommand{\rchi}{{\mathpalette\irchi\relax}}

\makeatletter
\newcommand*{\tp}{{\mathpalette\@tp{}}}
\newcommand*{\@tp}[2]{\raisebox{\depth}{$\m@th#1\intercal$}}
\makeatother

\def\0#1#2{\frac{#1}{#2}}

\usepackage{xspace}
\usepackage{siunitx}
\usepackage{hyperref}
\usepackage[nameinlink]{cleveref}
\usepackage{appendix}

\hypersetup{
	colorlinks,
	linkcolor={red!75!black},
	citecolor={blue!75!black},
	urlcolor={blue!75!black},
	pdftitle={\papertitle},
	pdfauthor={T\"opfel, Pawlowski, Braun},
	bookmarksopen=true,
	bookmarksopenlevel=2,
	bookmarksnumbered=true
}

\newcommand{\orcid}[1]{\href{https://orcid.org/#1}{\includegraphics[height=1.9ex,width=1.9ex]{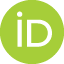}}}

\begin{document}

\title{\papertitle}

\author{Sebastian T\"opfel~\orcid{0000-0002-0459-4382}}
\affiliation{
	Technische Universit\"at Darmstadt, Department of Physics, Institut f\"ur Kernphysik, Theoriezentrum,\\
	Schlossgartenstra\ss e 2, D-64289 Darmstadt, Germany
}
\affiliation{
		Helmholtz Research Academy Hesse for FAIR, Campus Darmstadt,\\
		D-64289 Darmstadt, Germany
}

\author{Jan M. Pawlowski~\orcid{0000-0003-0003-7180}}
\affiliation{
	Institut f\"ur Theoretische Physik, Universit\"at Heidelberg, Philosophenweg 16, 69120 Heidelberg, Germany
}
\affiliation{ExtreMe Matter Institute EMMI, GSI, Planckstra{\ss}e 1, D-64291 Darmstadt, Germany}

\author{Jens Braun~\orcid{0000-0003-4655-9072}}
\affiliation{
	Technische Universit\"at Darmstadt, Department of Physics, Institut f\"ur Kernphysik, Theoriezentrum,\\
	Schlossgartenstra\ss e 2, D-64289 Darmstadt, Germany
}
\affiliation{
		Helmholtz Research Academy Hesse for FAIR, Campus Darmstadt,\\
		D-64289 Darmstadt, Germany
}
\affiliation{ExtreMe Matter Institute EMMI, GSI, Planckstra{\ss}e 1, D-64291 Darmstadt, Germany}

\begin{abstract}

We compute meson spectral functions at finite temperature and density in the quark-meson model, supplemented with a computation of the phase diagram. In particular, we provide a detailed analysis of the non-analytic structure of the meson two-point functions which is of great relevance for phenomenological applications, such as moat regimes and inhomogeneous phases. Furthermore, it is also relevant from a field-theoretical standpoint as it provides an insight into the applicability of derivative expansions of the effective action to studies of general fermion-boson models, both at zero and finite chemical potential. Our computation is based on a functional renormalization group setup that preserves causality, all spacetime symmetries, and the Silver-Blaze property. The combination of these properties can only be achieved by a Callan-Symanzik regulator. Instead of momentum shell integrations, renormalization group flows generated by such a regulator describe the change of the theory induced by a change of the masses of the mesons and quarks. A particular focus of our work lies on the construction of controlled Callan-Symanzik flows in the presence of spontaneous and explicit chiral symmetry breaking by means of chiral Ward-Takahashi identities. 
\end{abstract}
\maketitle

\section{Introduction}
The phase structure of strongly coupled systems remains a subject of intense investigation across various fields of physics, ranging from cosmology and high-energy physics to condensed-matter physics. Quantum chromodynamics (QCD) hosts a particularly rich phase structure whose understanding is crucial for the physics of the early universe and the structure of compact astrophysical objects such as neutron stars. 

A comprehensive analysis of the phase structure of QCD at low energies asks for a combined analysis of the order parameters and momentum-dependent correlation functions at finite temperatures and density. Moreover, these quantities are the fundamental building blocks in the computation of transport coefficients, and the latter establish a direct link between theory and experiment. In our present work, we take a step into this direction by performing a combined analysis of chiral order parameters and two-point correlation functions of mesons in the quark-meson model. Specifically, we use, for the first time, a functional renormalization group (fRG) framework that preserves the full spacetime symmetry as well as the important Silver-Blaze property set up in~\cite{Braun:2022mgx, Fehre:2021eob}. For a collection of works on non-perturbative spectral properties of complex systems, we refer the reader to  
~\cite{Gasenzer:2007za, Gasenzer:2010rq, Floerchinger:2011sc, Strodthoff:2011tz, Kamikado:2013sia, Tripolt:2013jra, Pawlowski:2015mia, Yokota:2016tip, Kamikado:2016chk, Jung:2016yxl, Pawlowski:2017gxj, Yokota:2017uzu, Wang:2017vis, Tripolt:2018jre, Tripolt:2018qvi, Corell:2019jxh, Huelsmann:2020xcy, Jung:2021ipc, Tan:2021zid, Heller:2021wan, Fehre:2021eob, Roth:2021nrd, Roth:2023wbp, Horak:2023hkp} as well as the comprehensive review \cite{Dupuis:2020fhh} and references therein.

The fRG setup developed in~\cite{Braun:2022mgx} is based on the Callan-Symanzik (CS) regulator whose flow describes the change of the theory at hand with a change of the mass. 
This regulator is particularly convenient because it not only respects spacetime symmetries (both in Minkowski space and Euclidean space) but also preserves the Silver-Blaze property of the theory. The latter is the requirement of the absence of chemical potential dependences in observables below the corresponding density onset. For the description in terms of correlation functions within functional approaches, see \cite{Marko:2014hea, Khan:2015puu,Braun:2020bhy}. These advantageous properties come at a twofold cost: firstly, this setup requires some care concerning its finiteness, and this intricacy has been resolved and thoroughly discussed in~\cite{Braun:2022mgx}. Secondly, the evolution of dynamical and explicit chiral symmetry breaking with the flow of the fermion mass has to be incorporated judiciously, using chiral Ward identities, which we shall discuss in the present work. 

We use the quark-meson model with its interesting phenomenological applications in the phase structure of QCD as a tailor-made example theory for a first application of renormalized spectral flows in combination with chiral Ward-Takahashi identities to the phase structure of a strongly correlated system. This also allows us to further the discussion of spectral properties of QCD at larger densities, including the signatures of the moat regime \cite{Rennecke:2021ovl, Rennecke:2023xhc} or inhomogeneous phases~\cite{Schon:2000qy,Buballa:2014tba}.

This work is organized as follows:
In \Cref{sec:framework}, we briefly remind the reader of the quark-meson model, and introduce our computational approach, renormalized spectral flows with the CS regulator. 
In \Cref{sec:sym_constr_CS}, we then discuss in detail how the flow of explicit chiral symmetry breaking with the flow of the fermion mass is controlled or rather adapted with the aid of Ward-Takahashi-type identities. With this symmetry-adapted CS framework at hand, we compute the effective potential, the curvature masses of the sigma mode and the pions, the mesonic  two-point functions as well as the corresponding spectral functions in the large-$\Nc$ limit. The latter limit is chosen as it permits a detailed analytic understanding of the quantities considered in our present work. In consequence, it offers an ideal benchmark case for our CS framework, that also offers novel physics insights. The corresponding results discussed in detail in \Cref{sec:res}.
We close with a brief summary and outlook in \Cref{sec:summary}. 

\section{Framework}\label{sec:framework}
In this section we present the general framework underlying our present work. 
We begin with a brief introduction of our low-energy model for QCD in \Cref{sec:QMmodelintro}, including a more general discussion of an intriguing shift symmetry of Yukawa-type models with a classical action that is only bilinear in the fields. 
In \Cref{sec:spec_reg}, we then discuss CS regulators which underlie spectral RG flows and highlight some important differences to Wilsonian flows. Moreover, we already explain briefly how this class of regulators affects the chiral symmetry of models which is most relevant for our computations of mesonic two-point functions and the phase diagram of our model. 
The aforementioned shift symmetry together with the properties of CS regulators can be used to derive an adapted version of the Wetterich equation for CS flows which allows to provide strong constraints on the effective action of chiral fermion-boson theories, as detailed in \Cref{sec:ShiftfermionicCS}. 
The reader may skip the latter subsection in a first reading.

\subsection{Quark-meson model and low energy QCD}
\label{sec:QMmodelintro}
In order to illustrate the effect of the explicit chiral symmetry breaking induced by a mass-like CS regularization, we consider a frequently used low-energy model of QCD, the quark-meson model (QM model). Phenomenologically speaking, this model describes the interaction of quarks via an exchange of the scalar $\sigma$-mode and pseudo-scalar pions. In Euclidean spacetime, the corresponding classical action can be written as follows:
\begin{align}
S[\Phi] =&\, \int_x \Big[ \vphantom{\frac{1}{2}}\psib (\iu \dslash-\iu \gamma^0 \mu) \psi + \iu h\psib \left(\sigma + \iu \gamma^5 \vec{\tau}\cdot \vec{\pi}  \right) \psi \nn\\
&\hspace{.3cm}+\frac{1}{2} \phi^\tp \left(- \partial^2 + m^2_\phi \right) \phi - H\sigma \Big]\ ,
\label{eq:action}
\end{align}
where $\int_x =\int {\rm d}^{\rm 4}x$, $\mu$ is the chemical potential, $\psi$ denotes the quark field with $\Nf=2$ flavor and~$\Nc=3$ color degrees of freedom, and $\phi$ is a $\Nf^2$-component bosonic field $\phi^\tp = (\sigma, \vec{\pi}^\tp)$.
For convenience, we also introduced  the (super)field 
\begin{align} 
\Phi^\tp = (\psi^\tp, \psib, \phi^\tp)\,,
\label{eq:Phi}
\end{align}
which contains all fields of our low-energy effective theory. From a phenomenological standpoint, the bosonic particles can be considered quark composites which carry the quantum numbers of the $\sigma$-mode and the pions, respectively:
\begin{align}
\label{eq:fermion content of mesons}
\sigma \sim \psib \psi,\ \qquad \vec{\pi} \sim \psib \gamma^5 \vec{\tau} \psi\,,
\end{align}
where the matrices $\tau^i$ with $i \in \{1,2,3 \}$ denote the Pauli matrices. 
In other words, we understand the bosons of our model as effective degrees of freedom which describe multi-fermion interactions. 
In contrast to the quark fields, the bosonic fields do not carry a color charge. 
The quark-meson interaction channel in \labelcref{eq:action} is associated with a dimensionless Yukawa coupling $h$. 

As a consequence of their relations to the quark fields as given in \labelcref{eq:fermion content of mesons}, a chiral $SU(2)_L \times SU(2)_R$ transformation of the quark fields implies a corresponding transformation of the meson fields. 
More specifically, due to the isomorphism $SU(2)_L \times SU(2)_R \cong O(4)$, chiral transformations of the fermions induce rotations among the mesons. Note that the term linear in the $\sigma$ field explicitly breaks the $O(4)$ symmetry and thus allows us to tune the masses of the pions and the quarks. In fact, this term can be related to the standard current quark mass term  $\sim \psib m_c \psi$ by means of a Hubbard-Stratonovich transformation. In the limit $H \rightarrow 0$, the chiral symmetry of our quark-meson model is intact.

We close this section with an important remark that the classical action of the QM model in \labelcref{eq:action} exhibits an interesting shift symmetry: if we would add a fermionic mass term 
to the classical action, 
\begin{align}
\label{eq:action_with_massterm} 
& S[\Phi] + \iu \int_x \psib\, m_\psi\, \psi\,, 
\end{align}
this term can be absorbed by a shift of the $\sigma$ field,
\begin{align} 
	\sigma \to \sigma^{\prime}=\sigma + \frac{m_\psi}{h}\,,
\label{eq:ShiftSigma} 
\end{align}
in the path integral. 
This shift symmetry on the quantum level is sourced in the following transformation property of the classical action (associated with an absorption property of the parameter~$H$):
\begin{align}
\label{eq:ShiftSymmetry} 
& S[\Phi] + \iu \int_x \psib\, m_\psi\, \psi 
=  S^{(\chi)}[\Phi^\prime] - H^\prime \int_x \sigma^\prime 
+ \text{const.}
\end{align}
with~$H^{\prime}=H + \frac{m_\phi^2 m_\psi}{h}$.
The superscript in $S^{(\chi)}$ indicates the chiral limit, meaning that we consider the classical action with $H = m_\psi = 0$.
As the term $\sim H \sigma$ is nothing but a source term for the $\sigma$-mode, it does not change the dynamics of the theory which is that in the chiral limit.   
Indeed, the effective action $\Gamma[\Phi]$ of the theory \labelcref{eq:action} is simply given by 
\begin{align}
	\Gamma[\Phi]=\Gamma^{(\chi)}[\Phi] - H \int_x \sigma\,.
	\label{eq:Gammachi+H}
\end{align}
This is a standard identity which results from a redefinition of sources in the path integral description of a system with linear symmetry breaking.
The implications of the shift symmetry \labelcref{eq:ShiftSymmetry} have been discussed in detail in \cite{Braun:2022mgx} and will be furthered in 
\Cref{sec:ShiftfermionicCS}.

\subsection{Spectral regularization}
\label{sec:spec_reg}
For a calculation of the phase structure and spectral functions, we employ the fRG approach which allows for studies of non-perturbative phenomena. At the same time, it can be directly related to the standard loop expansion of the effective action~\cite{Litim:2001ky,Litim:2002xm}, see also~\cite{Geissel:2024nmx} for a recent discussion. 
This approach is based on the Wetterich equation~\cite{Wetterich:1992yh} for the scale-dependent quantum effective action~$\Gamma_k$:
\begin{align}
\label{eq:Wetterich}
\dt \Gamma_k[\Phi] = \frac{1}{2} \STr\left\lbrace \dt R_k \left(\Gamma^{(2)}_k[\Phi] + R_k\right)^{-1}\right\rbrace\,.
\end{align}
Here, $t=\ln(k/\Lambda_0)$ is the so-called RG time, $k$ stands for the RG scale, and $\Lambda_0$ denotes some reference scale at which the initial condition~$\Gamma_{k=\Lambda_0}$ for the scale-dependent effective action~$\Gamma_k$ is fixed. 
Note that the supertrace $\STr$ not only contains the trace over all internal indices and momenta but also includes the field metric, which provides a minus sign for the fermionic subspace.

The actual computation of the effective action requires the specification of the regulator function $R_k$. In general, the structure of this function in field space is given by
\begin{align}
R_k(p,q) = \begin{pmatrix}
0 & -R^\psi_k(-p)^\tp & 0\\
R^\psi_k(p) & 0 & 0\\
0 & 0 & R^\phi_k(p)
\end{pmatrix} (2\pi)^4 \delta^{(4)}(p-q) .
\label{eq:general regulator}
\end{align}
In the infrared limit, $k \to 0$, we require that~$R_k\to 0$ such that we arrive at the full quantum effective action~$\Gamma \equiv \Gamma_0$.
Ideally, the regulator should be chosen such that it respects the analytic properties as well as all the symmetries of the theory under consideration. 
However, if an explicit breaking of a symmetry cannot be avoided in practice, then it is principally required to restore this symmetry throughout the entire RG flow. We emphasize that a regulator-induced breaking of symmetry affects every RG step such that the infrared endpoint of the RG trajectory does not align with the true result, although the regulator is explicitly removed for $k \to 0$. A symmetry restoration at every scale then ensures that physical observables are not spoiled by such an artificial symmetry breaking. In case of gauge theories, for example, this can be ensured by modified Ward-Takahashi identities, see, e.g., \cite{Pawlowski:2005xe, Gies:2006wv, Dupuis:2020fhh} for reviews. We shall come back to this issue below.

At this stage it is worth mentioning that the term linear in the $\sigma$ field in \labelcref{eq:action} belongs to the definition of the model and represents a physical source of explicit symmetry breaking. 
Therefore, loosely speaking, all efforts to restore the chiral symmetry should only be targeted at unphysical sources such as the regulator. 
Since the symmetry breaking term in our quark-meson model is linear in the sigma field, it drops out of the RG flow of the effective average action. 
In fact, the right-hand side of the Wetterich equation only depends on the second functional derivative of the effective action, see \labelcref{eq:Wetterich}. 
In other words, the information about physical explicit symmetry breaking does not enter the RG flow. 
In practice, this means that results for the case of finite explicit chiral symmetry breaking, $H >0$, can simply be obtained by adding a linear tilt to the effective action which has been computed in the limit~$H \to 0$.

In the present work, we employ a standard mass-like CS regulator which can be parametrized as
\begin{align}
& R^\psi_{k_\psi}(p)  = Z^\psi_{k_\psi}\, \iu k_\psi\quad\text{and}\quad
R^\phi_{k_\phi}(p)  = Z^\phi_{k_\phi}\, k_\phi^2\,,
\label{eq:CSregsFerm+Bos} 
\end{align}
for fermions and bosons respectively.  
Here, we have distinguished the fermionic from the bosonic cutoff scales but we shall keep in mind that the common choice is 
\begin{align}
k_\psi=k_\phi=k\,.
\label{eq:Commonk}
\end{align}
In general, this class of regulators provides an IR regularization as it introduces an explicit mass in the propagators of the quantum fields. 
We would like to highlight that the CS regulators have no momentum dependence in contradistinction to other canonically used regulator functions. 
As a consequence, our regularization scheme does not implement a Wilsonian-type momentum-shell integration for which the flow of $\Gamma_k$ would be dominated by fluctuations with momenta~$p^2\simeq k^2$. 
Instead, the CS flow is one, connecting theories with different masses. 
All momentum shells are already integrated out at every scale $k$. 
Because of this property, the initial condition for the Wetterich equation at a given mass scale~$k=\Lambda_0$ is the full effective action of this theory and should not be confused with a UV action in the Wilsonian sense.

It has been discussed in \cite{Ihssen:2023xlp} that a relative scaling between $k_\psi$ and $k_\phi$ may reduce the momentum transfer in the loops. Indeed, for Wilsonian regulators with an exponential momentum decay for large momenta it has been found that optimized flows require $k_\phi \propto \Lambda( k/\Lambda)^{\gamma}$ with $\gamma \approx 0.4 - 0.6$. Hence, a relative scaling may be more adapted to the derivative expansion, which is local in momentum space. Moreover, keeping the cutoff scales different allows us to discuss properties of the CS regulators  that are specific to fermions (or bosons).
 
In any case, in our explicit computations of the phase diagram and mesonic two-point functions of the quark-meson model in the large-$\Nc$ limit we only take into account purely fermionic loops. Hence, we only deal with the fermionic CS regulator and identify~$k_\psi$ with~$k$. 
From the definition of the fermionic regulator $R^\psi_k$, we deduce that it only affects the $\sigma$-direction of the meson field space but leaves the pion directions unaltered.
Since the ground state of the system is always aligned with the $\sigma$-direction by convention, the additional shift of the order of $k$ makes the explicit chiral symmetry breaking apparent.
Without applying a procedure that removes this artificial breaking of the chiral symmetry, the regulator-induced symmetry breaking will significantly contaminate results for physical observables.
We shall address this in more detail in \Cref{sec:sym_constr_CS,sec:res}. 

As detailed in~\cite{Braun:2022mgx}, the standard CS regulator is advantageous for several reasons.
It preserves the analytic properties of the theory and yields a propagator with a simple pole structure in momentum space since it does not generate additional poles in the complex plane associated with the time-like component of the four-momentum. 
This property ensures a spectral representation of the regularized propagator, which is of great relevance for the computation of real-time correlation functions. Furthermore, the CS regulator does not interfere with a multitude of symmetries of the theory. 
Specifically, the regulator not only respects Lorentz symmetry in the vacuum limit but also the Silver-Blaze symmetry in the presence of a finite chemical potential. These advantages come at a twofold cost: 

First of all, it only lowers the degree of the UV divergences by two, which is not always sufficient to render all loop integrals in the RG flow finite. Hence, in general an additional UV regularization is required, which has been evaluated thoroughly in \cite{Braun:2022mgx}. 

Secondly, the flow with the fermionic mass changes the amount of explicit chiral symmetry breaking in the systems and this flow must be taken into account accurately as both, explicit and dynamical chiral symmetry breaking are at the heart of the dynamics of QCD. 

\subsection{Yukawa theories, shift symmetries and the fermionic CS regulator} 
\label{sec:ShiftfermionicCS}
As indicated above, we shall only consider the large-$\Nc$ limit in our explicit computations of the phase diagram and the mesonic two-point functions of the quark-meson model. As an addition, however, we would like to discuss some important properties of the CS flow of the {\it full} theory in the following which are relevant in studies beyond the large-$\Nc$ limit. To this end, we shall take up the detailed discussion in \cite{Braun:2022mgx} and go beyond it. While in the latter work also combinations of chiral CS-type regulators have been discussed, we will concentrate on the (scalar) fermionic mass regulator \labelcref{eq:CSregsFerm+Bos} here. Still, the following 
arguments straightforwardly generalize to vector-type Yukawa models as also discussed in \cite{Braun:2022mgx}, as these models also carry a shift symmetry in the vector field.\footnote{Chiral CS-type regulators have been considered in \cite{Otto:2022jzl} for mapping out the phase structure of the QM model in the absence of vector modes. It would be interesting to augment this model with vector modes and to apply the shift-symmetry arguments in~\cite{Braun:2022mgx} and the further advances in the present section to this case.}

As stated above, our CS regulators provide an IR regularization as they introduce an explicit mass in the propagators of the quantum fields which entails that these regulators change the mass parameters of the theory.
Using the shift symmetry \labelcref{eq:ShiftSymmetry} of Yukawa-type theories, the regularized action in the {\it full} path integral can be written by absorbing the fermonic regulator in a term linear in the $\sigma$ field up to field-independent terms: 
\begin{align}
	S[\Phi] &+ \int_x \Big[\psib\, \iu k_\psi\, \psi + \frac{1}{2} \phi^\tp\, k^2_\phi\, \phi\Big] \nn\\
	\to&\, \int_x \Big[ \vphantom{\frac{1}{2}}\psib (\iu \dslash-\iu \gamma^0 \mu ) \psi + \iu h\psib \left(\sigma + \iu \gamma^5 \vec{\tau}\cdot \vec{\pi}  \right) \psi \nn\\
	&\hspace{.3cm}+\frac{1}{2} \phi^\tp \left(- \partial^2 + m^2_{\phi,k_\phi}\right) \phi - H_{k_\phi, k_\psi}\,\sigma \Big]\ \,,
	\label{eq:Action+DeltaS}
\end{align}
with 
\begin{align} 
	m_{\phi,k_\phi}^2 = m_{\phi}^2+k_\phi^2\,,\qquad H_{k_\phi,k_\psi} = H + \frac{m_{\phi,k_\phi}^2 k_\psi}{h}\,. 
\label{eq:kdepMasses}
\end{align}	
The property required for a Yukawa theory in order to exhibit the shift symmetry  \labelcref{eq:ShiftSymmetry} is a scalar coupling of the fermions to the scalar field and the absence of a scalar self-interaction. 
Clearly, this condition is satisfied in the Yukawa model discussed here, leading to \labelcref{eq:Action+DeltaS}. 
Note that this property also holds true for the fRG approach to QCD with emergent composites, and hence the present analysis extends also to QCD, see \cite{Gies:2002hq, Braun:2008pi, Mitter:2014wpa, Braun:2014ata, Cyrol:2017ewj, Fu:2019hdw, Ihssen:2024miv}. 

Considering the classical action in \labelcref{eq:Action+DeltaS} with an additional standard fermionic mass term, see \labelcref{eq:action_with_massterm}, we observe that we can absorb this mass into the fermionic CS regulator by $k_\psi +m_\psi \to k_\psi $. Then, the fermionic cutoff simply constitutes the full current quark mass of the model. 
Hence, in the present two-flavor low-energy effective theory in the isospin-symmetric limit with identical up and down current quark masses, $m_u=m_d =m_l$, we may simply stop the flow at the infrared cutoff value $k_{\psi,\textrm{min}}= m_l$ with the light current quark mass $m_l$. Then,
the entire (explicit) chiral symmetry breaking is now carried by the fermionic cutoff. 
This has several important consequences:
First of all, it follows from \labelcref{eq:Gammachi+H,eq:kdepMasses} that the fermionic CS flow is given by 
\begin{align}
k_\psi	\partial_{k_\psi} \Gamma_{k_\phi,k_\psi}[\Phi] &= - k_\psi	\partial_{k_\psi} H_{k_\phi, k_\psi} \int_x \sigma \nn\\
&=  -\frac{m_{\phi,k_\phi}^2 k_\psi}{h} \int_x \sigma\,.   
\label{eq:ChiralSymBreakingFlow} 
\end{align}
Note that we have dropped scale-dependent yet field-independent terms. 
We emphasize that~\labelcref{eq:ChiralSymBreakingFlow} is a subtle equation. It only holds true for the {\it full} path integral and hence requires the presence of mesonic fluctuations. 
Indeed, with the identification \labelcref{eq:Commonk}, the full flow reads 
\begin{align}
	\dt \Gamma_k[\Phi] = \frac{1}{2} \Tr\left[\frac{k^2 }{\Gamma^{(2)}_k[\Phi] + R_k}\right]_{\phi\phi} - \frac{(m_{\phi}^2+3 k^2) k}{h} \int_x \sigma\,.
\label{eq:ShiftSymmetricCSFlow}
\end{align}
Here, the trace only sums over the mesonic degrees of freedom, as indicated by the subscript of the propagator on the right-hand side. In \labelcref{eq:ShiftSymmetricCSFlow} we have left out the renormalization terms derived in \cite{Braun:2022mgx} that are required to make the flow finite, and made all $k$-dependences explicit. 

Here, an important comment is in order: \Cref{eq:ShiftSymmetricCSFlow} can be viewed as a novel flow equation for CS flows derived from the Wetterich equation. An immediate consequence of \labelcref{eq:ChiralSymBreakingFlow} and~\labelcref{eq:ShiftSymmetricCSFlow} is that the bosonic contributions are fully determined by the scale dependence of the fermionic contribution to the full path integral. While this may appear as a purely conceptional statement with no practical consequences at first glance, it indeed provides powerful constraints for truncations taking into account bosonic fluctuations as they have to satisfy \labelcref{eq:ChiralSymBreakingFlow} and \labelcref{eq:ShiftSymmetricCSFlow}. This entails that the quality of any results in the full system and in particular the realization of the inherent symmetries depend crucially on the judicious implementation and control of the fermionic contributions to the effective action.  
Of specific importance in the present system is the chiral symmetry: any calculation of loop integrals with the fermionic CS regulator must accommodate quantitatively the change of explicit and dynamical chiral symmetry breaking within the flow towards smaller current quark masses. Consequently, the development of a suitable framework for such computations of fermionic contributions to the full path integral via chiral Ward-Takahashi identities is of high revelance and is one of the main aspects of the present work.

Our setup is completed by the following interesting relation that has been observed and used in \cite{Gao:2021vsf}. The equation of motion (EoM) for the $\sigma$-mode is defined as
\begin{align}
\left. \frac{\delta \Gamma_{k_\phi,k_\psi}[\Phi]}{\delta \sigma(x)} \right|_{\sigma = \sigma_\textrm{\tiny{EoM}}}= 0\,,
	\label{eq:eom_def}
\end{align}
which, as a consequence of \labelcref{eq:Gammachi+H}, leads to
\begin{align} 
	\frac{\delta \Gamma_{k_\phi}^{(\chi)}[\Phi]}{\delta \sigma(x)} \Bigg|_{\sigma = \sigma_\textrm{\tiny{EoM}}}= \frac{m_{\phi,k_\phi}^2 k_\psi}{h}\,.
	\label{eq:FullEoM}
\end{align}
Assuming now that~$\sigma_{\text{EoM}}$ is homogeneous, an integration of \Cref{eq:FullEoM} allows us to compute the effective potential in the chiral limit up to a spacetime volume factor~\cite{Gao:2021vsf}:
\begin{align} 
	U_{k_\phi}^{(\chi)}(\sigma) =\int_0^{k_{\psi}(\sigma)} {\rm d}k_{\psi} \frac{{\rm d}\sigma_{\text{EoM}}}{{\rm d} k_{\psi} }\frac{m_{\phi,k_\phi}^2k_\psi}{h}\,,
\label{eq:Vchi}
\end{align}	
where $\sigma$ now represents constant meson fields and the quarks are evaluated at zero. Note that we have kept the mesonic cutoff $k_\phi$ fixed and used \labelcref{eq:FullEoM} twofold: \textit{(i)} We solve it for the solution $\sigma_\textrm{EoM}(k_\psi)$ as a function of the fermionic RG scale. \textit{(ii)} We solve it for the solution $k_\psi(\sigma)$ for a given solution $\sigma$ of the EoM.

For example, \labelcref{eq:Vchi} can be used to extract the full effective potential from the quark-mass dependence of the chiral condensate 
as done in \cite{Gao:2021vsf}. It also can be used to compute the full field dependence  of the effective potential from that obtained in a Taylor expansion about the minimum, see \cite{Braun:2023qak}. Moreover, if we initialize the integration at $k_\psi\to \infty$ with a trivial initial effective action, \labelcref{eq:Vchi} provides us with yet another representation of the flow equation. In this case, the renormalization or rather the large-mass limit has to be taken with care. This idea has been pursued in functional QCD studies with Dyson-Schwinger equations for thermodynamic properties in~\cite{Isserstedt:2020qll}. 

We close our general analysis of CS flows by adding that~\labelcref{eq:ChiralSymBreakingFlow,eq:Vchi} illustrate very impressively that a change in $k_\psi$ is nothing but a change of the amount of chiral symmetry breaking. 
Moreover, it provides a remarkable perspective on the fermionic part of the flow, whose consequences will be discussed elsewhere.

\section{Symmetry-constrained Callan-Symanzik regularization}
\label{sec:sym_constr_CS}

In the present section we introduce and discuss the chiral Ward-Takahashi identities that are chiefly important for the CS setup. To that end, we compute the effective action of our quark-meson model in a one-loop approximation in the large-$\Nc$ with the fermionic CS regulator as introduced in \Cref{sec:spec_reg}. 
To be more specific, we focus on the computation of the effective potential and the two-point functions of the sigma mode and the pions.
Note that the wavefunction renormalization of the quarks and the vertex renormalization associated with the quark-meson interaction are constant in the large-$N_{\rm c}$ approximation, i.e., they do not depend on the RG scale. Only purely fermionic loops are taken into account in this limit whereas boson fluctuation effects are neglected. 
However, we emphasize that this setup is adequate and sufficient to discuss the issue associated with the regulator-induced breaking of the chiral symmetry as well as to demonstrate how it can be systematically cured with the aid of chiral Ward-Takahashi identities. Beyond our results for phenomenologically relevant quantities, this aspect is essential for future CS computations of the effective action with meson fluctuation effects as indicated in \Cref{sec:ShiftfermionicCS}.

\subsection{Effective potential}
By expanding the meson fields about a homogeneous background, we arrive at the following result for the scale-dependent effective action:
\begin{align}
\label{eq:mf result for Gamma}
\frac{1}{V_4}\Gamma_k(\phi) = \frac{1}{V_4}\Gamma_{\Lambda_0}(\phi)  - 8 \Nc L_k(\Lambda_0, \phi)\ ,
\end{align}
which is equivalent to the effective potential $U_k(\phi)$. Here, $V_4$ is the spacetime volume and $\Lambda_0$ denotes a reference scale at which the RG flow of the effective action is initialized. 
The initial condition $\Gamma_{\Lambda_0}$ is usually used to tune the results for a given set of low-energy observables such that they assume their physical values in the vacuum limit.
We shall discuss the determination of $\Gamma_{\Lambda_0}$ in more detail in \Cref{subsec: UV ansatz} when we consider concrete applications. 
The auxiliary function $L_k$ parameterizes the properly regularized loop integral
\begin{align}
\label{eq:loop function CS full}
L_k(\Lambda_0, \phi) = L^{(\text{CS})}_k(\Lambda_0, \phi) + \ct_k(\Lambda_0, \phi)
\end{align}
{\allowdisplaybreaks
with
\begin{widetext}
\begin{align}
\label{eq:loop function CS}
L^{(\text{CS})}_k(\Lambda_0, \phi) 
&=\frac{1}{2} \int_{\vec{p}}\ \frac{1}{\beta} \sum_{n \in \mathbb{Z}} \ln\left( \tilde{p}^2_n + h^2 \vec{\pi}^2 + (h \sigma + k^\prime)^2\right)\Big|^{k^\prime = k}_{k^\prime = \Lambda_0}  \\
&= \frac{1}{(2\pi)^2} \int_0^\infty \d p\ p^2 \left. \left\lbrace \sqrt{p^2 + h^2 \vec{\pi}^2 + (h \sigma +k^\prime)^2}+ \frac{1}{\beta} \sum_{\pm} \ln\left(1 + \e^{-\beta (\sqrt{p^2 + h^2 \vec{\pi}^2 + (h \sigma +k^\prime)^2} \pm \mu)} \right)\right\rbrace \right|^{k^\prime = k}_{k^\prime = \Lambda_0}\ .\nn
\end{align}
\end{widetext}
Here, we} introduced the inverse temperature $\beta = 1/T$ and the generalized four-momentum $\pt^\tp_n = (\nu_n + \iu \mu, \vec{p}^{\,\tp})$ with the Matsubara frequency~$\nu_n=(2n+1)\pi T$.
The counterterm $\ct$ in \labelcref{eq:loop function CS full} depends on the scale~$\Lambda_0$ and is required to render the loop corrections ultraviolet finite since the CS regulator by itself does not provide enough UV regularization for the effective potential. 
These counterterms have been chosen such that they respect the symmetries of our model. For details, we refer to \Cref{app: counterterm UV reg}.

As discussed earlier, a suitable IR regularization is implemented through the CS regulator by construction.
This can be readily recognized from the appearance of the scale $k^\prime$ which acts like a fermion mass in the loop integral, see \labelcref{eq:loop function CS}.
In this regard, we emphasize that the scale~$\Lambda_0$ should {\it not} be confused with a UV momentum cutoff of the loop integral.
It represents a mass-like scale associated with the CS regulator and enters the loop integral because the RG flow is initialized at $k^{\prime}=\Lambda_0$.
Therefore, this scale in general also appears in the counterterms. 

The expression in \labelcref{eq:loop function CS} shows that the chiral symmetry is broken explicitly by the presence of the CS regulator since the field content of the loop function is not invariant under $O(4)$ transformations. 
Specifically, an expansion of the loop function in the meson fields shows that the regulator induces not only terms linear in the sigma field but terms of arbitrarily high order in that field.
In particular, terms with odd powers of the sigma field are generated, spoiling the $O(4)$ symmetry of the effective action. 
Nevertheless, it also becomes clear from \labelcref{eq:loop function CS} that the $O(3)$ symmetry in the pion subspace is still intact. 

Let us now discuss how the regulator-induced breaking of the chiral symmetry can be removed in a controlled and systematic fashion. 
To this end, we would like to note that the Lie group $O(N)$ and its subgroup $SO(N)$ have the same generators. 
Thus, an $O(N)$-invariant quantity is also invariant under $SO(N)$ transformations and vice versa. 
Let us now take a look at infinitesimal $SO(4)$ transformations in the mesonic subspace. 
Intact chiral symmetry implies that all mesonic quantities should be invariant with respect to such rotations, i.e.,
\begin{align}
\CO_k^{\supsc}[\phi] = \CO_k^{\supsc}[(1+ \alpha^j X_j) \phi]\ ,
\end{align}
where $\alpha^j \in \mathbb{R}\; (j \in \{1,\dots,6\})$ is the $j$-th rotation angle and $X_j$ is the corresponding generator of~$SO(4)$. This implies that
\begin{align}
\label{eq:SO(4) WTI}
 \frac{\delta \CO^{\supsc}_k[\phi]}{\delta \sigma}\ \pi_i - \frac{\delta \CO^{\supsc}_k[\phi]}{\delta \pi_i}\ \sigma = 0\,
\end{align}
for $i \in \{1,2,3\}$. 
We can consider this relation a Ward-Takahashi identity (WTI) which describes the invariance under rotations in the mesonic subspace. 
Let us now consider a quantity~$\CO_k$, which is not invariant with respect to $O(N)$ transformations of the meson fields.
For example, such a quantity may have been obtained from the effective potential  \labelcref{eq:mf result for Gamma} as generated by a CS flow.
In order to build a correspondingly symmetric quantity from $\CO_k$, we make the ansatz
\begin{align}
\label{eq:symmetry ansatz}
 \CO_k^{\supsc}[\phi]  = \CO_k[\phi] + c_k[\phi]\ .
\end{align}
In other words, we assume that any quantity, which is invariant with respect to $O(N)$ transformations, can be suitably decomposed into two asymmetric parts.
The remaining task now is to construct the additional term $c_k$ in accordance with the WTI in \labelcref{eq:SO(4) WTI}. 
We emphasize that this symmetrization procedure should not alter the dependence of $\CO_k$ on the pion fields since this subspace is not affected by the CS regulator. 
It then follows that \labelcref{eq:SO(4) WTI} has to be solved for $c_k$ with the initial condition
\begin{align}
\qquad c_k[\sigma=0,\vec{\pi}] = 0
\end{align}
for all~$k$.
Note that our ansatz implies that $c_k$ can be considered a counterterm for the contributions in $\CO_k$ which explicitly break the rotation symmetry. 
In order to ensure that the regulator-induced symmetry breaking is removed even in the presence of external parameters, e.g., temperature
or chemical potential, the counterterm must in general also carry a corresponding dependence on such parameters.

For convenience, let us now introduce the symmetrization operator $\CS$ such that
\begin{align}
\label{eq:symmetrization}
\CO_k^{\supsc} \equiv \CS \CO_k \coloneqq \CO_k + c_k\,.
\end{align}
We shall employ this operator for our studies below. 
Throughout this work, it is sufficient to consider the symmetrization of functions rather than functionals such that a generic solution for $c_k$ can be readily given. 
For details, we refer the reader to \Cref{app: symmetrization}.

Next, we apply our considerations to the loop integral in \labelcref{eq:mf result for Gamma} to remove the regulator-induced breaking of the chiral symmetry. 
This eventually leads us to a physically meaningful expression with an $O(N)$-symmetric loop:
\begin{align}
\frac{1}{V_4}\Gamma^{(\text{phys})}_k(\phi)  = \frac{1}{V_4}\Gamma_{\Lambda_0}(\phi) - 8 \Nc L^{\supsc}_k(\Lambda_0, h^2 \phi^2)\, ,
\end{align}
where $L^{\supsc}_k=\CS L^{(\text{CS})}_k$ is given by
\begin{widetext}
\begin{align}
\left(\CS L^{(\text{CS})}_k\right)(\Lambda_0, \rchi) 
&= \frac{1}{(2\pi)^2} \int_0^\infty \d p\ p^2 \left. \left\lbrace \sqrt{p^2 + \rchi + k^{\prime\,2}}+ \frac{1}{\beta} \sum_{\pm} \ln\left(1 + \e^{-\beta (\sqrt{p^2 + \rchi + k^{\prime\,2}} \pm \mu)} \right)\right\rbrace \right|^{k^\prime =k}_{k^\prime = \Lambda_0}\ .
\end{align}
\end{widetext}
A comparison with \labelcref{eq:loop function CS} shows that the symmetrization procedure has successfully removed the symmetry breaking terms $\sim \sigma k$ from the loop integral. 
We add that the term~$\Gamma_{\Lambda_0}$ can still contain a term which is linear in the sigma field, controlling the amount of physical explicit symmetry breaking.
Otherwise, the initial condition to the Wetterich equation is invariant with respect to $O(N)$ transformations of the meson fields, see our discussion above.

\subsection{Euclidean two-point functions}
Correlation functions can be obtained from an expansion of the effective action in terms of fluctuation fields $\Phi^{\vphantom{\tp}}_{\text{fl}}(x) = \Phi(x) - \Phi_0$ about $\Phi^{\vphantom{\tp}}_{\text{fl}}(x) = 0$, where~$\Phi_0$ denotes a constant background. 
In our truncation scheme, the expansion of the effective action in momentum space reads
\begin{align}
\label{eq:vertex expansion}
\Gamma_k[\phi] \sim \frac{1}{2} \int_P \int_Q \phifl^\tp(-P)\ \Gamma^{(2)}_k(P,Q)\ \phifl^{\vphantom{\tp}}(Q) + \dots\,,
\end{align}
where we dropped all terms which are irrelevant for the two-point function of the mesons. 
The latter is given by
\begin{align}
\label{eq:two-point projection}
\Gamma^{(2)}_k(P,Q) &= \Gammat^{(2)}_k(Q)\ \dmom{P+Q} \nn \\
 &=\left.\left(\frac{\delta}{\delta \phifl^\tp(-P)} \frac{\delta}{\delta \phifl^{\vphantom{\tp}}(Q)} \Gamma_k[\phi]\right)\right|_{\phifl^{\vphantom{\tp}}=0}
\end{align}
with
%
\begin{align}
\label{eq:full two-point}
\Gammat^{(2)}_k(Q) 
&= \Gammat^{(2)}_{\Lambda_0}(Q) +  \int^k_{\Lambda_0} \frac{\d k^\prime}{k^\prime} \partial_{t^\prime} \Gammat^{(2)(\text{CS})}_{k^\prime}(Q) \nn\\ 
& \qquad - 8 \Nc \left.\left( \frac{\partial}{\partial \phi^\tp}\frac{\partial}{\partial \phi} \ct_k(\Lambda_0,\phi)\right)\right|_{\phi = \phi_0}
\end{align}
%
and
\begin{align}
\label{eq:CS flow of two-point}
&\dt \Gammat^{(2)(\text{CS})}_k(Q) = - 8 h^2 \Nc\ \times \\
&\times\!\! \int_{\vec{p}} \frac{1}{\beta}\sum_{n \in \mathbb{Z}} \dt \frac{\pt_n^\tp(\pt_n + Q) \uM{4} -(m_q + k)^2\, \eta}{\left(\pt_n^2 + (m_q + k)^2 \right) \left((\pt_n+Q)^2 + (m_q + k)^2 \right)}\,.
\nn 
\end{align}
Here, we have introduced the matrix $\eta = \text{diag}(1,-1,-1,-1)$ for convenience.
The ground state of the effective potential is given by $\phi_0^\tp = (\sigma_0, \vec{0}^{\,\tp})$ which determines the dynamical quark mass as $m_q = h |\sigma_0|$.\footnote{
For a system with a physical source of explicit symmetry breaking, the ground state is always finite, i.e., $\sigma_0 >0$.
However, when using the CS regulator, the global minimum of the effective potential may in fact lie at negative values in field space.
}
Since we are only interested in a one-loop calculation in our present exploratory study with the CS regulator, we use the value of~$m_q$ at $k=0$ when we evaluate the two-point function.
From the action of our quark-meson model, we would expect that the meson two-point functions agree identically in the limit $m_q \to 0$. 
However, because of the presence of the CS regulator, this is clearly not the case here.
We would like to stress that the CS regulator also affects the results for~$m_q \neq 0$.
In fact, the CS regulator always gives rise to a unphysical chiral symmetry breaking.

As in the case of the effective potential, we can systematically remove the regulator-induced breaking of the chiral symmetry to obtain meson two-point functions which respect the global $O(4)$ symmetry.
To this end, we begin with a general analysis of the structure of the effective action.
Intact chiral symmetry implies, that there exists a $\hat{\Gamma}$ such that 
\begin{align}
\Gamma^{(\text{phys})}_k[\phi] = \hat{\Gamma}_k[\phi^2]\,.
\end{align}
Considering for a moment a generalized homogeneous meson background field,
\begin{align}
\varphib^\tp = (\sigmab, \vec{\pib}^\tp)\,,
\end{align}
we find that the generalized two-point function can be cast into the form
\begin{align}
\Gamma_{k,\text{gen}}^{(\text{phys})(2)}(\varphib) &= \left.\left(\frac{\delta}{\delta \phi^\tp} \frac{\delta}{\delta \phi} \Gamma_k^{(\text{phys})}[\phi] \right)\right|_{\phi = \varphib}\nn\\
&= A_k(\varphib^2)\ \uM{4} + B_k(\varphib^2)\ \varphib \varphib^\tp\, ,
\label{eq:chiral two-point form}
\end{align}
where we have dropped any arguments indicating dependences on external momenta for convenience. The coefficients $A_k$ and~$B_k$ are given by
\begin{align}
A_k(\varphib^2) &= 2 \left.\left(\frac{\delta}{\delta \Theta} \hat{\Gamma}_k[\Theta]\right)\right|_{\Theta = \varphib^2}\ ,\\
B_k(\varphib^2) &= 4  \left.\left(\frac{\delta^2}{\delta \Theta^2} \hat{\Gamma}_k[\Theta]\right)\right|_{\Theta = \varphib^2}\, .
\end{align}
We observe that an intact chiral symmetry translates into an $O(4)$ symmetry among the field directions of the generalized background within the coefficients $A_k$ and $B_k$. 
Imposing chiral symmetry is then equivalent to requiring that the two-point function assumes the form  \labelcref{eq:chiral two-point form}. 
In order to derive an expression for the physical two-point function (i.e., the two-point function with removed regulator-induced chiral symmetry breaking), we compute the corresponding coefficients for our CS scheme,
{\allowdisplaybreaks
\begin{align}
B_k^{(\text{CS})}(\varphib) &= \frac{h}{(h \sigmab +k) \pib_{i}} \Gamma^{(2)}_{k,\sigma\pi_i}(\varphib)\ ,
\end{align}
\begin{align}
A_k^{(\text{CS})}(\varphib) &= \Gamma^{(2)}_{k,\sigma\sigma}(\varphib) - \frac{(h \sigmab +k)^2 }{h^2} B_k^{(\text{CS})}(\varphib) \nn \\
 &= \Gamma^{(2)}_{k,\pi_i\pi_i}(\varphib) - \pib^2_{i}\ B_k^{(\text{CS})}(\varphib)\ ,
\end{align}
and symmetrize} their dependence on the background according to \labelcref{eq:symmetrization},
\begin{align}
A_k &= \CS A_k^{(\text{CS})} , \quad B_k = \CS B_k^{(\text{CS})}\, .
\end{align}
From this, we can then reconstruct the generalized and symmetrized two-point function according to \labelcref{eq:chiral two-point form}. Finally, setting the generalized background equal to the ground state yields the symmetrized version of \labelcref{eq:full two-point}, 
\begin{align}
\Gamma_k^{(\text{phys})(2)} = \Gamma_{k, \text{gen}}^{(\text{phys})(2)}(\varphib=\phi_0)\ .
\end{align}
Concretely, we obtain
\begin{align}
\label{eq:symCS two-point}
\Gammat^{(\text{phys})(2)}_k(Q) 
&= \Gammat^{(2)}_{\Lambda_0}(Q) +  \int^k_{\Lambda_0} \frac{\d k^\prime}{k^\prime} \partial_{t^\prime} \Gammat^{(2)(\text{sym})}_{k^\prime}(Q) \nn \\
& - 8 \Nc \left.\left( \frac{\partial}{\partial \varphib^\tp}\frac{\partial}{\partial \varphib} \left( \CS \ct_k \right)(\Lambda_0,\varphib^2)\right)\right|_{\varphib = \phi_0}\ ,
\end{align}
where
\begin{align}
\label{eq:symCS flow of two-point}
&\dt \Gammat^{(2)(\text{sym})}_k(Q) = - 8 h^2 \Nc\ \times \\
&\times \int_{\vec{p}} \frac{1}{\beta}\sum_n \dt \frac{\left( \pt_n^\tp(\pt_n + Q)+ k^2 \right) \uM{4} -m^2_q\ \eta}{\left(\pt_n^2 + m^2_q + k^2 \right)\ \left((\pt_n+Q)^2 + m^2_q + k^2 \right)}\ .\nn 
\end{align}
Note that, because of the restoration of the global $O(4)$ symmetry, the two-point functions are now invariant under $m_q \rightarrow -m_q$. 

With respect to our results presented in \Cref{sec:res}, we would like to add that we always evaluate the Matsubara sum analytically as we also did for the effective potential. 
As we do not discuss the analytic form of the two-point functions in the following, we refrain from showing these expressions here. 
However, explicit expressions for the full two-point function \labelcref{eq:symCS two-point} at zero temperature but finite quark chemical potential together with related quantities can be found in \Cref{app: Explicit zero-temperature results} as we require them for our discussion below. 

\subsection{RG consistency}
Our formal expressions for the effective potential and the two-point functions depend on the scale~$\Lambda_0$.
At this scale, we fix the initial condition of the Wetterich equation.
In a first step, this is done by choosing a specific ansatz for~$\Gamma_{k=\Lambda_0}$.
Then, the values of the couplings at this scale are in general tuned such that physical values of a specific set of (low-energy) observables are recovered in the vacuum limit. 
To give an example, this set may contain pole masses of the particles under consideration.
Recall that the scale-dependent action~$\Gamma_k$ of our CS scheme should be considered a full quantum effective action of a theory, in which  the quark mass scales with~$k$.
In our analysis, we take $k=0$ as the point at which each observable is assigned its corresponding physical value.

We now assume that we have fixed the scale-dependent effective action~$\Gamma_k$ at some scale~$\Lambda_0$ in the vacuum limit. 
With this initial condition at hand, the Wetterich equation then defines a unique RG trajectory through a space of theories from the scale $k=\Lambda_0$ to the scale $k=0$. 
In our truncation scheme, the RG flow also allows us to determine~$\Gamma_k$ at scales~$k>\Lambda_0$.
It follows that we can initialize the CS flow at any scale~$k=\Lambda$ along this trajectory without changing the effective action at~$k=0$:
\begin{align}
\label{eq:RGc condition vac}
\Lambda \frac{\d}{\d \Lambda} \Gamma_{k=0}= 0\,.
\end{align}
This is the RG consistency condition which expresses the {\it invariance} of the physics described by the quantum effective action at~$k=0$ on our choice for the initialization scale~$\Lambda$.

In general, this invariance principle should also hold in the presence of finite external parameters $\mext$, e.g., temperature $T$ and quark chemical potential $\mu$, see~\cite{Braun:2018svj} for a detailed discussion.
Therefore, we will now show explicitly how to construct a general initial condition~$\Gamma_{\Lambda}$ in accordance with \labelcref{eq:RGc condition vac}.
To this end, we consider a generic RG flow of the effective action which can at least formally be integrated to obtain~$\Gamma_k$:
\begin{align}
\label{eq:generic flow}
\dt \Gamma_k(\mext) &= f_k(\mext) \nn\\\Leftrightarrow \quad \Gamma_k(\mext) &= \Gamma_{\Lambda_0} + \int^k_{\Lambda_0} \frac{\d k^\prime}{k^\prime} f_{k^\prime}(\mext)\,.
\end{align}
As mentioned above, $\Gamma_{\Lambda_0}$ denotes the initial condition of the Wetterich equation, which has been fixed at some reference values $\mextvac$ for the external parameters.
For our concrete studies of observables as presented in \Cref{sec:res}, we always consider $\mext^{(0)}$ to refer to the vacuum values of the external parameters.
To achieve an RG-consistent extension of our initial condition, we set $\mext = \mextvac$ and increase the initialization scale to values $\Lambda > \Lambda_0$ such that all external parameters remain much smaller than $\Lambda$, i.e., 
\begin{align}
\label{eq:RGc restriction on parameters}
T/\Lambda \ll1 \qquad  \text{and} \qquad  \mu/\Lambda \ll 1\ . 
\end{align}
To be concrete, we obtain:
\begin{align}
\label{eq:RGc Initial condition}
\Gamma_{\Lambda}= \Gamma_{\Lambda_0} + \int_{\Lambda_0}^\Lambda \frac{\d k}{k}\ f_k(\mextvac)\ .
\end{align}
At finite external parameters, the RG-consistent effective action then reads
\begin{align}
\label{eq:RGc Gamma}
\Gamma_k(\mext) = \Gamma_{\Lambda} + \int_{\Lambda}^k \frac{\d k^\prime}{k^\prime} f_{k^\prime}(\mext)\ .
\end{align}
As it should be, we recover \labelcref{eq:generic flow} in the limit~$\mext\to \mextvac$.
Since the value for $\Lambda$ has been chosen large compared to all external parameters of interest, the theory at scales~$k$ close to $\Lambda$ and beyond can effectively always be considered a vacuum theory. 
As a result, the RG consistency condition  \labelcref{eq:RGc condition vac} is fulfilled for all $\Lambda$ which are sufficiently large relative to the external parameters. 
Conversely, for a fixed value of $\Lambda$, RG consistency is ensured for all values of external parameters which satisfy \labelcref{eq:RGc restriction on parameters}.
In the following, we shall employ this procedure to minimize regulator artifacts that potentially emerge if the external parameters are increased to values which are comparable to the value of ~$\Lambda_0$, see also~\cite{Braun:2018svj} for a concrete discussion of various examples.

\section{Results}
\label{sec:res}
In this section we present our results for the curvature masses, two-point functions, and spectral functions of the sigma mode and the pions as obtained from our quark-meson model.
Moreover, we discuss the phase diagram in the light of our results for the two-point functions.
Since the focus of our present work lies on the discussion of a meaningful implementation of the CS regulator in theories with fermions, we shall illustrate in detail the necessity of symmetry constraints as presented in the previous section and also demonstrate the relevance of an RG-consistent construction of the effective action for studies at finite temperature and chemical potential. 

\subsection{Initial condition}\label{subsec: UV ansatz}
In order to properly assess the effects of symmetrization and also RG consistency on the unconstrained
CS scheme, we need to make sure that the various schemes are indeed comparable.  
To this end, we first make an ansatz for the initial condition $\Gamma_{\Lambda_0}$ in the symmetrized CS scheme and then adapt the initial conditions in the other schemes such that the physics in the vacuum limit is the same for all cases:
\begin{align}
\label{eq:UV ansatz}
\Gamma^{(\text{scheme})}_{\Lambda_0} &= \Gamma_{\Lambda_0} - 8 \Nc \left.\left[  L^{\supsc}_0 -  L^{(\text{scheme})}_0 \right]\right|_{T=\mu=0}\ .
\end{align}
Note that the vacuum physics in RG-consistent and symmetrized CS calculations is already identical to the one in the symmetrized only scheme. 
By construction, differences in the results from the various schemes can only appear in our finite-temperature or finite-density studies. 
Specifically, in case of the unconstrained  CS calculations, we shall see that the regulator-induced symmetry breaking interferes with effects induced by the external parameters~$T$ and~$\mu$. 

For the initial condition $\Gamma_{\Lambda_0}$ of the Wetterich equation we employ the following ansatz in case of our symmetrized CS studies:
\begin{align}
\label{eq:initial condition}
\Gamma_{\Lambda_0}[\Phi]  = \int_x &\Big[ \psib (\iu \dslash-\iu \gamma^0 \mu) \psi 
+  \iu h\psib \left(\sigma + \iu \gamma^5 \vec{\tau}\cdot \vec{\pi}  \right)\psi  \nn\\
&+ \frac{1}{2} \phi^\tp \left(- Z_{\Lambda_0} \partial^2 + m_{\Lambda_0}^2 \right) \phi - H \sigma \Big]\,.
\end{align}
We choose the wavefunction renormalizations for the scalar and pseudo-scalar mesons to be identical at the scale~$\Lambda_0$, i.e., $Z_{\sigma,\Lambda_0} = Z_{\pi,\Lambda_0} \equiv Z_{\Lambda_0} = 1$. 
For the initial condition corresponding to the effective potential~$U_{\Lambda_0}=\Gamma_{\Lambda_0}/V_4$ and the two-point functions, we then obtain
\begin{align}
\frac{1}{V_4}\Gamma_{\Lambda_0}(\phi) &= \frac{1}{2} m_{\Lambda_0}^2 \phi^2 - H \sigma\ ,\\
\Gammat^{(2)}_{\Lambda_0}(Q) &= (Z_{\Lambda_0} Q^2 + m_{\Lambda_0}^2) \uM{4}\ ,
\end{align} 
respectively. 
In our numerical studies below, the parameter $m_{\Lambda_0}^2$, the Yukawa coupling $h$ as well as the symmetry breaking parameter $H$ are tuned such that a given set of low-energy observables is recovered.

\subsection{Curvature masses and spectral functions}\label{subsec: study I}
In the following we compare our numerical results for the curvature masses and the spectral functions as obtained from different treatments of the CS flow. 
To this end, we set $Z_{k}=Z_{\Lambda_0} = 1$. 
Effects from scale-dependent wavefunction renormalization factors are then discussed in \Cref{sec:reneffects_po}. 

\subsubsection{Parameter fixing}\label{subsubsec: scale fixing I}
In our numerical studies, we shall always consider the physical number of color degrees of freedom, i.e., $\Nc=3$. 
For the initial scale, we choose $\Lambda_0 = 500 \MeV$. 
We can then fix the couplings in the initial action~$\Gamma_{\Lambda_0}$ such that we obtain specific values for the (constituent) quark mass and the pion decay constant at $T=\mu=H=0$.
In the context of our quark-meson model, these quantities are assumed to satisfy the following general relations:
\begin{align}
\label{eq:mc and fpi} 
m_q = h |\sigma_0|\ , \qquad f_\pi = |\sigma_0|\,. 
\end{align}
Again,~$\sigma_0$ is associated with the ground state of the system and is assumed to be homogeneous.
The specific values are chosen to be
\begin{align}
\hat{m}_q &= 265\MeV\quad\text{and}\quad \hat{f}_\pi = 90\MeV
\label{eq:values for m_c and f_pi}
\end{align}
for the quark mass and the pion decay constant in the chiral limit at $k=0$, respectively.
They determine the Yukawa coupling as $h = \hat{m}_{q}/\hat{f}_{\pi}$ and also the 
mass parameter~$m^2_{\Lambda_0}$:
\begin{align}
\label{eq:scale fixing of msUV}
m^2_{\Lambda_0} &= 16 \Nc\ h^2 \left.\left(\frac{\d}{\d \rchi} L^{\supsc}_0(\Lambda_0,\rchi)\right)\right|_{\rchi=\hat{m}_q^2}\\
&=\frac{h^2 \Nc}{2 \pi^2}
\left[ \hat{m}^2_q \ln\left( \frac{\hat{m}^2_q}{\Lambda_0^2}\right)\! -\!  (\hat{m}^2_q\!+\! \Lambda_0^2) \ln\left(1\!+\! \frac{\hat{m}^2_q}{\Lambda_0^2}\right)\right]\,. \nn
\end{align}
For the explicit symmetry breaking parameter~$H$ in the initial action~$\Gamma_{\Lambda_0}$, we choose 
\begin{align}
H/ \Lambda_0^3 \approx 0.0294\, , 
\label{eq:values for Lambda_0 and H}
\end{align}
which yields a pion pole mass of $\hat{m}_{\text{pole},\pi}\approx 138 \MeV$ in the vacuum limit.
In general, the pole mass is determined by
\begin{align}
\Gammat^{(2)}_{\pi} (\iu m_{\text{pole},\pi}, \vec{0}^{\,}) =0\,,
\end{align}
where~$\Gamma^{(2)}_{\pi}$ is the pion two-point function.
Note that our prescription for the determination of the model parameters simply corresponds to the standard formulation of renormalization conditions in perturbation theory. 
\begin{figure*}[t]
  \begin{minipage}[b]{0.49\linewidth}
    \includegraphics{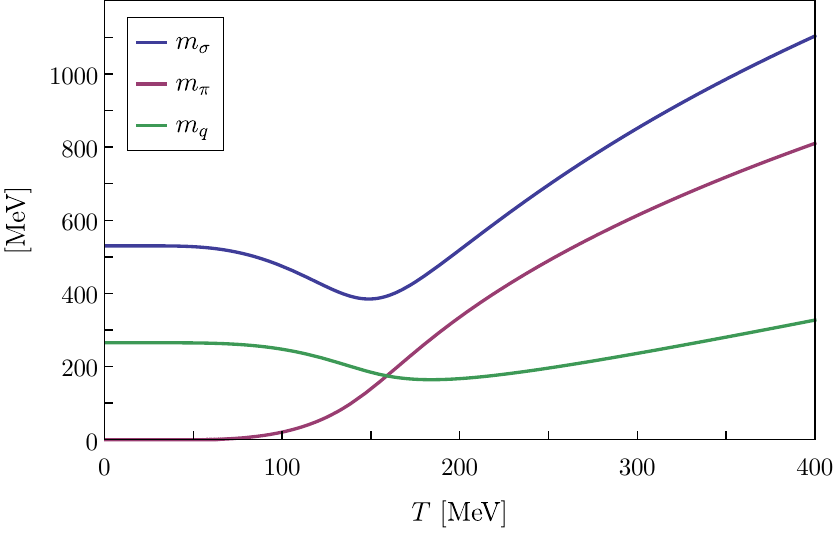}
  \end{minipage}
 \hfill%
 \begin{minipage}[b]{0.49\linewidth}
    \includegraphics{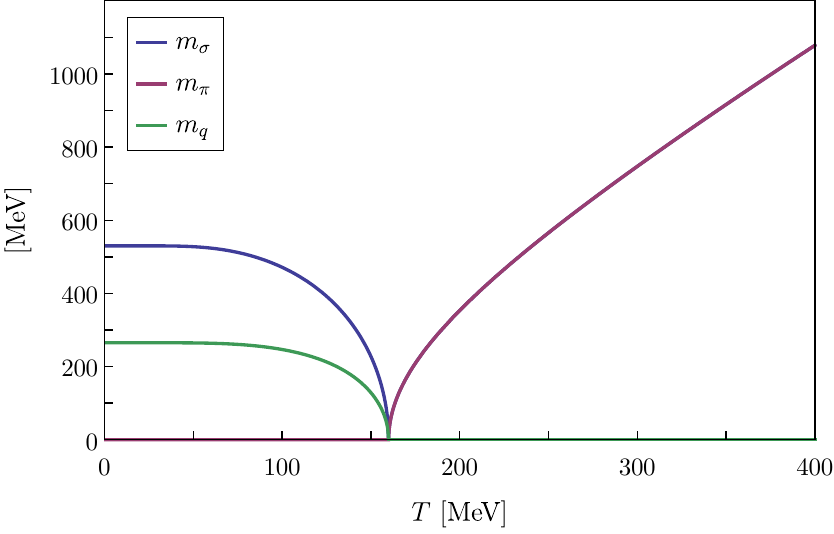}
  \end{minipage}
   \caption{Meson curvature masses and the quark mass as functions of temperature $T$ at $\mu =0$ in the chiral limit $H \rightarrow 0$ as obtained from the unconstrained CS scheme (left panel) and the RG-consistent and symmetrized CS scheme (right panel).}
  \label{fig:Mass_collection_in_different_schemes_for_chiral_limit}
\end{figure*}

We would like to add that our ansatz \labelcref{eq:initial condition} for the functional form of the effective action~$\Gamma_{\Lambda_0}$, which is only bilinear in the fields, is not bounded from below since we have~$m^2_{\Lambda_0}<0$ in \labelcref{eq:scale fixing of msUV}. 
In principle, we could also add a term~$\sim \lambda_{\phi}( \phi^\tp\phi)^2$ with~$\lambda_{\phi}>0$ to our ansatz,
rendering the effective action bounded from below at the initial RG scale~$\Lambda_0$.
In the following, however, we shall restrict ourselves to an initial action bilinear in the fields as given by \labelcref{eq:initial condition}. 
Solving the Wetterich equation with this initial condition, we then find that the $O(4)$-symmetrized effective action is bounded from below for all scales~$k<\Lambda_0$. 
Thus, in particular, in the limit~$k\to 0$, which is associated with physical values for observables in our setting, the effective action is bounded from below. 

Recall that, as discussed in \Cref{sec:spec_reg}, the effective action at the initial scale~$\Lambda_0$ should not be confused with a UV action describing the theory at some high momentum scale, as it would be the case for Wilsonian-type RG flows.\footnote{To illustrate the relation of CS flows and Wilsonian-type flows, it is instructive to consider a Wilsonian-type flow associated with an RG scale~$\tilde{k}$ and a CS flow with a scale~$k$. As a rule of thumb, the parameter $m^2_{\tilde{k}=\tilde{\Lambda}}$ in the Wilsonian setting may then be related to the CS result for the meson two-point function evaluated at~$k=0$ and~$Q^2\sim \tilde{\Lambda}^2$: $m^2_{\tilde{\Lambda}}\sim \Gamma^{(2)}_{k=0}(Q^2 \sim \tilde{\Lambda}^2)$. In accordance with this loose analogy, we indeed observe that the meson two-point functions are positive for sufficiently large momenta corresponding to~$m^2_{\tilde{\Lambda}}>0$, see also below.}
Within our CS setting, the scale-dependent effective action always describes an IR action in which the masses of the particles are increased when we increase the scale~$k$. 
From Wilsonian-type RG flows, it is well known that the fermion dynamics tend to shape the effective potential into one with a non-trivial minimum and thereby generate a finite ground state as more and more momenta are integrated out from the path integral.
In case of CS flows, all of those momenta are already integrated out for all $k$ such that our effective potential has a negative curvature at all RG scales.
Using an ansatz at $k = \Lambda_0$ which is bilinear in the fields thus necessarily renders the theory unstable at that scale.
However, an RG step towards any lower scale $k<\Lambda_0$ generates terms of higher order in the fields.
These quantum fluctuations then render the theory stable for $k<\Lambda_0$.
\subsubsection{Curvature masses}
Aiming at a phenomenological study of the QCD phase diagram, we start by computing the curvature masses of the sigma mode and the pions. 
These masses may then be used to pinpoint the crossover from the phase governed by chiral symmetry breaking to the chirally symmetric phase. 
The curvature masses can be extracted from the effective action as follows:\footnote{We define the curvature mass of a meson to be the static limit of the the corresponding two-point function which is identical to the curvature of the effective potential at the ground state, see~\cite{Helmboldt:2014iya} for a discussion.} 
\begin{align}
\label{eq:meson curvature masses projection}
m^{2}_{\phi_\nu} = \frac{1}{V_4} \left.\left( \frac{\partial^2}{\partial \phi^2_\nu} \Gamma_0(\phi)\right)\right|_{\phi=\phi_0}\,,
\end{align}
where $\nu \in \{0,1,2,3\}$ and $\phi^\tp_0=(\sigma_0, \vec{0}^{\,\tp})$. 
The value of~$\sigma_0$ can be obtained from a minimization of the effective action:
\begin{align}
\label{eq:ground state projection}
\left.\left(\frac{\partial}{\partial \sigma} \Gamma_0(\phi)\right)\right|_{\phi=\phi_0} \overset{!}{=}0\ .
\end{align}
Since observables are always computed at the ground state of the system, differences in the results for observables as obtained from our different CS schemes are strongly correlated with the differences in the behavior of the ground state as a function of the external parameters. 
With our symmetrization procedure at hand, we formally replace $\Gamma$ by $\Gamma^{(\text{phys})}$ in our symmetrized CS computations of observables.
\begin{figure*}[t]
  \begin{minipage}[b]{0.49\linewidth}
    \includegraphics{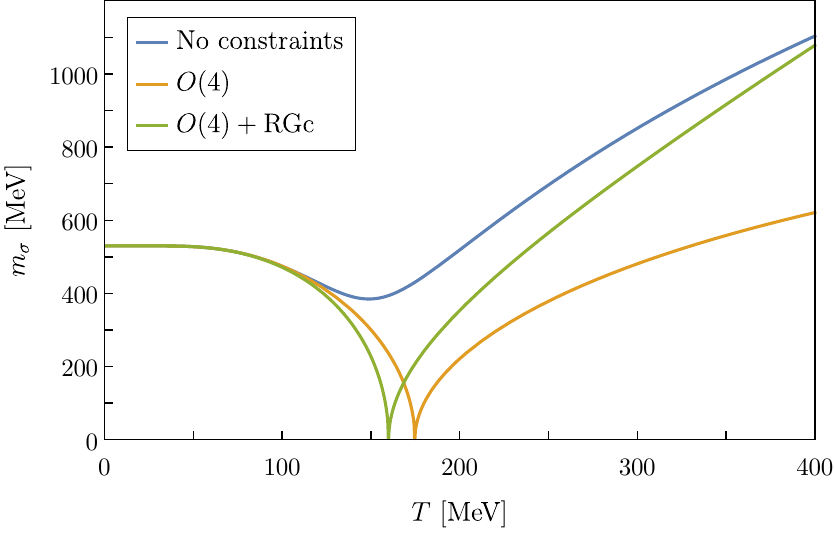}
  \end{minipage}
 \hfill%
 \begin{minipage}[b]{0.49\linewidth}
    \includegraphics{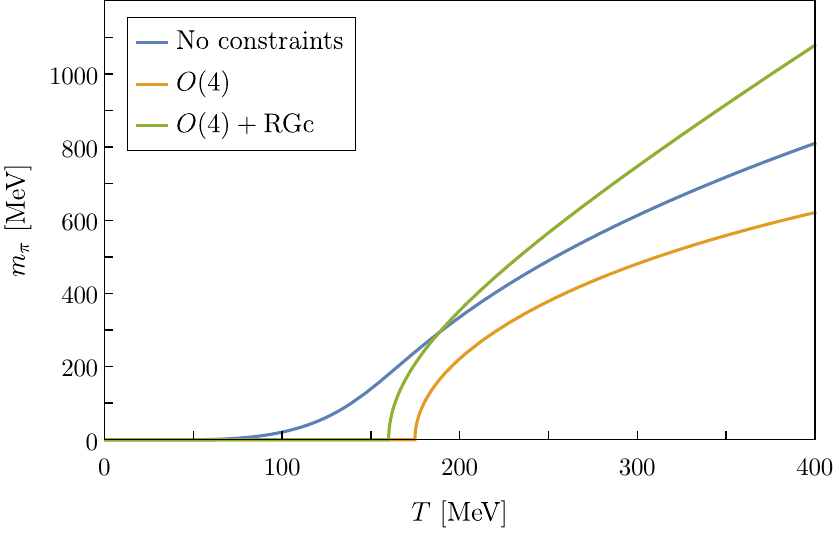}
  \end{minipage}
   \caption{
The curvature mass of the sigma mode (left panel) and the pions (right panel) as a function of temperature $T$ at $\mu=0$ in the chiral limit $H \rightarrow 0$.
The different colors correspond to our different CS schemes:
Whereas the symmetrized CS calculation (\glqq$O(4)$\grqq) and the symmetrized and RG-consistent calculation (\glqq$O(4)+\text{RGc}$\grqq) indicate that the system undergoes a second-order phase transition, the unconstrained CS calculation (\glqq No constraints\grqq) is plagued by strong regularization artifacts and appears to indicate that the pions are massive for all temperatures and that the system undergoes a crossover even in the absence of a physical explicit symmetry breaking.}
 \label{fig:curvature_masses_chiral_limit}
\end{figure*}

Let us start with the discussion of curvature masses at finite temperature and zero chemical potential in the chiral limit $H \rightarrow 0$. 
Since we have tuned our model parameters such that the dynamics of the system is governed by a finite ground state, we expect the pions to be massless whereas the sigma modes and the quarks acquire a finite mass.
In the absence of explicit chiral symmetry breaking, thermal fluctuations then drive the ground state continuously towards smaller and smaller values until it exactly vanishes at the critical temperature~$\tc$. 
Above the critical temperature, the ground state vanishes, which leads to massless quarks in this phase whereas the masses of the mesons are finite and degenerate. 
Phenomenologically speaking, this temperature then marks the chiral phase transition. 
The expected temperature dependence of the various masses implies that the masses of both the mesons and the quarks should vanish identically at the transition, provided that the transition is of second order.
For example, the quark mass can then serve as an order parameter to distinguish between the two separate phases of matter. 
For finite explicit chiral symmetry breaking, however, the quark mass is continuous and remains finite for all temperatures and the transition turns into a crossover.
In this case, we shall use the minimum of the curvature mass of the sigma mode to define a crossover temperature. 
Note that the inverse of this mass can be related to the correlation length in the system.
Therefore, it represents a meaningful definition for the crossover as it is associated with a maximum in the correlation length.  
In the chiral limit, this minimum coincides with the definition of the critical temperature in terms of the quark mass.

In \Cref{fig:Mass_collection_in_different_schemes_for_chiral_limit}, we show our results for the curvature masses as functions of the temperature for~$\mu=0$ as obtained from CS calculations without symmetry constraints (left panel) and from a symmetrized as well as RG-consistent CS calculation (right panel). 
By comparing the two panels, it becomes apparent that the regulator-induced breaking of the chiral symmetry severely spoils the results for the curvature masses, even though there is by construction no regulator-induced chiral symmetry breaking in the unconstrained CS calculations at $T=\mu=0$.
In fact, our ansatz \labelcref{eq:UV ansatz} for the effective action at the initial UV scale has been chosen such that the results from all schemes agree identically with the symmetrized CS scheme in the vacuum limit.
To be specific, we observe that the quark mass at finite temperature as obtained from the unconstrained CS calculation is always finite and even increases at high temperatures, indicating that there is no phase transition at all.
Contrary to that, the mass of the sigma mode as a function of the temperature exhibits a minimum which would indicate a crossover at $\tpc \approx 149\MeV$. 
Moreover, not only that the meson masses are not degenerate at high temperatures, the difference between them actually increases at high temperatures.
Since an increase of the temperature is expected to suppress at least the effect of a physical explicit symmetry breaking such that the meson curvature masses should approach each other as functions of temperature and the quark mass should continuously tend to zero, these observations can be traced back to the regulator-induced breaking of the chiral symmetry. 
Once this regulator-induced chiral symmetry breaking is removed and the global $O(4)$ symmetry of the effective action is restored, we observe the expected behavior of the curvature masses, see \Cref{fig:curvature_masses_chiral_limit}.
For example, in this case, we find that the quark mass tends to zero at $\tc \approx 160 \MeV$ and remains zero above this temperature, indicating a second-order chiral phase transition.
Moreover, the meson masses also tend to zero at this temperature and then become degenerate for~$T>\tc$.
Note that the results from the symmetrized and RG-consistent CS calculation differ only quantitatively but not qualitatively from those obtained from the symmetrized only CS calculation.
For example, we find~$\tc \approx 175 \MeV$ for the phase transition temperature, if we do not take RG consistency into account. 
This is illustrated in \Cref{fig:curvature_masses_chiral_limit} for the meson masses.
From this figure, it also becomes clear that unconstrained CS calculations are pathological and do not have any predictive power.
\begin{figure*}[t]
  \begin{minipage}[b]{0.49\linewidth}
    \includegraphics{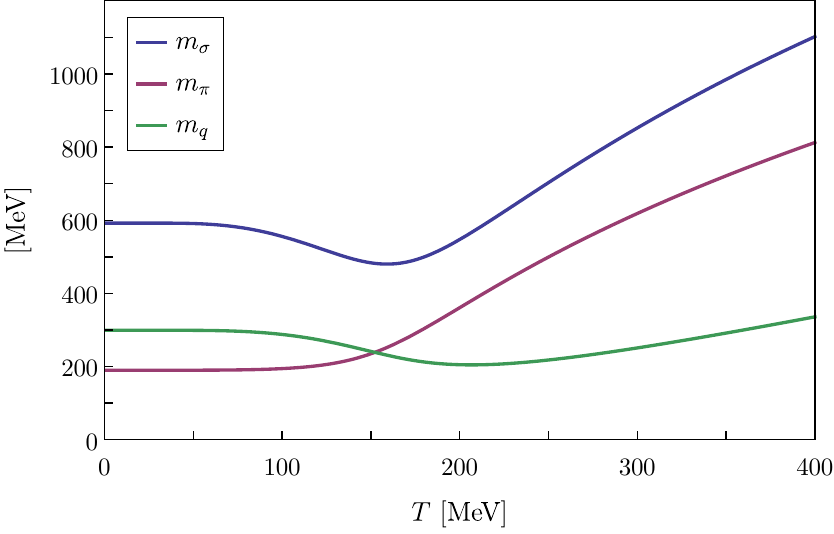}
  \end{minipage}
 \hfill%
 \begin{minipage}[b]{0.49\linewidth}
    \includegraphics{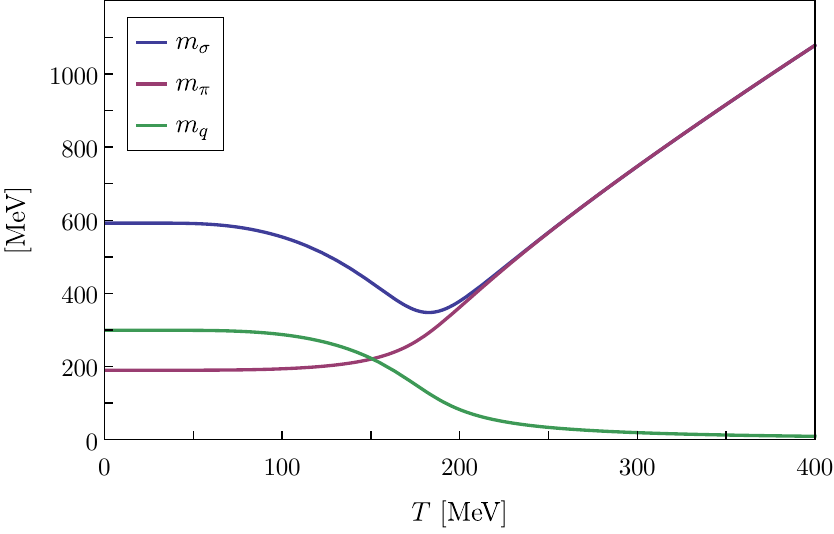}
  \end{minipage}
   \caption{Meson curvature masses and the quark mass from unconstrained CS calculations (left panel) and from symmetrized as well as RG-consistent CS calculations (right panel) as functions of temperature $T$ at $\mu =0$ for a finite (physical) explicit symmetry breaking, $H>0$.}
  \label{fig:Mass_collection_in_different_schemes}
\end{figure*}

\begin{figure*}[t]
  \begin{minipage}[b]{0.49\linewidth}
    \includegraphics{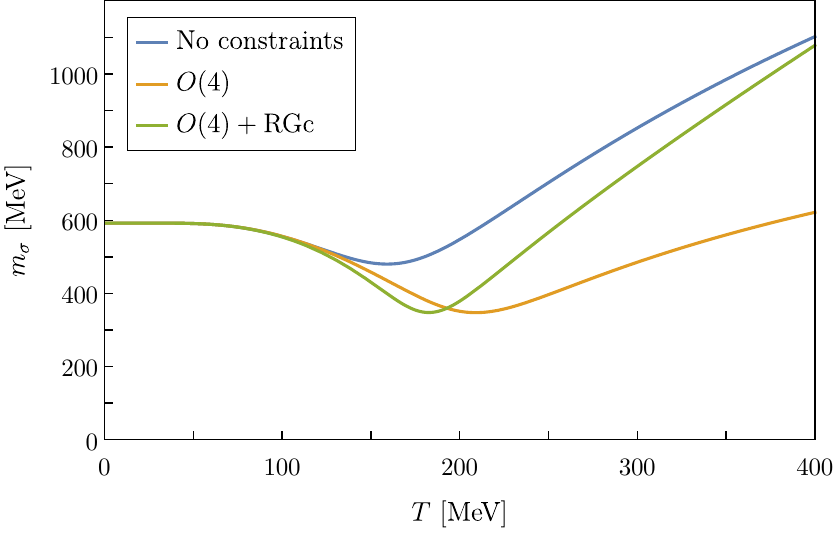}
  \end{minipage}
 \hfill%
 \begin{minipage}[b]{0.49\linewidth}
    \includegraphics{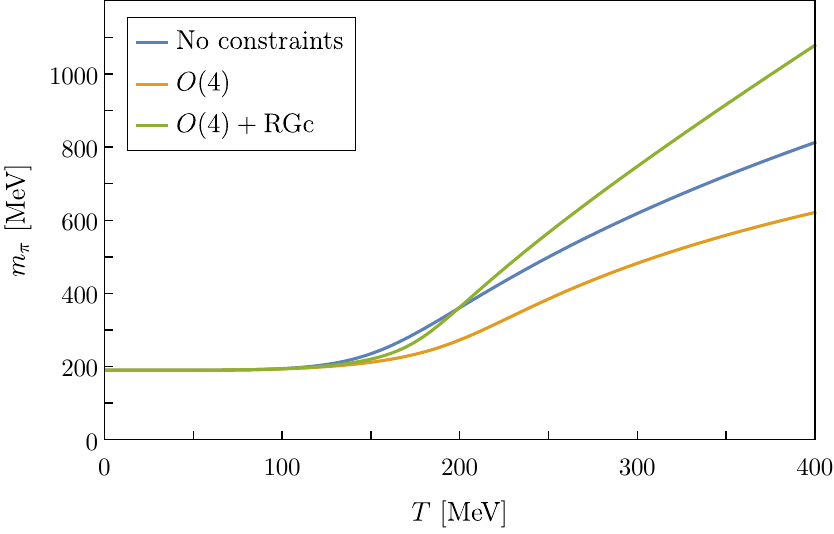}
  \end{minipage}
   \caption{The curvature mass of the sigma mode (left panel) and the pions (right panel) as a function of temperature $T$ at $\mu=0$ for a finite (physical) explicit symmetry breaking, $H>0$. The different colors correspond to our different treatments of the CS flow: symmetrized CS calculations (\glqq$O(4)$\grqq), symmetrized and RG-consistent calculations (\glqq$O(4)+\text{RGc}$\grqq), and unconstrained CS calculations (\glqq No constraints\grqq). We observe again that the unconstrained CS calculations suffer from strong regularization artifacts.}
  \label{fig:curvature_masses}
\end{figure*}

We now turn to the case of a finite physical explicit symmetry breaking term in the action, i.e., we consider~$H > 0$. 
In this case, the ground state of the system should always be finite. 
As a consequence, we expect to find a crossover rather than a phase transition. 
For sufficiently high temperatures, however, we still expect that the quark mass approaches zero. 
In the same way, we expect to find that the difference between the sigma and the pion masses tends to zero.  
As can be deduced from Figs.~\ref{fig:Mass_collection_in_different_schemes} and \ref{fig:curvature_masses}, the results from the unconstrained CS calculations are again severely spoiled by the explicit regulator-induced symmetry breaking. 
For example, the quark mass does not tend to zero at high temperature but rather increases, in contrast to the symmetrized CS calculation, as can be seen in \Cref{fig:Mass_collection_in_different_schemes}.
At least at first glance, the meson masses may appear less affected by the regularization artifacts.
However, the comparison with the results from the symmetrized calculations make apparent that the predictions for the meson masses suffer strongly from the regulator-induced explicit symmetry breaking, in particular at high temperature. 
This is also illustrated in \Cref{fig:curvature_masses}.
Looking at the results from the symmetrized as well as RG-consistent calculations in the right panel of \Cref{fig:Mass_collection_in_different_schemes}, we observe that the system undergoes a crossover which is associated with a pseudo-critical temperature~$\tpc \approx 183 \MeV$ as defined by the minimum of the sigma mass.
At high temperature, we then find that the quark mass tends to zero continuously and the difference between the meson masses decreases, as expected. 
Note that we find $m_\pi \approx 190 \MeV$ for the pion curvature mass in the vacuum limit which translates into a pole mass of about $m_{\text{pole}, \pi} \approx 138 \MeV$, as we shall see below in our studies of spectral functions.

In the zero-temperature limit at finite chemical potential, the dynamics of the system is governed by the Silver-Blaze property~\cite{Cohen:2003kd}.
This refers to the fact that the partition function of a physical system at $T = 0$ does not exhibit any dependence on the chemical potential~$\mu$, provided that the chemical potential remains smaller than a critical value $\muSB$, see, e.g., \cite{Marko:2014hea,Khan:2015puu, Braun:2017srn, Braun:2020bhy}.
As a consequence, the chemical potential needs to be increased beyond the threshold $\muSB$ in order to excite the system at zero temperature.
In our case, the critical value for the chemical potential is determined by the pole mass of the lowest lying state with non-vanishing baryon number, i.e., with the mass associated with the quark state.

As the CS regulator respects the Silver-Blaze symmetry of the system, our choice for the initial conditions, see \labelcref{eq:UV ansatz}, implies that our three CS schemes at zero temperature yield the same effective action, provided that the chemical potential is smaller than its critical value.
Even more, we find that all schemes are in fact identical at $k=0$ as long as the chemical potential does not exceed the initial scale~$\Lambda_0$.
We shall illustrate this in the following.

At zero temperature, the loop corrections exhibit the feature that they can be separated into a vacuum part without explicit dependence on the external parameters, $L_{k,\text{vac}}$, and a chemical potential dependent part $M_k$. 
{\allowdisplaybreaks
From our construction of initial conditions for the different schemes in \labelcref{eq:UV ansatz} it then follows that
\begin{align}
\Gamma^{(\text{scheme})}_{k=0} &=\ \Gamma^{(\text{scheme})}_{\Lambda_0} - 8 \Nc \left[L^{(\text{scheme})}_{0,\text{vac}} - M^{(\text{scheme})}_0\right] \nn\\
&=\ \Gamma_{\Lambda_0} - 8 \Nc \left[L^{(\text{sym})}_{0,\text{vac}} - M^{(\text{scheme})}_0\right] \nn\\
\overset{\mu < \Lambda_0}&{=} \Gamma_{\Lambda_0} - 8 \Nc \left[L^{(\text{sym})}_{0,\text{vac}} - M_0\right]\ .
\label{eq:gammaschemet0mu}
\end{align} 
By restricting} ourselves to chemical potentials $\mu<\Lambda_0$,\footnote{We restrict ourselves to chemical potentials~$\mu<\Lambda_0$ as our model is not expected to have predictive power for chemical potentials beyond that scale where, e.g., diquark condensation may become relevant, see, e.g., \cite{Braun:2019aow, Leonhardt:2019fua}.} the matter part~$M_k$ becomes independent of the mass scale $\Lambda_0$. As a result, all the considered schemes are identical at~$k=0$. 
In the following we may therefore drop the superscript of the effective action at zero temperature.
Furthermore, \labelcref{eq:gammaschemet0mu} eventually implies the absence of regularization artifacts for~$0 \leq \mu < \Lambda_0$ in the three CS schemes.

Let us now consider the explicit form of the effective action at zero temperature and finite chemical potential as it follows from an evaluation on an homogeneous background:
\begin{align}
\label{eq:effective action zeroT}
\frac{1}{V_4}\Gamma_0(\phi) = \frac{1}{2} m_{\Lambda_0}^2 \phi^2 - H \sigma - 8 \Nc  L^{(\text{sym})}_0(\Lambda_0, h^2 \phi^2)
\end{align}
with
\begin{align}
L^{(\text{sym})}_0(\Lambda_0, \rchi) = L^{(\text{sym})}_{0,\text{vac}}(\Lambda_0, \rchi) - M_0(\mu,\rchi)\ .
\end{align}
To be more concrete, the vacuum and chemical potential dependent part of the loop function are given by
\begin{widetext}
\begin{align}
L^{(\text{sym})}_{0, \text{vac}}(\Lambda, \rchi) 
&= \frac{1}{(2 \pi)^2} \frac{1}{32}\left[ \Lambda^2 (\Lambda^2+ 2 \rchi) + 2\rchi^2 \ln\left(\frac{\rchi}{\Lambda^2}\right) - 2 (\Lambda^2+\rchi)^2 \ln\left(1 + \frac{\rchi}{\Lambda^2}\right) \right] 
\end{align}
and
\begin{align}
M_0(\mu,\rchi)  
&= \frac{1}{(2\pi)^2} \frac{\theta(\mu - \sqrt{\rchi})}{24} \left(\mu \sqrt{\mu^2-\rchi}\ (5\rchi - 2\mu^2) - 3 \rchi^2 \arsinh\left(\sqrt{\frac{\mu^2}{\rchi}-1} \right) \right)\,,
\end{align}
\end{widetext}
respectively.

\begin{figure*}[t]
  \begin{minipage}[b]{0.49\linewidth}
    \includegraphics{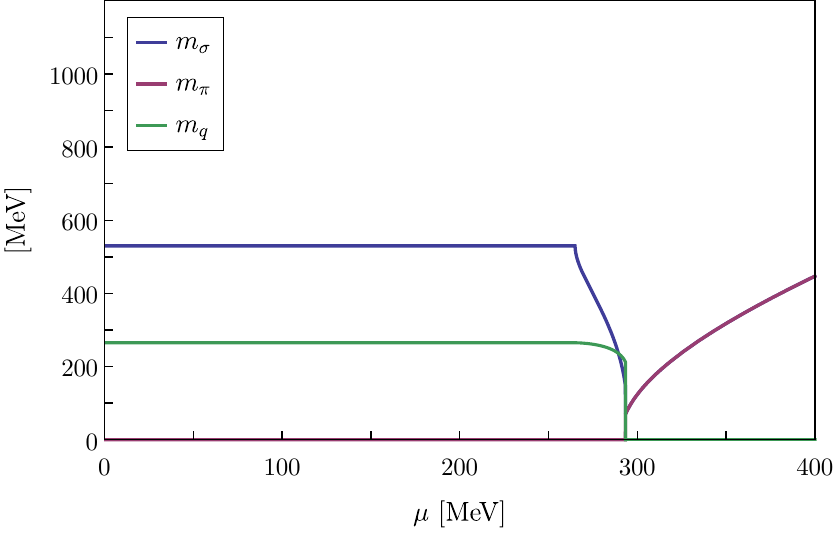}
  \end{minipage}
 \hfill%
 \begin{minipage}[b]{0.49\linewidth}
    \includegraphics{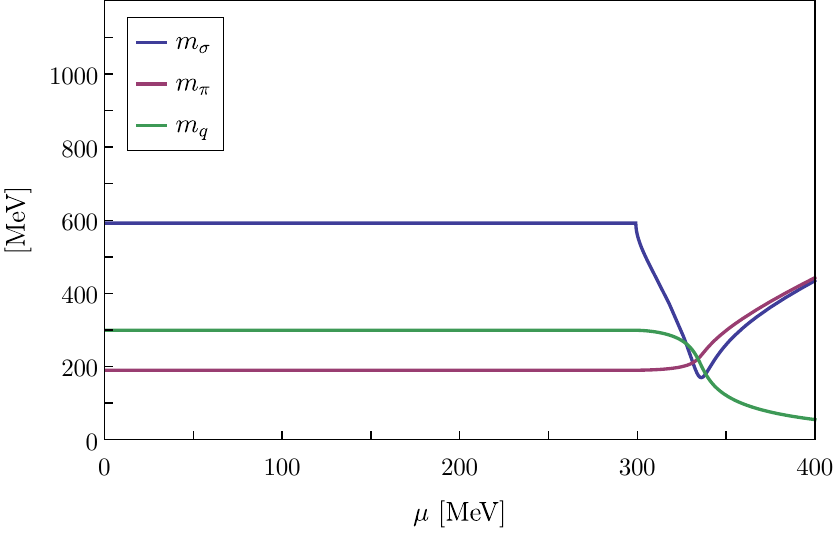}
  \end{minipage}
   \caption{The meson curvature masses and the quark mass in the chiral limit (left panel), $H \to 0$, and with physical explicit symmetry breaking (right panel), $H > 0$, at zero temperature. For~$H\to 0$, the behavior of the masses indicate a first-order phase transition and a crossover for~$H>0$.}
  \label{fig:zero-temperature_masses}
\end{figure*}

In \Cref{fig:zero-temperature_masses}, we show the curvature masses as obtained from the effective action \labelcref{eq:effective action zeroT}. 
In accordance with our discussion of the Silver-Blaze property above, we find that all masses agree identically with their respective values in the vacuum limit for~$\mu <\muSB$.
Here, the Silver-Blaze threshold is set by the constituent quark mass at $T=\mu=0$.
Since the value of the quark mass in the vacuum limit depends on whether or not we consider explicit symmetry breaking, we generally find $\muSB(H) = m_q(H)$, where we have $m_q(0)=\hat{m}_q$ by construction, see \Cref{subsubsec: scale fixing I}. 
In the chiral limit, i.e., for $H\to 0$, the behavior of the quark mass as a function of the chemical potential indicates a first-order phase transition at the critical value $\muc \approx 293 \MeV$.
Note that, in contrast to the situation at the second-order phase transition discussed above, the meson curvature masses do not tend to zero at the critical chemical potential but assume a finite value:
\begin{align}
\text{min}\left(m_\sigma\right)  = \sqrt{m^2_{\Lambda_0} + \frac{1}{\pi^2} \Nc h^2\muc^2}\ .
\end{align}
Choosing now~$H$ such that the pion pole mass assumes its physical value, we observe that the first-order transition turns into a crossover at $\mupc \approx 336 \MeV$ as indicated by the continuous behavior of the masses as functions of the chemical potential, see \Cref{fig:zero-temperature_masses}. 
Here,~$\mupc$ is defined by the minimum of the curvature mass of the sigma mode.
These observations imply that there is no critical point in the phase diagram of the quark-meson model for physical pion masses, at least in our present approximation. 
\subsubsection{Spectral functions}
In the following we construct meson spectral functions from the Euclidean two-point function. 
The CS regulator ensures the existence of the Käll\'{e}n-Lehmann (KL) spectral representation for the regularized propagator at every scale $k$, i.e.,
\begin{align}
\label{eq:KL spectral representation}
\frac{1}{\tilde{\Gamma}^{(2)}_k(Q)} = \int_{\mathbb{R}} \frac{\d \lambda}{2 \pi} \frac{\lambda}{Q_0^2 + \lambda^2}\ \rho_k(\lambda, \vec{Q})
\end{align}
with the matrix-valued spectral (density) function $\rho$. 
Note that the KL representation directly implies the spectral function to be an odd function in its first argument, $\rho_k(-\lambda, \vec{Q}) = -\rho_k(\lambda, \vec{Q})$. 
Given \labelcref{eq:KL spectral representation}, the spectral function can be computed as follows:
\begin{align}
\label{eq:definition spectral function}
\rho_k(\omega, \vec{Q}) = 2\lim_{\varepsilon \rightarrow 0^+} \Im{\frac{1}{\tilde{\Gamma}^{(2)}_k(\iu(\omega + \iu \varepsilon), \vec{Q})}}\,.
\end{align}
Here, $\omega$ is defined along the imaginary axis of the complex $Q_0$-plane such that it can be associated with an energy variable in Minkowski spacetime.
Note that \labelcref{eq:definition spectral function} can always be used as a definition of the spectral function but \labelcref{eq:KL spectral representation} only holds if the regulator admits a spectral representation on all scales. 
For all numerical calculations, the real-valued parameter $\varepsilon$ is kept small but finite to render the computation of spectral functions numerically stable.
To be concrete, we use $\varepsilon = 3 \MeV$.
\begin{figure}[b]
    \includegraphics{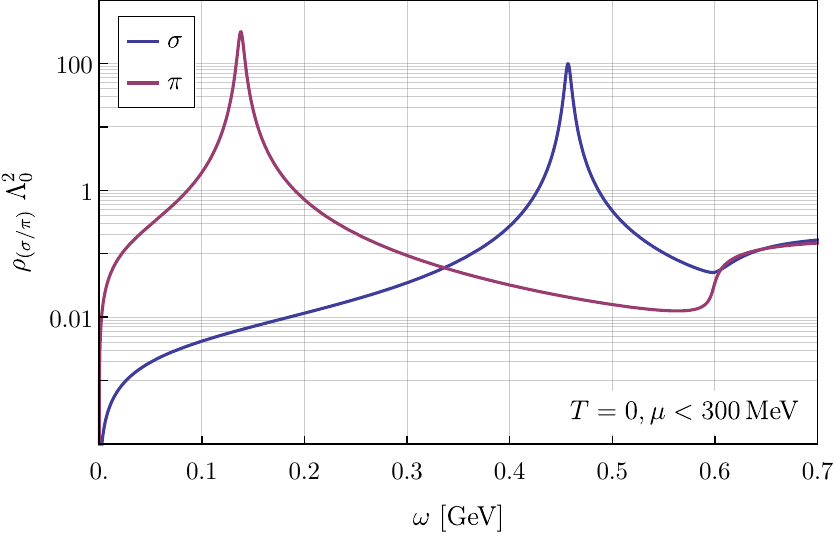}
   \caption{Spectral functions of the mesons at~$T=0$ and $\mu< \muSB$ with $\muSB = m_q \approx 300 \MeV$ as obtained from the symmetrized CS calculation.}
  \label{fig:quark_masses_and_vacuum_spectral function}
\end{figure}

Let us now present our results for the spectral functions of the mesons as function of the energy~$\omega$ in the presence of a physical explicit symmetry breaking, $H >0$. 
As above, we start by comparing the results from different treatments of the CS flow at finite temperature and zero chemical potential. 
In all our numerical studies, we shall set the spatial external momenta~$\vec{Q}$ of the two-point function to zero, i.e., $\vec{Q} = 0$, such that the energy $\omega_{\text{res}}$ associated with a resonance peak in the spectral function can be identified exactly with the mass of the corresponding meson. 

In \Cref{fig:quark_masses_and_vacuum_spectral function}, we show the spectral functions of the sigma mode and the pions in the vacuum limit. 
The pole masses of these two mesons can be extracted from a localization of the peaks in these functions.
Recall that we do not take the limit~$\varepsilon \to 0$ but only set~$\varepsilon$ to a sufficiently small value such that
poles in the propagator for $\varepsilon = 0$ now give rise to finite resonances instead of divergences.   
We find $m_{\text{pole},\sigma} = \omega_{\text{res},\sigma} \approx 457 \MeV$ for the sigma mode. 
For the pion spectral function, the position of the pole at $m_{\text{pole},\pi} = \omega_{\text{res},\pi} \approx 138 \MeV$ is not a prediction in our present study as we have used the parameters of our model to tune the pion pole mass in the vacuum limit. 
In any case, the spectral functions exhibit more structure than the typical peaks associated with the pole masses. 
For example, at $\omega \approx 600 \MeV$ the kink and turning point in the spectral functions of the two mesons is commonly interpreted as a decay of an excited state of the respective meson into an energetically more favorable state.
Within our model study, we can identify these decays as the processes $\sigma^\prime \rightarrow \psib \psi$ and $\pi^\prime \rightarrow \psib \psi$, where the primes indicate the excited state. 
In other words, when we approach the energy $\omega_{\text{decay}} = 2 m_q$ associated with a pair of two constituent quarks with zero center-of-mass energy, it is favorable for the mesons to decay into these quark states.\footnote{Recall that~$\Lambda_0$ is not a UV momentum cutoff but represents a mass scale in the CS scheme. Therefore, the CS regularization does not restrict the range of the external four-momenta of correlation functions. Still, this does not imply that our results necessarily have predictive power at arbitrarily high external momenta as the quark-meson model does not contain the correct degrees of freedom in these regimes.}
\begin{figure*}
  \begin{minipage}[b]{0.49\linewidth}
    \includegraphics{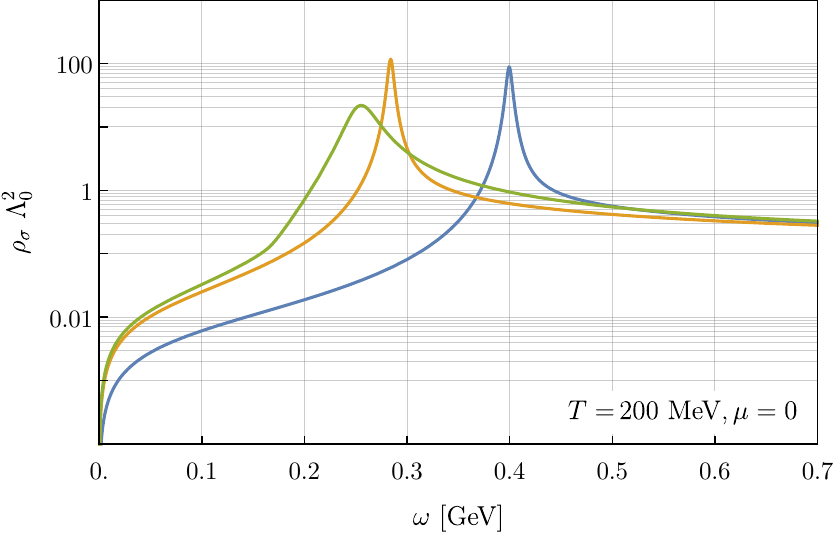}
  \end{minipage}
 \hfill%
 \begin{minipage}[b]{0.49\linewidth}
    \includegraphics{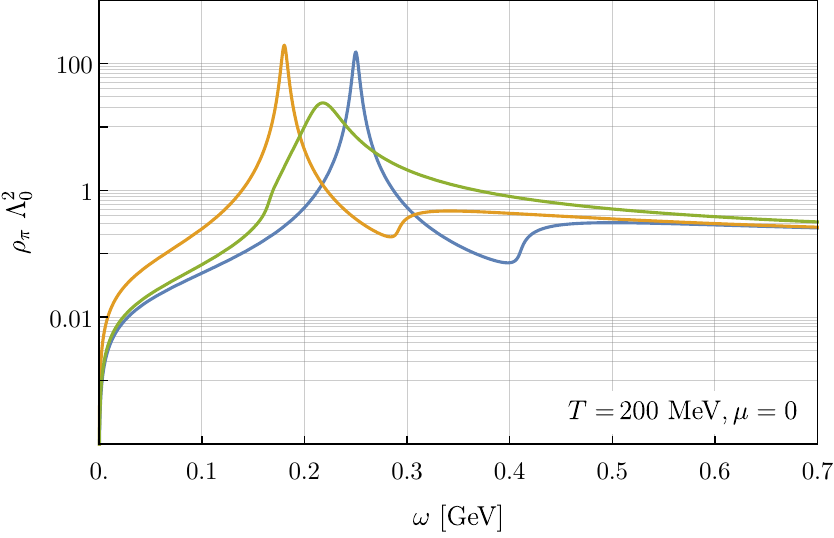}
  \end{minipage}
   \caption{Spectral function of the sigma mode (left panel) and the pion (right panel) at $T = 200 \MeV$ and $\mu = 0 \MeV$ as obtained from a CS calculation with no constraints (blue line), an $O(4)$-symmetrized CS calculation (orange line), and an RG-consistent $O(4)$-symmetrized CS calculation (green line).}
     \label{fig:comparison_of_spectral_functions_at_finite_temperature}
\end{figure*}

By increasing the temperature, we can deform the spectral functions. 
To be more specific, as the quark mass decreases with increasing temperature, the structures associated with the decay of the mesons in two quark states are continuously shifted to lower temperatures and also become suppressed.
For temperatures associated with $\omega_{\text{decay}} < \omega_{\text{res}}$, the width of the resonance peaks increases with increasing temperature and their height decreases.\footnote{Note that this is not a numerical artifact associated with our choice for $\varepsilon$ but can be traced to thermal fluctuations. In fact, at high temperatures, our results for the spectral functions converge quickly as a function of $\varepsilon$.}
The broadening of the peak in a spectral function can be traced back to the fact that thermal fluctuations then induce a screening for the pole which is present in the propagator of the vacuum theory.
Accordingly, the mass associated with this peak is no longer a pole mass at these temperatures but should rather be considered a resonance mass. 
Moreover, the spectral functions of the sigma mode and the pions as extracted from the $O(4)$-symmetrized CS caclulations (orange and green lines in \Cref{fig:comparison_of_spectral_functions_at_finite_temperature}) become more and more degenerate with increasing temperature. 
As this phenomena can be traced back to an almost vanishing quark mass at high temperatures, this behavior of the spectral functions cannot be observed in our unconstrained CS caclulations (blue lines in \Cref{fig:comparison_of_spectral_functions_at_finite_temperature}).

Apart from these more phenomenological aspects, we would like to point to differences in the results as obtained from our different CS schemes. 
As can be deduced from \Cref{fig:comparison_of_spectral_functions_at_finite_temperature}, the differences are clearly not only qualitative but also quantitative. 
For example, at $T=200 \MeV$ and $\mu = 0 \MeV$, we observe that the results from the unconstrained CS calculation (blue lines) receive significant corrections from the symmetrization (orange lines).
By ensuring RG consistency in addition to the symmetrization (green lines), the spectral functions receive additional corrections. For example, we observe a broadening of the peak.

\begin{figure*}[t]
\centering
  \begin{minipage}[b]{0.49\linewidth}
    \includegraphics{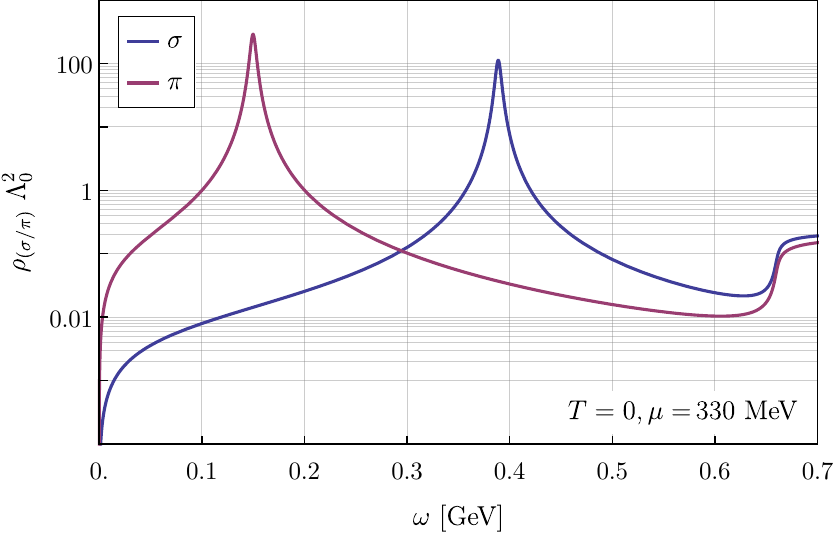}
  \end{minipage}
 \hfill%
 \begin{minipage}[b]{0.49\linewidth}
    \includegraphics{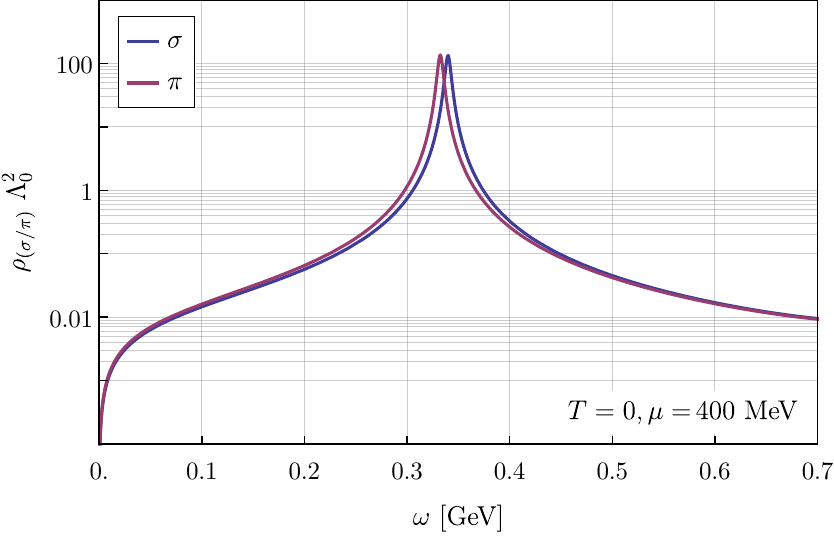}
  \end{minipage}
   \caption{Meson spectral function at zero temperature and $\mu =330 \MeV$ (left panel) as well as $\mu = 400 \MeV$ (right panel).}
  \label{fig:spectral_function_at_finite_mu}
\end{figure*}

At zero temperature, the dynamics is governed by the Silver-Blaze property over a wide range of values for the chemical potential, i.e., the meson spectral functions remain unchanged compared to their form in the vacuum limit up to $\muSB \approx 300 \MeV$. 
Exceeding the chemical potential beyond the threshold~$\muSB$, the kink-like structures of the meson spectral functions observed at $\omega_{\text{decay}} = 2 m_q$ in the vacuum limit are now found at $\omega_{\text{decay}} = 2 \mu$.
Thus, the position of these structures is continuously shifted to larger values by increasing the chemical potential. 
In this regime, these kink-like structures can then be associated with a decay into two quarks where each quark carries the Fermi energy~$\mu$. 
Note that, compared to the case of finite temperature and zero chemical potential, the structures of the zero-temperature spectral functions, which are associated with meson decays, do not follow the $\mu$-dependence of the ground state as quark states with energy less than $\mu$ cannot be occupied due to the Pauli exclusion principle. 
In other words, the mesons can only decay into their quark content if there is enough energy to create a pair of quarks ``sitting" at the Fermi surface or above. 
Moreover, the kink of the sigma spectral function quickly transforms into a turning point for $\mu > \muSB$,  whereas the analogous structure in the pion spectral function does not change qualitatively. 
Meson spectral functions computed exactly in the zero-temperature limit are shown in \Cref{fig:spectral_function_at_finite_mu} for different values of the chemical potential beyond the regime governed by the Silver-Blaze property. 
When increasing the chemical potential beyond the point associated with the first-order transition observed in the chiral limit, the sigma and pion spectral functions become degenerate, similar to the case of finite temperature and zero chemical potential, but with the difference that the spectral functions exhibit sharp peaks instead of broad resonances.
This is due to the fact that the decay energy $\omega_{\text{decay}}$ at zero temperature increases with the chemical potential. 
Recall that the poles in the propagator of the vacuum theory remain poles at finite external parameters as long as $\omega_{\text{decay}} > \omega_{\text{res}}$. 
\subsection{Renormalization effects on physical observables}
\label{sec:reneffects_po}
In the previous subsection, we have discussed that regularization artifacts and, in particular, the regulator-induced explicit breaking of the chiral symmetry affect physical observables on both the qualitative and quantitative level. 
Therefore, we shall only consider the symmetrized and RG-consistent CS scheme from here on.

The focus of the following study is on the computation of renormalized curvature masses and renormalized spectral functions which requires to compute the wavefunction renormalization factors.
Throughout this section, we shall only present results for the phenomenologically most relevant case, which involves finite physical explicit symmetry breaking.

\subsubsection{Scale fixing}
Let us again begin by fixing the UV parameter of our model such that we obtain specific values for the (constituent) quark mass and the pion decay constant at $T=\mu=H=0$. 
These quantities are defined as~\cite{Klevansky:1992qe}
\begin{align}
\label{eq:mc and fpi2}
m_q = \hb_\sigma |\sigmab_0|\ , \quad f_\pi = \frac{m_q}{\hb_\pi}\, .
\end{align}
Thus, we use the pion decay constant to fix the renormalized Yukawa coupling in the vacuum limit. 
The definitions of the renormalized Yukawa coupling and ground state~$\sigmab_0$ read
\begin{align}
\label{eq:renormalized couplings}
\hb_{(\sigma/\pi)} = (Z_{(\sigma/\pi)}^{\perp})^{-\frac{1}{2}} h\,, \qquad \sigmab_0 = (Z_\sigma^{\perp})^{\frac{1}{2}} \sigma_0\,.
\end{align} 
Note that the quark mass is invariant under renormalization in our one-loop calculation in the large-$\Nc$ limit.

From \labelcref{eq:symCS two-point} we obtain the bosonic wavefunction renormalizations by a projection involving derivatives with respect to the momenta. 
In the vacuum limit, we have
\begin{align}
Z_{(\sigma/\pi)}^\perp = Z_{(\sigma/\pi)}^\parallel \equiv Z_{(\sigma/\pi)}
\end{align}
 with 
\begin{widetext}
\begin{align}
\begin{pmatrix}
Z_\sigma & \vec{0}^\tp\\
\vec{0} & Z_\pi \uM{3}
\end{pmatrix}
&= \frac{1}{2} \lim_{Q \rightarrow 0} \frac{{\rm d}^2}{{\rm d} |Q|^2}\ \tilde{\Gamma}^{(\text{phys})(2)}_0(Q)
\nn\\
&= Z_{\Lambda_0} \uM{4} + \frac{h^2 \Nc}{4 \pi^2} \left(\uM{4}\ \ln\left(1+\frac{\Lambda_0^2}{m_q^2}\right)- \frac{\uM{4}+\eta}{3}\frac{\Lambda^2_0}{m_q^2 + \Lambda_0^2}\right)\,
\label{eq:vacuum Z}
\end{align}
\end{widetext}
in the limit~$k\to 0$.

Before we discuss the determination of the model parameters in the present case, a comment on the wavefunction renormalizations in \labelcref{eq:vacuum Z} is in order: 
First of all, we deduce from our exact one-loop results for the meson two-point functions in the large-$\Nc$ limit as given in \Cref{app: Explicit zero-temperature results} that these functions are non-analytic in~$Q$ in the limit~$k\to 0$ and~$m_q\to 0$.
Because of this non-analytic behavior, an expansion of these functions in~$Q$ about~$Q=0$ does not exist in these limits and the corresponding wavefunction renormalizations as defined by the projection above are not well-defined.\footnote{This also holds true at finite temperature in the limit~$k\to 0$ and~$m_q\to 0$.}
In \labelcref{eq:vacuum Z}, this manifests itself by the fact that the wavefunction renormalizations diverge in the limit~$m_q\to 0$. 
In addition, we observe in \labelcref{eq:vacuum Z} that~$Z_{\sigma}- Z_{\pi} \neq 0$ in the limit~$m_q\to 0$, falsely indicating a broken chiral symmetry. 
The latter is an artifact of our CS regularization in this limit.
We emphasize that our exact, fully momentum-dependent results for the two-point functions do not depend on the order of the limits~$k\to 0$ and~$m_q\to 0$, see \Cref{app: Explicit zero-temperature results}. 
Moreover, the sigma and pion two-point functions agree in the limit~$m_q\to 0$, as it should be. 
For~$m_q >0$, an expansion of these functions in~$Q$ about~$Q=0$ exists which leads us to the expression in \labelcref{eq:vacuum Z}. 
Note that, at finite chemical potential, the two-point functions exhibit additional non-analyticities, see below and also \Cref{app: Explicit zero-temperature results}.

Our result for the wavefunction renormalizations together with the chosen values $\hat{m}_q$ and $\hat{f}_\pi$ for the quark mass and the pion decay constant at $T=\mu=H=0$ then consistently determine the values of the model parameters~$m^2_{\Lambda_0}$ and~$h$,
\begin{align}
m^2_{\Lambda_0}&= 16 \Nc\ h^2 \left.\left(\frac{\d}{\d \rchi} L^{\supsc}_0(\Lambda_0,\rchi)\right)\right|_{\rchi=\hat{m}_q^2}\, , \\
h^2 &= \frac{Z_{\Lambda_0}}{\left(\frac{\hat{f}_\pi}{\hat{m}_q}\right)^2 - \frac{\Nc}{4\pi^2}\ln\left(1+\frac{\Lambda_0^2}{\hat{m}_q^2}\right)}\ .
\end{align}
Note that the second term in the denominator of this expression is strictly positive since $\Lambda_0 > \hat{m}_q$. 
As a consequence, the condition $0 < h^2 < \infty$ restricts our choices for $\hat{m}_q$, $\hat{f}_\pi$ and $\Lambda_0$. 
Since our phenomenologically inspired choice for the latter quantities already used in \Cref{subsubsec: scale fixing I} is compatible with these restrictions, we choose the same values here which also simplifies direct comparisons of results obtained with and without an inclusion of the wavefunction renormalization factors of the mesons. 
To be specific, choosing~$\Nc=3$, we set $\hat{m}_q = 265 \MeV$, $\hat{f}_\pi = 90 \MeV$ and $\Lambda_0 = 500 \MeV$ in the vacuum and chiral limit.
We add that our results for the bare parameters and the wavefunction renormalizations imply that renormalized couplings do not depend on~$Z_{\Lambda_0}$. 
\begin{figure}
\includegraphics{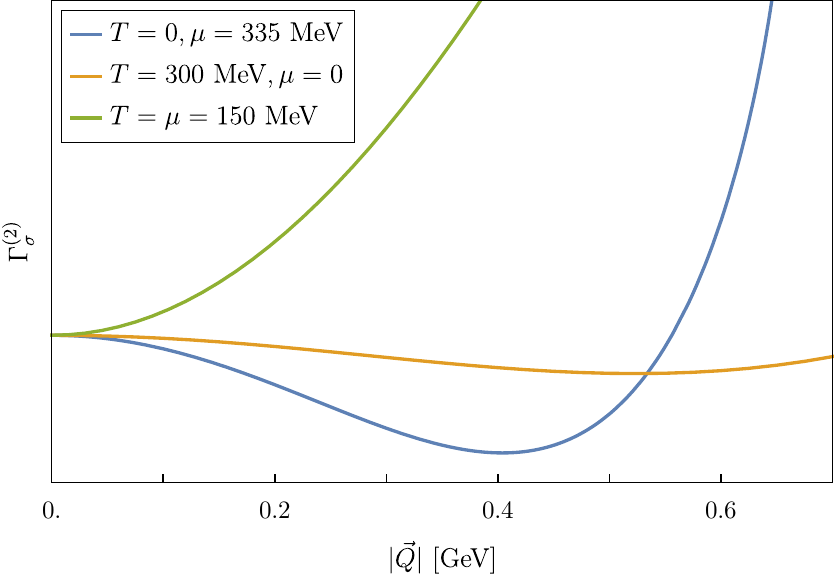}
\caption{Two-point correlator of the $\sigma$-mode as a function of spatial external momenta at $Q_0 =0$ for several temperatures and chemical potentials, normalized such that they agree at~$|\vec{Q}|=0$.  }
\label{fig:two-point_correlator_at_finite_external_momenta}
\end{figure}

The renormalized symmetry breaking parameter
\begin{equation}
\Hb = \frac{H}{\sqrt{Z_\sigma^\perp}}
\end{equation}
is fixed such that we obtain a pion pole mass of $\hat{m}_{\text{pole},\pi} =  138 \MeV$ in the vacuum limit, 
\begin{equation}
\Hb / \Lambda_0^3 \approx 0.018  
\end{equation}
For the pion decay constant~$f_{\pi}$, we then find~$f_{\pi}\approx 92.7 \MeV$. 
Furthermore, our choice for the model parameters yields~$m_q \approx 282 \MeV$ for the (constituent) quark mass. 
\begin{figure*}[t]
  \begin{minipage}[b]{0.49\linewidth}
    \includegraphics{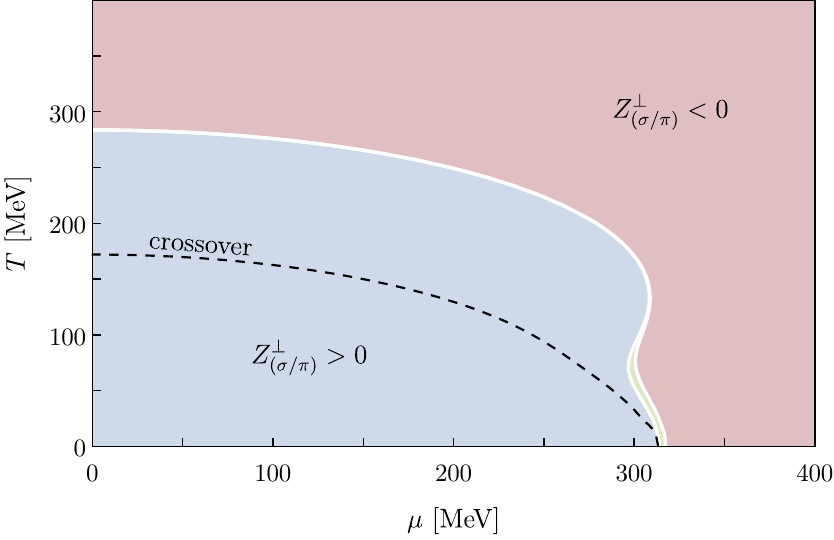}
  \end{minipage}
 \hfill%
 \begin{minipage}[b]{0.49\linewidth}
    \includegraphics{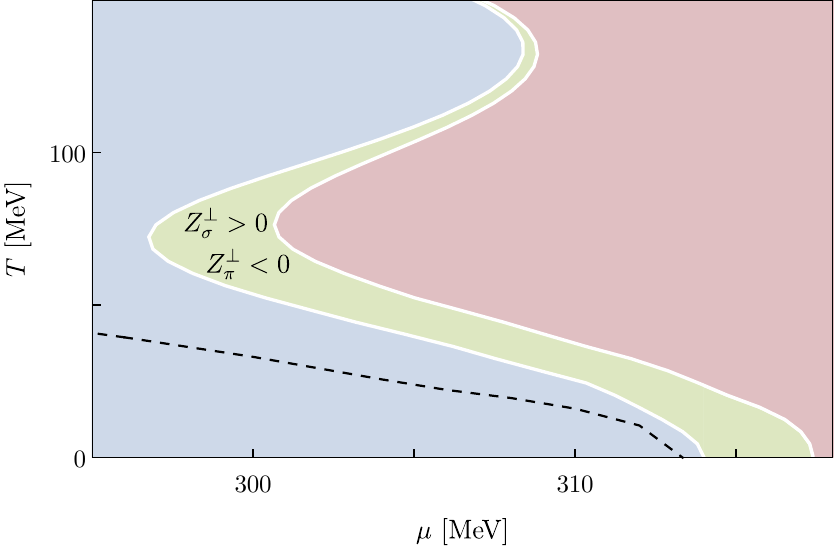}
  \end{minipage}
   \caption{Phase diagram in the plane spanned by the temperature~$T$ and the chemical potential~$\mu$: 
   In the blue-colored region, both wavefunction renormalizations are positive whereas the red-colored area indicates that both wavefunction renormalizations are negative. 
   In the small green-colored region at high densities, we find $Z^\perp_\sigma>0$ and $Z^\perp_\pi<0$. 
   A zoom into the latter region is shown in the right panel. 
   The dashed line represents the crossover line which fully lies in the region of $Z^\perp_{(\sigma/\pi)}>0$. The phase structure shown here is  corroborated by a computation in full functional QCD \cite{FPPRWY}.}
  \label{fig:Sign_diagram_of_Z}
\end{figure*}

We would like to mention that the renormalized Yukawa coupling associated with quark-pion interactions at $T=\mu=H=0$ is readily determined as $\hb^{(0)}_\pi = \hat{m}_q/\hat{f}_\pi$. It follows that \labelcref{eq:renormalized couplings} and  \labelcref{eq:vacuum Z} can then be used to determine the corresponding value for quark-sigma interactions:
\begin{align}
\hb^{(0)}_\sigma = \hb^{(0)}_\pi \sqrt{\frac{Z_\pi^\perp}{Z_\sigma^\perp}} = \frac{1}{\sqrt{\left(\frac{\hat{f}_\pi}{\hat{m}_q}\right)^2- \frac{\Nc}{6\pi^2} \frac{\Lambda_0^2}{\hat{m}^2_q+\Lambda_0^2}}}\,.
\end{align}
We obtain $\hb^{(0)}_\sigma/\hb^{(0)}_\pi \approx 1.23$ within our present approximation.
For $H>0$, we find $\hb_\sigma/\hb_\pi \approx 1.25$ in the vacuum limit.
Note that, in QCD studies, we have~$\hb_\sigma/\hb_\pi \approx 1.5$, see~\cite{Fu:2019hdw}. 
In any case, we deduce from these considerations that the minimum~$\sigmab_0$ of the effective potential should in general not be identified with the pion decay constant.

Having fixed the parameters of our model in the vacuum and chiral limit, we can now study the dynamics of the system at finite temperature and chemical potential. 
Interestingly, we observe negative values of the wavefunction renormalizations in certain regimes of the phase diagram. 
Note that, at finite temperature and/or chemical potential, we define them as the second derivative of the corresponding two-point functions with respect to the spatial momenta evaluated at vanishing momenta, see \labelcref{app: eq:Z projection}. 
A zero-crossing of a wavefunction renormalization in general indicates that the expansion of the effective action in external momenta (i.e., a derivative expansion in position space) has not been performed about the correct expansion point. 
In fact, according to our definition, the wavefunction renormalization associated with a given field is nothing but the curvature of the corresponding momentum-dependent two-point function at vanishing momentum. 
This is illustrated in \Cref{fig:two-point_correlator_at_finite_external_momenta}, where we show the dependence of the two-point function of the sigma mode on the spatial momentum for several values of the temperature and the chemical potential.

We add that a negative curvature of bosonic two-point functions as a function of the external momentum may be considered an indication of an instability associated with the formation of an inhomogeneous ground state at low temperatures~\cite{Braun:2014fga,Roscher:2015xha,Braun:2015fva}, see also \cite{Buballa:2014tba,Tripolt:2017zgc,Motta:2023pks}.
In addition, such a feature also underlies the discussion of the existence of moat regimes in the QCD phase diagram which are defined by~$Z_{(\sigma/\pi)}^{\perp}<0$, see \cite{Rennecke:2021ovl,Pisarski:2021qof,Rennecke:2023xhc,Haensch:2023sig,Pannullo:2024sov}. 
We shall briefly come back to this below when we discuss the position of the boundaries of regimes associated with negative wavefunction renormalizations relative to the position of the chiral phase boundary in our model.
In any case, we indeed observe in \Cref{fig:two-point_correlator_at_finite_external_momenta} that, for temperatures and chemical potentials beyond a certain threshold, the curvature of the sigma two-point function at the origin becomes negative.
Nevertheless, the two-point correlator always remains bounded from below with a minimum at finite spatial external momentum. 

The pion two-point function behaves similarly to the sigma two-point function, as can be seen in \Cref{fig:Sign_diagram_of_Z}, where we show the phase diagram of our model divided into regimes with positive and negative wavefunction renormalizations. 

Finally, we would like to add that we observe a non-analytic behavior of the meson two-point functions in the zero-temperature limit at $Q_0 = 0$ and $|\vec{Q}| = 2(\mu^2-m_q^2)^{\frac{1}{2}}$ (for~$k=0$). 
This value of $|\vec{Q}|$ corresponds precisely to twice the Fermi momentum for massive quarks, $\pF = (\mu^2-m_q^2)^{\frac{1}{2}}$. 
More generally speaking, as can be deduced from our analytic results for the zero-temperature two-point functions in \Cref{app: Explicit zero-temperature results}, these correlation functions have branch cuts for $\mu > m_q$. 
The existence of the latter give rise to what is known as Friedel oscillations~\cite{doi:10.1080/14786440208561086,doi:10.1080/00018735400101233,Friedel:1958} in the context of condensed-matter physics, see~\cite{FetterAndWalecka} for an introduction. 
For a discussion of this aspect in the context of moat regimes, we refer the reader to~\cite{Rennecke:2024}.

In contrast to the non-analytic, yet finite behavior of the meson correlation functions that we observe at zero temperature, a divergence has been found at $Q_0 = m_q = 0$ and $|\vec{Q}|=2 \mu$ in low-dimensional Gross-Neveu-type models~\cite{Koenigstein:2021llr,Koenigstein:2023yzv,Koenigstein:2024cyq}. 
The emergence of such a different behavior in low-dimensional models may not be too surprising since the number of spatial dimensions crucially controls the behavior of the loop integrals at those values of the external parameters (temperature, chemical potential, masses) where the corresponding integrand has a pole along the contour of integration. 
In other words, the origin of the observed non-trivial behavior of the correlation functions is the same but the pole structure of the underlying integrand manifests itself differently at the level of the integral, depending on the dimensionality of the theory. 
\begin{figure*}[t]
  \begin{minipage}[b]{0.49\linewidth}
    \includegraphics{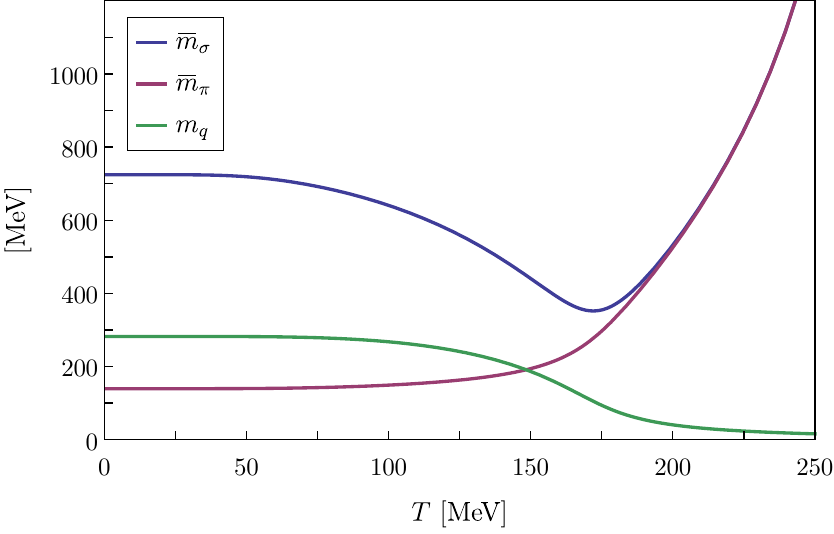}
  \end{minipage}
 \hfill%
 \begin{minipage}[b]{0.49\linewidth}
    \includegraphics{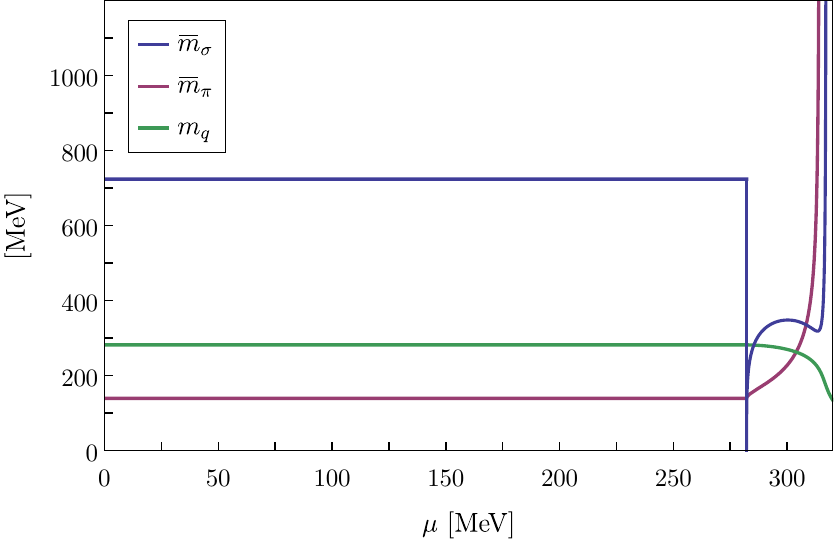}
  \end{minipage}
  \caption{Renormalized curvature masses of the sigma mode and the pion as well as the quark mass as functions of the temperature~$T$ at zero chemical potential (left panel) and as functions of the chemical potential at zero temperature (right panel).}
  \label{fig:renormalized_curvature_masses}
\end{figure*}
\subsubsection{Renormalized curvature masses and phase diagram}
\label{subsubsec:renmass_pd}
Let us now discuss the renormalized curvature masses of the mesons which are readily obtained from the bare curvature masses:
\begin{align}\label{eq:renormalized curvature masses}
\mb^2_{(\sigma/\pi)} = \frac{m^2_{(\sigma/\pi)}}{Z^\perp_{(\sigma/\pi)}}\,.
\end{align}
As discussed above, the wavefunction renormalization factors exhibit a zero-crossing for sufficiently high temperatures and/or large chemical potentials, see \Cref{fig:Sign_diagram_of_Z}. 
In the left and right panel of \Cref{fig:renormalized_curvature_masses}, we therefore only show the renormalized curvature masses of the mesons in regimes where the wavefunction renormalizations remain positive.\footnote{In regimes, where the two-point function assumes a finite non-trivial minimum at~$\vec{Q}=\vec{Q}_{\text{min}}$, we could in principle use the following generalization of the definition of the wavefunction renormalizations:
\begin{align}
\begin{pmatrix}
Z^\perp_\sigma & \vec{0}^\tp\\
\vec{0} & Z^\perp_\pi\ \uM{3}
\end{pmatrix}(k) &= \lim_{\vphantom{\vec{0}}Q_0 \rightarrow 0}  \lim_{\vec{Q} \rightarrow\vec{Q}_{\text{min}}}\frac{1}{2}\frac{\d^2}{\d |\vec{Q}|^2} \Gammat_k^{(2)}(Q)\,.
\nn
\end{align}
Still, even if this definition is employed, there would exist points in the space of external parameters at which the wavefunction renormalizations become zero.
Instead of employing derivatives with respect to the spatial momenta, we may alternatively define the wavefunction renormalization factors via the second derivative with respect to the time-like momentum~$Q_0$ (see, e.g., \cite{Braun:2009si,Braun:2021uua}) which may then be free of a sign change. In the vacuum limit, both definitions should yield the same result. We shall not discuss this aspect further in the present work.}
In addition to the meson curvature masses, we also show the quark mass in these figures. 

Let us begin our discussion with the renormalized curvature masses at finite temperature but zero chemical potential, see left panel of \Cref{fig:renormalized_curvature_masses}. 
From the minimum of the renormalized curvature mass of the sigma mode as a function of the temperature, we deduce that the temperature-induced chiral crossover occurs at $\tpc \approx 172 \MeV$. 
Above the pseudo-critical temperature, the differences in the curvature masses of the sigma mode and pion become rapidly suppressed by thermal fluctuations. 
At (very) high temperatures, the curvature masses are eventually found to be dominated by the zero-crossing of the wavefunction renormalizations at~$T \approx 284 \MeV$. 
\begin{figure}[b]
\centering
\includegraphics{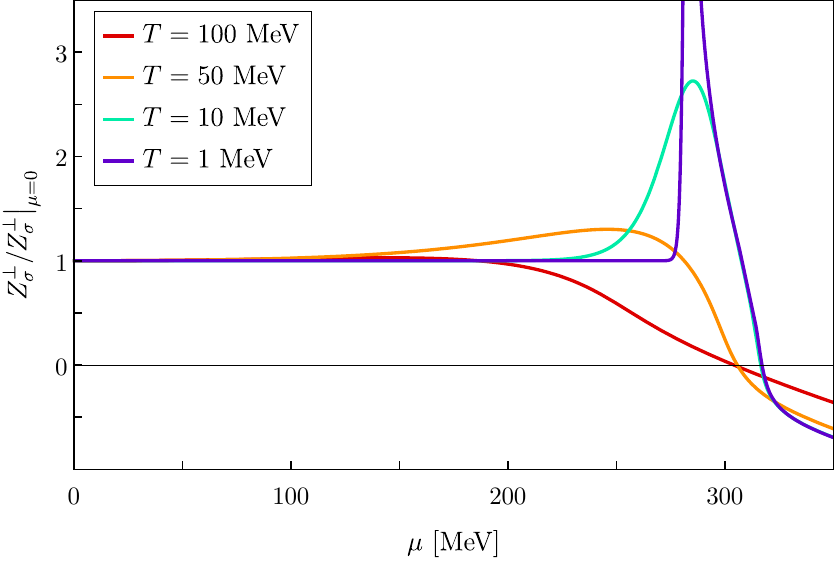}
 \caption{Wavefunction renormalization $Z^\perp_\sigma$ of the sigma field as a function of the quark chemical potential. By lowering the temperature, we observe how the pole at $T=0$ and $\mu = \muSB = m_q \approx 282 \MeV$ builds up.}
  \label{fig:Z_sigma_diverging}
\end{figure}

Computing now the chiral crossover temperature as a function of the chemical potential, we obtain the phase boundary of our model in the plane spanned by the temperature and the chemical potential. 
Following this boundary to larger values of the chemical potential, we observe that there is no critical endpoint associated with a first-order phase transition line coming from low temperatures and large chemical potentials.\footnote{Such a critical endpoint at finite temperature and chemical potential may possibly be generated by the inclusion of additional terms in our ansatz \labelcref{eq:initial condition} as suggested by previous model studies, see, e.g., \cite{Schaefer:2008hk, Otto:2022jzl} for discussions of this aspect.}
Indeed, even in the zero-temperature limit, we find a crossover at $\mupc \approx 313 \MeV$, see \Cref{fig:Sign_diagram_of_Z,fig:renormalized_curvature_masses}. 
From \Cref{fig:Sign_diagram_of_Z}, we also deduce that the crossover line is fully located within the region where both wavefunction renormalizations are positive. 
However, at low temperatures, the crossover line lies very close to the edge of the region defined by $Z^\perp_{\pi} <0$ and~$Z^{\perp}_{\sigma}>0$, see the green region in the right panel of \Cref{fig:Sign_diagram_of_Z}. 
Note that this region quickly merges into a region where the wavefunction renormalizations associated with both mesons become negative. 
At low temperatures and $\mu>\mupc$, the dependence of the curvature masses of the mesons on the chemical potential is therefore completely dominated by the zero-crossings of the wavefunction renormalizations. 
\begin{figure*}[t]
  \begin{minipage}[b]{0.49\linewidth}
    \includegraphics{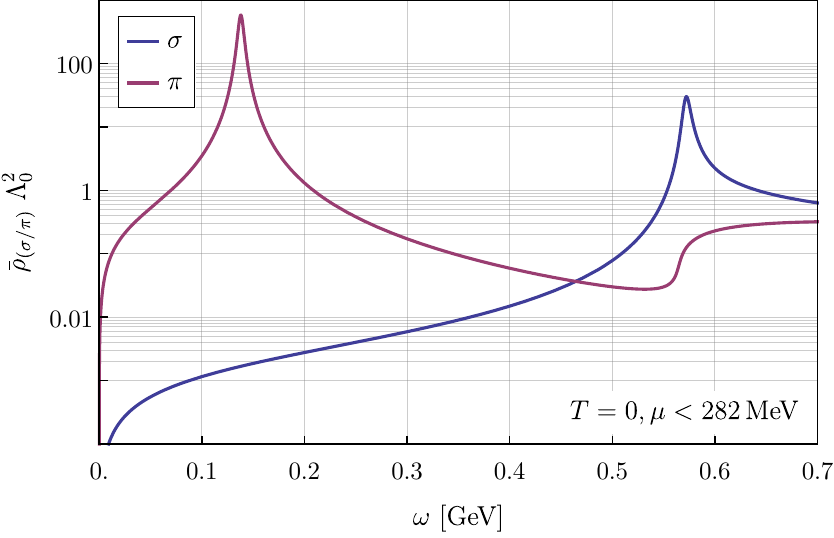}
  \end{minipage}
 \hfill%
 \begin{minipage}[b]{0.49\linewidth}
    \includegraphics{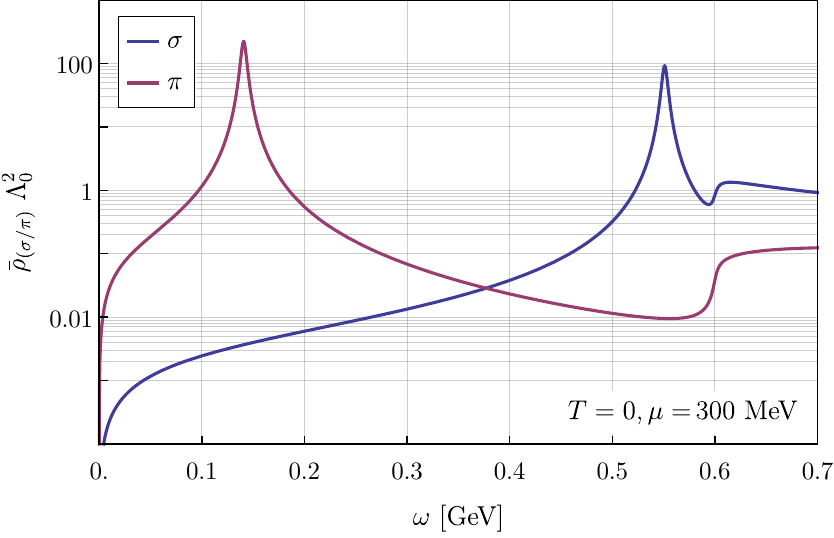}
  \end{minipage}
  \caption{Renormalized zero-temperature spectral functions of the mesons for $\mu< \muSB$ with $\muSB = m_q \approx 282\MeV$ (left panel) and~$\mu = 300 \MeV$ (right panel).
   Recall that the results are invariant under a shift of the chemical potential for~$\mu<m_q$ at $T=0$ because of the Silver-Blaze symmetry.}
  \label{fig:renormalized_spectral_function_T=0}
\end{figure*}
\begin{figure*}[t]
  \begin{minipage}[b]{0.49\linewidth}
    \includegraphics{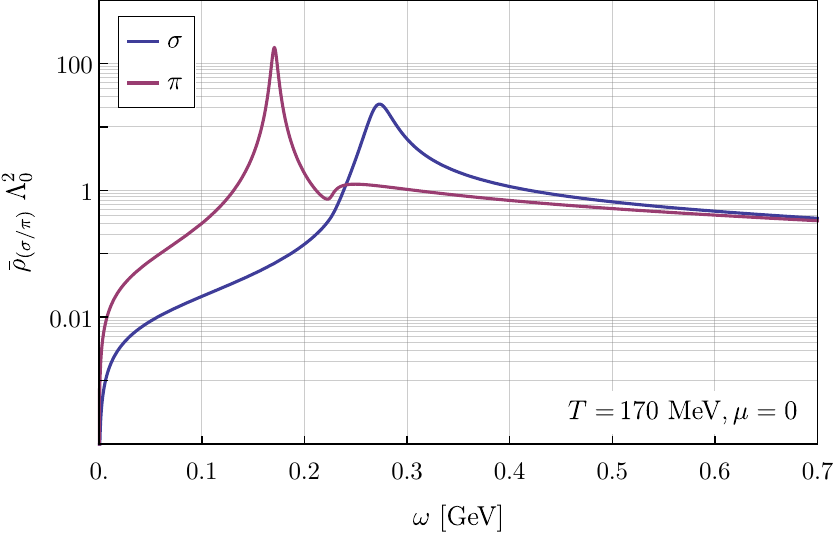}
  \end{minipage}
 \hfill%
 \begin{minipage}[b]{0.49\linewidth}
    \includegraphics{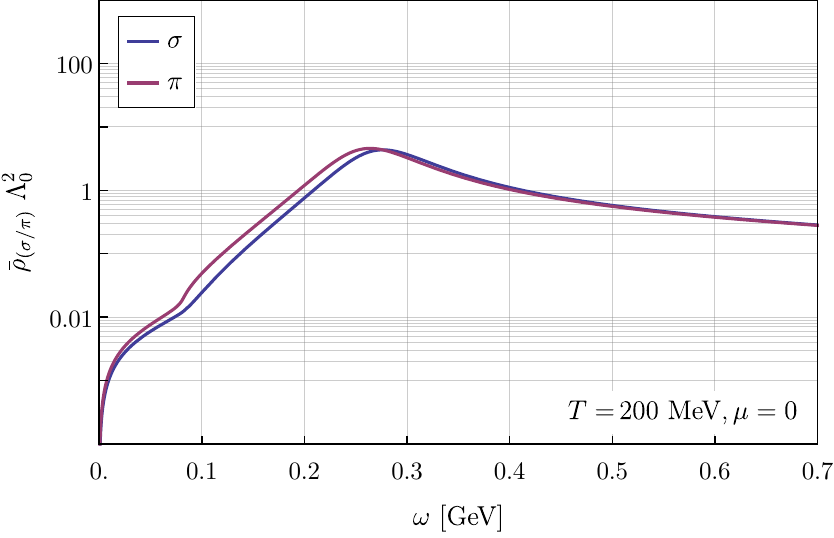}
  \end{minipage}
  \caption{Renormalized meson spectral functions at zero chemical potential and $T=170\MeV$ (left panel) and $T=200\MeV$ (right panel), respectively.}
  \label{fig:renormalized_spectral_function_T}
\end{figure*}

Interestingly, in addition to the aforementioned crossover point at~$\mupc \approx 313 \MeV$, we observe a point associated with a vanishing (renormalized) sigma mass at~$\mu = m_q$ in the zero-temperature limit. 
This point is associated with a diverging wavefunction renormalization of the sigma field. 
To be specific, the chemical potential dependent contributions to $Z^\perp_\sigma$ generate a divergence at the Fermi surface:\footnote{Strictly speaking, our wavefunction renormlization~$Z^\perp_\sigma$ is not well-defined at~$T=0$ and~$\mu=m_q$. This results from the fact that the projection used for the computation of~$Z^\perp_\sigma$ cannot be applied for all values of~$\mu$ because of non-analyticities of the two-point function in the zero-temperature limit at finite chemical potential, see our exact results for the two-point functions at zero temperature in \Cref{app: Explicit zero-temperature results} and our comment on Friedel oscillations above.} 
\begin{align}\label{eq:Z_sigma diverging}
\quad Z^\perp_\sigma (T=0,\mu) \propto \frac{h^2 \Nc}{6 \pi^2} \frac{\mu}{\sqrt{\mu^2 -m_q^2}} \theta(\mu - m_q)\, .
\end{align}
Since the bare sigma mass remains finite at~$\mu=m_q$, the renormalized sigma mass $\mb_\sigma$ jumps to zero at this point. 
The dependence of $Z^\perp_\sigma$ on the chemical potential is shown for various temperatures in \Cref{fig:Z_sigma_diverging} below.
We add that the pionic wavefunction renormalization does not exhibit such a behavior. 
In fact, the renormalized curvature mass of the pion as a function of the chemical potential at $T=0$ remains constant for $\mu<m_q$, as dictated by the analytic properties of quantum field theories together with the  Silver-Blaze symmetry.
Beyond the Silver-Blaze threshold, the pion curvature mass increases with the chemical potential, see right panel of \Cref{fig:renormalized_curvature_masses}. 

Last but not least, we would like to mention that the renormalized curvature mass of the pion in the vacuum limit agrees almost identically with the pole mass of $138 \MeV$ whereas the result for the unrenormlized curvature mass of the pion is found to deviate by almost $40\%$ from the value of the pole mass, see our discussion in \Cref{subsec: study I}. 
For the sigma mode, however, we find that the difference of the renormalized curvature mass and the corresponding pole mass is greater than the one of the unrenormalized curvature mass and the pole mass.

\subsubsection{Renormalized spectral functions}\label{subsubsec: renormalized spectral functions}
In accordance with our previous renormalization prescription for the fields, we define the renormalized spectral functions as follows:
\begin{align}
\overbar{\rho}_{(\sigma/\pi)}(\omega, \vec{Q}) = Z^\perp_{(\sigma/\pi)}\ \rho_{(\sigma/\pi)}(\omega, \vec{Q})\, .
\end{align}
Note that the wavefunction renormalizations do not change the energy of a resonance peak, i.e., pole masses are inherently 
renormalization-invariant quantities. 

As before, we shall only discuss the meson spectral functions for vanishing external momenta, $\vec{Q} = 0$. 
In the vacuum limit, we observe the sigma mass resonance to be at $\omega_{\text{res},\sigma} \approx 572 \MeV$, see left panel of \Cref{fig:renormalized_spectral_function_T=0}. 
From the pion spectral function we can read off a pole mass of $m_{\text{pole},\pi} \approx 138 \MeV$. 
Recall that we tuned our model parameters such that we obtain this value for the pion pole mass. 
Moreover, the pion spectral function shows the familiar structure for the decay $\pi^\prime \rightarrow \psib \psi$ at $\omega_{\text{decay}} = 2 m_q \approx 564 \MeV$ whereas the corresponding structure in the sigma spectral function for the process $\sigma^\prime \rightarrow \psib \psi$ is highly suppressed and therefore not visible on the scale of the plot in \Cref{fig:renormalized_spectral_function_T=0}.
We would like to add that the resonance peak in the spectral function of the sigma mode does not correspond to a Dirac-delta peak and hence cannot be associated with a pole mass since~$\omega_{\text{res},\sigma} > \omega_{\text{decay}}$. 
Note that this observation should be considered with caution as it may be an artifact of our parameter choice.
In other words, this may be changed by, e.g., introducing additional couplings (parameters) in the initial action~$\Gamma_{\Lambda_0}$.
At zero temperature and above the Silver-Blaze threshold, $\mu > m_q$, the situation changes rapidly as the kink-like structures associated with a decay now scale as $\omega_{\text{decay}} = 2 \mu$, see right panel of \Cref{fig:renormalized_spectral_function_T=0} for an illustration of this shift.
As mentioned in \Cref{subsec: study I}, the mesons can only decay if the excitation energy is high enough to create a pair of quarks of energy $\omega_{\text{F}} = (\pF^2 + m_q^2)^{\frac{1}{2}} = \mu$ each. 

At finite temperature and zero chemical potential, the resonance at $\omega_{\text{decay}} = 2 m_q$ is shifted as a function of the temperature because of the temperature dependence of the quark mass.
Moreover, we find that the resonance peak in the sigma spectral function broadens rapidly with increasing temperature and its height is continuously lowered. 
For temperatures $T\gtrsim 180 \MeV$, the renormalized meson spectral functions eventually become more and more degenerate, see \Cref{fig:renormalized_spectral_function_T}. 

\section{Summary and Outlook}
\label{sec:summary}
In the present work we discussed CS flows of chiral fermion-boson models within the functional RG framework.
Such flows are generated by the use of a corresponding IR regulator in the Wetterich equation.  
Unlike in the case of RG flows of the Wilsonian type, CS flows do not describe the evolution of a given theory from some high-momentum regime down to a low-momentum regime but rather specify the dependence of a given effective action on a mass scale, e.g., the fermion mass. 
In principle, the initialization of CS flows therefore requires the knowledge of the effective action of a given theory at some (mass) scale. 
In practice, an ansatz for the effective action must be chosen at some initial mass scale.

A CS regulator is advantageous as it gives access to Lorentz-invariant renormalization group flows in the vacuum limit and does not violate the Silver-Blaze symmetry of quantum field theories in the presence of a finite chemical potential~\cite{Braun:2022mgx}. 
Besides the fact that CS flows require an additional UV regularization, however, the introduction of an explicit fermion mass in form of an RG scale~$k$ as done by a CS regulator naturally breaks the chiral symmetry.  
For example, in case of our low-energy QCD model, $k$-dependent chiral-symmetry violating terms of arbitrarily high order in the meson fields are in general generated such that the effective action can no longer be split into a part with intact chiral symmetry and a source term depending only linearly on the sigma field, as naturally expected from a more general standpoint~\cite{Zinn-Justin:2002ecy}.
In order to systematically restore this regulator-induced explicit chiral symmetry breaking, we derived Ward-Takahashi identities which allow to restore the global chiral symmetry of the effective action in the CS flow. 
In other words, these identities systematically remove the unphysical symmetry breaking terms from the flow, rendering the computation of physical quantities from the corresponding effective action meaningful.
In our framework, physical explicit symmetry breaking must then be introduced by a linear source term. 
 
We demonstrated the application of our CS framework by computing different quantities, such as curvature masses, the phase diagram, momentum-dependent two-point functions of the mesons, and the spectral functions of the latter within a quark-meson-type model. 
To this end, we restricted ourselves to calculations at one-loop order in the large-$\Nc$ limit as the RG flow can be solved analytically in this case and therefore provided us with a clean test for our framework. 
In particular, it allowed us to demonstrate the strong effect of the regulator-induced explicit breaking of the chiral symmetry on physical observables. 

With our symmetrized CS framework at hand, we then analyzed the phase diagram of our quark-meson model in the plane spanned by the temperature and the quark chemical potential together with the meson two-point functions. 
From the latter, we also derived the corresponding spectral functions. 
In general, we find that our results are in accordance with previous results from studies of related models. 
In addition, our exact one-loop results provide the opportunity for an analytic understanding of intriguing phenomena potentially existing in different regimes of the phase diagram, such as the emergence of moat regimes and Friedel oscillations at finite chemical potential. 
For a combined analysis of the latter two phenomena, we refer the reader to~\cite{Rennecke:2024}. 
From a field-theoretical standpoint, our results for the meson two-point functions allow us to analyze the limits of the application of derivative expansions of the effective action, both at vanishing and finite chemical potential.

Of course, the CS framework developed in our present work only sets the methodological stage for future non-perturbative studies of thermodynamic and spectral properties of chiral fermion-boson models. 
Since our framework does not rely on a specific approximation of the effective action, a natural next step is to employ it for computations which include effects from bosonic fluctuations. 
Whereas such effects may be suppressed at very low temperatures along the chemical potential axis, they are known to be very relevant for an analysis of the phase diagram at finite temperature and chemical potential.   \\[-1ex]

{\it Acknowledgments.--} We thank Fabian Rennecke and Shi Yin for discussions and comments on the manuscript. 
As members of the fQCD collaboration~\cite{fQCD}, the authors also would like to thank the members of this collaboration for discussions. 
This work is supported in part by the Deutsche Forschungsgemeinschaft (DFG, German Research Foundation) through the Collaborative Research Center TransRegio CRC-TR 211 ``Strong-interaction matter under extreme conditions" -- project number 315477589 -- TRR 211. This work is also funded by the Deutsche Forschungsgemeinschaft (DFG, German Research Foundation) under Germany’s Excellence Strategy EXC 2181/1 - 390900948 (the Heidelberg STRUCTURES Excellence Cluster) and the Collaborative Research Centre SFB 1225 (ISOQUANT). It is also supported by EMMI.

\appendix

\section{UV regularization and counterterms}\label{app: counterterm UV reg}
In order to render the RG flows in the presence of a CS regulator finite, we employ an additional UV regularization by introducing a momentum cutoff~$\lambda$.
Afterwards, we then add corresponding counterterms to cancel those UV divergences which arise in the limit $\lambda \to \infty$. 
As in the case of ordinary regulator functions, the construction of counterterms is in principle constrained by the symmetries of the underlying quantum field theory. 
For our present study, the counterterm for the effective action reads
\begin{widetext}
\begin{align}
\ct_k(\Lambda,\phi) &=\frac{1}{32\pi^2} (k-\Lambda)(k+\Lambda + 2 h \sigma) \lim_{\lambda \rightarrow \infty} \Big(\left[ (k+h\sigma)^2 + (\Lambda + h \sigma)^2 + 2 h^2 \vec{\pi}^2\right] \ln\lambda - \lambda^2 \Big)\,.
\label{eq:counterterm_app}
\end{align}
\end{widetext}
The counterterms for $1$PI $n$-point functions with $n \geq 1$ are then readily obtained by differentiating this result $n$~times with respect to the fields.
Note that our analysis of UV divergences takes Lorentz symmetry into account, leading to counterterms where the cutoff parameter $\lambda$ represents a momentum scale for all components of the internal four-momenta.

Note that the expression in \labelcref{eq:counterterm_app} does not exhibit any dependence on the temperature or the chemical potential since loop contributions generated by these external parameters are naturally free of UV divergences. 
In other words, the counterterms provide an UV regularization for the vacuum theory but leave the behavior of the IR-regularized system with respect to temperature and chemical potential unaffected. 
Therefore and since the CS regulator respects the Silver-Blaze symmetry, our counterterms do also not violate the latter symmetry. 

It should be mentioned that our counterterms introduce an artificial scale dependence to the UV-regularized loop of the effective action.
In order to perform actual calculations, we have to explicitly choose a scale and this choice then belongs to the definition of our model. 
In our explicit calculations, we choose this scale to be $\Lambda_0$. 
We also add that the form of the counterterms is only unique up to finite terms in the limit~$\lambda\to\infty$. 
We have chosen them such that we have a non-trivial ground state for~$k\to 0$. 
For an ansatz of the effective action at the scale~$\Lambda_0$ which contains terms of higher orders in the fields, it is possible to adjust the counterterms such that an artificial scale dependence can be avoided. However, explicit calculations within such a setting are beyond the scope of this work.

Note that the different realization of the IR and UV regularization does not lead to an overall inconsistent regularization scheme in our case. 
As the counterterms respect the symmetries of the underlying theory and do not alter how momenta enter the loop integrals, they are consistent with the CS regulator which we use to regularize the IR.
In other words, the momentum modes of the vacuum and non-vacuum contributions to the loop integrals are treated consistently within our present setting.

Finally, we add that the counterterms inherit the symmetry and additivity properties of the associated loop integrals:
\begin{align}
\ct_\Lambda(k,\phi) = -\ct_k(\Lambda,\phi)
\end{align}
and
\begin{align}
\ct_k(\Lambda_0,\phi) + \ct_\Lambda(k,\phi) = \ct_\Lambda(\Lambda_0,\phi)\, .
\end{align}
\section{Symmetrization}\label{app: symmetrization}
The artificial breaking of the chiral symmetry induced by the CS regulator can be cured by employing a suitable symmetrization procedure.  
To this end, we begin by noting that the CS regulator leaves the $O(3)$ symmetry among the pion fields intact as it acts like a fermion mass term. 
The restoration of the $O(4)$ symmetry among the mesons can therefore be mapped onto the two-dimensional problem of restoring the circular symmetry in the plane spanned by $x=\sigma$ and $y = \sqrt{\vec{\pi}^2}$. 
The WTI in \labelcref{eq:SO(4) WTI} for a function symmetric in $x$ and $y$ then reads
\begin{align}
\left(y \frac{\partial}{\partial x} - x \frac{\partial}{\partial y} \right) g^{(\text{sym})}(x,y) = 0\, .
\end{align}
Our prescription for the restoration of the symmetry with respect to orthogonal transformations relies on the addition of an auxiliary function $c$ which serves as a ``counterterm" for symmetry-breaking terms. 
For an arbitrary function $g$, which is non-singular in $x=0$, we identify
\begin{equation}
g^{(\text{sym})} = g + c\, .
\end{equation}
The WTI can now be used to obtain a partial differential equation for $c$. 
Since the CS regulator does not affect the $O(3)$ symmetry in the pion subspace, our symmetrization procedure should leave this subspace unchanged as well. 
This constraint fixes the initial condition necessary to uniquely solve the WTI for~$c$:
\begin{align}
&\left(y \frac{\partial}{\partial x} - x \frac{\partial}{\partial y} \right) c(x,y) = f(x,y)
\end{align}
with
\begin{align}
&f(x,y) \coloneqq \left(x \frac{\partial}{\partial y} - y \frac{\partial}{\partial x} \right) g(x,y)
\end{align}
and
\begin{align}
c(0,y) = 0
\end{align}
for all $y \in \mathbb{R}$.
Using the fact that $g$ is a function of $y^2$ because of the intact $O(3)$ symmetry, we find
\begin{align}
&c(x,y) \\
&= F_{\sqrt{x^2+y^2}}\left(\arcsin\left(\frac{x}{\sqrt{x^2+y^2}}\right)\right) - F_{\sqrt{x^2+y^2}}(0)\ ,\nn
\end{align}
where the function~$F$ is defined as the following indefinite integral:
\begin{align}
&F_\alpha(s) = \int \d s\ f(\alpha \sin(s), \alpha \cos(s))\, .
\end{align}

In order to gain a better understanding of what exactly our symmetrization procedure does, let us finally consider the function
\begin{align}
\CO_k(\phi) = \widetilde{\CO}\left((h \sigma + k)^2 + h^2 \vec{\pi}^2 \right)\,,
\end{align}
which is supposed to be symmetrized in the field variables by the addition of a suitable function $c$. 
Following the steps above, we obtain
\begin{align}
c_k(\phi) = \widetilde{\CO}\left(k^2 + h^2 \phi^2 \right) - \widetilde{\CO}\left((h \sigma + k)^2 + h^2 \vec{\pi}^2 \right)\, .
\end{align}
Thus, the symmetrization eventually leads us to
\begin{align}
\CO^{(\text{sym})}_k(\phi) = \CO_k(\phi)  + c_k(\phi) = \widetilde{\CO}\left(k^2 + h^2 \phi^2 \right)\, .
\end{align}
\section{Analytic zero-temperature results}
\label{app: Explicit zero-temperature results}
In the following we present analytic results from calculations performed in the symmetrized CS scheme at zero temperature and finite chemical potential~$\mu \geq 0$. 
For the fully momentum-dependent two-point functions of the mesons, we obtain
{\allowdisplaybreaks
\begin{widetext}
\begin{align}
\label{app: eq:full 2-point correlator}
\Gammat^{(2)}_k(Q)&= \Gammat^{(2)}_{\Lambda_0}(Q) - 8 h^2 \Nc \left[\CI_{x}^{(2)}(Q) - \CM^{(2)}_x(Q) \right]\Big|^{x = \sqrt{k^2+m_q^2}}_{x = \sqrt{\Lambda_0^2+m_q^2}}\, ,\\
\CI_{x}^{(2)}(Q) &= \frac{1}{16 \pi^2}\left\lbrace \uM{4}\ x^2 \ln\left(\frac{x^2}{\Lambda_0^2} \right) + \frac{Q^2 \uM{4}+ 2m_q^2 (\uM{4} + \eta)}{4}\ \times \right.\nn\\
&\left.\times\ \left[ 2\ln\left(\frac{x^2}{\Lambda_0^2} \right)  - \frac{\sqrt{Q^2+ 4x^2}}{Q} \ln\left(\frac{(\sqrt{Q^2+ 4x^2}-Q)x^2}{(\sqrt{Q^2+ 4x^2}+Q)(Q^2 +x^2)+2Qx^2} \right)\right]\right\rbrace\, ,
\label{eq:vac_2point}
\\
\CM^{(2)}_x(Q) &= \frac{\theta(\mu - x)}{8\pi^2} \left\lbrace \uM{4} \left[\mu \sqrt{\mu^2-x^2}-x^2 \artanh\left(\frac{\sqrt{\mu^2-x^2}}{\mu}\right)\right] + \frac{Q^2 \uM{4} + 2m_q^2 (\uM{4} + \eta)}{4}\ \times \vphantom{\frac{4x^2 Q_0^2 + Q^2\left(Q \mu + \sqrt{\mu^2-x^2}\sqrt{Q^2+4x^2} \right)^2}{4x^2 Q_0^2 + Q^2\left(Q \mu - \sqrt{\mu^2-x^2}\sqrt{Q^2+4x^2} \right)^2}} \right. \nn \\
&\times \left[ \frac{\mu}{|\vec{Q}|} \ln\left(\frac{4 \mu^2 Q_0^2 + \left(Q^2 - 2|\vec{Q}| \sqrt{\mu^2-x^2} \right)^2}{4 \mu^2 Q_0^2 + \left(Q^2 + 2|\vec{Q}| \sqrt{\mu^2-x^2} \right)^2} \right) - \frac{|Q_0|}{|\vec{Q}|}\left(\arctan\left(\frac{Q^4 + 4 x^2 \vec{Q}^2 + 4\mu^2 (Q_0^2 - \vec{Q}^2)}{8 |Q_0| |\vec{Q}|\ \mu \sqrt{\mu^2 - x^2}}\right) -\frac{\pi}{2} \right) \right.\nn\\
&\left.\left. + \frac{\sqrt{Q^2 + 4x^2}}{2Q} \ln\left(\frac{4x^4 Q_0^2 + Q^2\left(Q \mu + \sqrt{\mu^2-x^2}\sqrt{Q^2+4x^2} \right)^2}{4x^4 Q_0^2 + Q^2\left(Q \mu - \sqrt{\mu^2-x^2}\sqrt{Q^2+4x^2} \right)^2} \right) -2 \ln\left(\frac{\mu+\sqrt{\mu^2-x^2}}{x} \right)  \right] \right\rbrace \, .
\label{eq:matt_2point}
\end{align}
\end{widetext}
With respect to} derivative expansions of the effective action and the computation of wavefunction renormalization factors, we note that already the vacuum contribution~$\CI_{x}^{(2)}$ to the two-point functions of the mesons can only be expanded in powers of~$Q$ about~$Q=0$ for~$x>0$, i.e., we must have either~$k>0$ or~$m_q>0$.  
For~$x=0$, i.e., $k=0$ and~$m_q=0$, the vacuum contribution is non-analytic in~$Q$ and wavefunction renormalizations can therefore not be defined by an expansion of the two-point functions in~$Q$ about $Q=0$ up to second order.\footnote{Note that the unexpanded expressions for the two-point functions are well-defined in the limits~$m_q\to 0$ and~$k\to 0$ and that the latter two limits commute.}
For example, if we first take the limit~$k\to0$ and then expand the vacuum contribution in \labelcref{eq:vac_2point} in powers of~$Q$ and finally take the limit~$m_q\to0$, the non-analytic behavior of these functions in~$Q$ manifests itself in logarithmically diverging wavefunction renormalizations for the mesons, see \labelcref{eq:vacuum Z}. 
For~$x>0$, an expansion of the two-point functions in~$Q$ about~$Q=0$ is well-defined and therefore allows to derive well-defined wavefunction renormalizations for the two meson fields. 
In particular, taking the limit~$k\to 0$ first, we can expand the two-point functions in~$Q$ which eventually leads us to the wavefunction renormalizations presented in \labelcref{eq:vacuum Z}. 
Similarly, we could also take the limit~$m_q\to 0$ first and then expand the two-point functions in~$Q$ to derive well-defined RG equations for the wavefunction renormalizations for~$k>0$. 
In any case, if both mass scales go to zero, an expansion of the two-point functions around~$Q=0$ breaks down and thus cannot provide us with a meaningful description of wavefunction renormalizations. 
With respect to computations of RG flows of wavefunction renormalizations of meson fields with smooth Wilsonian-type regulators, we note that an expansion of the meson two-point functions in powers of the momenta can in general be expected to be well-defined for any finite RG scale. 
However, the limit of vanishing RG scale remains problematic if we are considering the limit of vanishing quark mass as well.

In addition, the chemical potential dependent part $\CM^{(2)}$ of the two-point functions in \labelcref{eq:matt_2point} exhibits non-analyticities in the momenta which are not cured by a finite quark mass or a finite value of the RG scale. 
To be specific, the function $\CM^{(2)}$ at $Q_0=0$ is non-analytic in $|\vec{Q}|=2\sqrt{\mu^2 - x^2}$, which renders an expansion in spatial momenta~$\vec{Q}$ potentially ill-defined, see again our discussion in \Cref{sec:reneffects_po}. 
We shall come back to this issue below when considering the corresponding wavefunction renormalizations. 
We also note that the chemical potential dependent part of the two-point functions for values of the three-momentum $\vec{Q}$ with $0<|\vec{Q}|\leq 2\sqrt{\mu^2 - x^2}$ is non-analytic in $Q_0=0$. 
This property restricts the values of~$\vec{Q}$ for which an analytic continuation of the zero-temperature theory back to Minkowski space can be reliably performed. 
Of course, this aspect is of particular importance for the computation of spectral functions. 
In our case, taking the value of $\vec{Q}=0$ is the most convenient choice for the presentation of spectral functions.

Finally, we add that, for $\mu^2 > (k^2+m_q^2)$, the two-point functions are discontinuous at $(Q_0, \vec{Q}) = (0, \vec{0})$ such that the limit $Q \rightarrow 0$ does no longer exist. 
The case of vanishing four-momentum has then to be performed by an iterated limit. 
Accordingly, our explicit zero-temperature curvature masses read:
\begin{widetext} 
\begin{align}
\begin{pmatrix}
m^2_\sigma & \vec{0}^\tp\\
\vec{0} & m^2_\pi\ \uM{3}
\end{pmatrix}(k) &= \lim_{\vec{Q} \rightarrow \vec{0}} \lim_{\vphantom{\vec{0}}Q_0 \rightarrow 0} \Gammat_k^{(2)}(Q)\, ,\\
 \lim_{\vec{Q} \rightarrow \vec{0}}\lim_{\vphantom{\vec{0}}Q_0 \rightarrow 0} \CI^{(2)}_x(Q) &= \lim_{\vphantom{\vec{0}}Q_0 \rightarrow 0} \lim_{\vec{Q} \rightarrow \vec{0}} \CI^{(2)}_x(Q) = \frac{1}{16\pi^2} \left(\uM{4}\ x^2 \ln\left(\frac{x^2}{\Lambda_0^2}\right) + m_q^2\ (\uM{4}+\eta) \left[ \ln\left(\frac{x^2}{\Lambda_0^2}\right) -2 \right]\right)\, ,\\
  \lim_{\vec{Q} \rightarrow \vec{0}}\lim_{\vphantom{\vec{0}}Q_0 \rightarrow 0} \CM^{(2)}_x(Q) &= \frac{\theta(\mu -x)}{8\pi^2}\left(\uM{4}\left[ \mu \sqrt{\mu^2-x^2}-x^2 \artanh\left(\frac{\sqrt{\mu^2-x^2}}{\mu}\right) \right] \nn \right.\\ 
  &\left. \hspace{2cm} \qquad  - m_q^2 (\uM{4} + \eta) \ln\left(\frac{\mu + \sqrt{\mu^2-x^2}}{x} \right)\right)\, . 
\end{align}
\end{widetext}
The presence of a finite quark chemical potential explicitly breaks Lorentz invariance and therefore the two-point functions are sensitive to the specific order of the limits~$Q_0\to 0$ and~$\vec{Q}\to \vec{0}$. 
In order to obtain results for the curvature masses which are consistent with those derived from the effective potential, we have to take the static limit, where $Q_0$ is set to zero first~\cite{Topfel:2024cll}. 

The wavefunction renormalizations associated with the meson fields can be obtained from a projection of the corresponding two-point functions that includes taking derivatives with respect to momenta and taking limits. 
Note that, at zero temperature and $\mu^2 \geq (k^2+m_q^2)$, it is not allowed to interchange the projection onto the wavefunction renormalization factors and the integration over the time-like loop momenta~\cite{Topfel:2024cll}. 
In any case, we only consider wavefunction renormalizations associated with modes perpendicular to the heat bath:
\begin{align}
\begin{pmatrix}
Z^\perp_\sigma & \vec{0}^\tp\\
\vec{0} & Z^\perp_\pi\ \uM{3}
\end{pmatrix}(k) &= \lim_{\vec{Q} \rightarrow \vec{0}} \lim_{\vphantom{\vec{0}}Q_0 \rightarrow 0} \frac{1}{2}\frac{\d^2}{\d |\vec{Q}|^2} \Gammat_k^{(2)}(Q)\,. \label{app: eq:Z projection}
\end{align}
The computation of the right-hand side of this equation requires to compute derivatives of the vacuum contribution and the chemical potential dependent contribution to the two-point functions of the mesons, respectively:
\begin{widetext} 
\begin{align}
 \lim_{\vec{Q} \rightarrow \vec{0}}\lim_{\vphantom{\vec{0}}Q_0 \rightarrow 0} \frac{1}{2}\frac{\d^2}{\d |\vec{Q}|^2} \CI^{(2)}_x(Q) &=  
 \lim_{\vphantom{\vec{0}}Q_0 \rightarrow 0}\lim_{\vec{Q} \rightarrow \vec{0}} \frac{1}{2}\frac{\d^2}{\d |\vec{Q}|^2} \CI^{(2)}_x(Q)=
 \frac{1}{32\pi^2}\left( \uM{4}\left(2 + \ln\left(\frac{x^2}{\Lambda_0^2}\right)\right) + (\uM{4}+\eta) \frac{m_q^2}{3x^2} \right)\, ,
 \label{app: eq:Z projection I}\\
  \lim_{\vec{Q} \rightarrow \vec{0}}\lim_{\vphantom{\vec{0}}Q_0 \rightarrow 0} \frac{1}{2}\frac{\d^2}{\d |\vec{Q}|^2} \CM^{(2)}_x(Q) &= \frac{\theta(\mu -x)}{16\pi^2} \left( - \uM{4} \ln\left(\frac{\mu + \sqrt{\mu^2-x^2}}{x} \right) + (\uM{4}+\eta) \frac{m_q^2\ \mu}{6 x^2 \sqrt{\mu^2-x^2}} \right)\, .
  \label{app: eq:Z projection M}
\end{align}
\end{widetext}
As in our computations of the curvature masses, we again consider an iterated limit which is required because of the non-analyticities of the two-point functions discussed above. 
A few comments are still in order here: 
First of all, the chemical potential dependent part of $Z^\perp_\sigma$ has a pole of order $\frac{1}{2}$ at $\mu = x$. 
Assuming $\mu < \Lambda_0$, this implies that the wavefunction renormalization associated with the sigma field is ill-defined at the Fermi surface $\mu^2 = (k^2 + m_q^2)$. 
This is a direct consequence of applying a projection involving derivatives with respect to momenta to the full two-point correlator, although this correlator at $Q_0=0$ is not analytic in~$\vec{Q}$. 
Moreover, we observe that the first term in the bracket on the right-hand side of \labelcref{app: eq:Z projection M} is negative since the logarithm associated with this term is positive for all $\mu > x$. 
This leads to a continuous decrease of the wavefunction renormalizations when we increase the chemical potential beyond the Silver-Blaze threshold. 
Depending on the UV parameter $Z_{\Lambda_0}$ and in particular the Yukawa coupling, the wavefunction renormalizations can therefore become negative, which is indeed what we observe, see our discussion in \Cref{subsubsec:renmass_pd}. 

\bibliography{qcd}{}

\begin{thebibliography}{81}%
\makeatletter
\providecommand \@ifxundefined [1]{%
 \@ifx{#1\undefined}
}%
\providecommand \@ifnum [1]{%
 \ifnum #1\expandafter \@firstoftwo
 \else \expandafter \@secondoftwo
 \fi
}%
\providecommand \@ifx [1]{%
 \ifx #1\expandafter \@firstoftwo
 \else \expandafter \@secondoftwo
 \fi
}%
\providecommand \natexlab [1]{#1}%
\providecommand \enquote  [1]{``#1''}%
\providecommand \bibnamefont  [1]{#1}%
\providecommand \bibfnamefont [1]{#1}%
\providecommand \citenamefont [1]{#1}%
\providecommand \href@noop [0]{\@secondoftwo}%
\providecommand \href [0]{\begingroup \@sanitize@url \@href}%
\providecommand \@href[1]{\@@startlink{#1}\@@href}%
\providecommand \@@href[1]{\endgroup#1\@@endlink}%
\providecommand \@sanitize@url [0]{\catcode `\\12\catcode `\$12\catcode
  `\&12\catcode `\#12\catcode `\^12\catcode `\_12\catcode `\%12\relax}%
\providecommand \@@startlink[1]{}%
\providecommand \@@endlink[0]{}%
\providecommand \url  [0]{\begingroup\@sanitize@url \@url }%
\providecommand \@url [1]{\endgroup\@href {#1}{\urlprefix }}%
\providecommand \urlprefix  [0]{URL }%
\providecommand \Eprint [0]{\href }%
\providecommand \doibase [0]{https://doi.org/}%
\providecommand \selectlanguage [0]{\@gobble}%
\providecommand \bibinfo  [0]{\@secondoftwo}%
\providecommand \bibfield  [0]{\@secondoftwo}%
\providecommand \translation [1]{[#1]}%
\providecommand \BibitemOpen [0]{}%
\providecommand \bibitemStop [0]{}%
\providecommand \bibitemNoStop [0]{.\EOS\space}%
\providecommand \EOS [0]{\spacefactor3000\relax}%
\providecommand \BibitemShut  [1]{\csname bibitem#1\endcsname}%
\let\auto@bib@innerbib\@empty
\bibitem [{\citenamefont {Braun}\ \emph
  {et~al.}(2023{\natexlab{a}})\citenamefont {Braun} \emph
  {et~al.}}]{Braun:2022mgx}%
  \BibitemOpen
  \bibfield  {author} {\bibinfo {author} {\bibfnamefont {J.}~\bibnamefont
  {Braun}} \emph {et~al.},\ }\bibfield  {title} {\bibinfo {title}
  {{Renormalised spectral flows}},\ }\href
  {https://doi.org/10.21468/SciPostPhysCore.6.3.061} {\bibfield  {journal}
  {\bibinfo  {journal} {SciPost Phys. Core}\ }\textbf {\bibinfo {volume} {6}},\
  \bibinfo {pages} {061} (\bibinfo {year} {2023}{\natexlab{a}})},\ \Eprint
  {https://arxiv.org/abs/2206.10232} {arXiv:2206.10232 [hep-th]} \BibitemShut
  {NoStop}%
\bibitem [{\citenamefont {Fehre}\ \emph {et~al.}(2023)\citenamefont {Fehre},
  \citenamefont {Litim}, \citenamefont {Pawlowski},\ and\ \citenamefont
  {Reichert}}]{Fehre:2021eob}%
  \BibitemOpen
  \bibfield  {author} {\bibinfo {author} {\bibfnamefont {J.}~\bibnamefont
  {Fehre}}, \bibinfo {author} {\bibfnamefont {D.~F.}\ \bibnamefont {Litim}},
  \bibinfo {author} {\bibfnamefont {J.~M.}\ \bibnamefont {Pawlowski}},\ and\
  \bibinfo {author} {\bibfnamefont {M.}~\bibnamefont {Reichert}},\ }\bibfield
  {title} {\bibinfo {title} {{Lorentzian Quantum Gravity and the Graviton
  Spectral Function}},\ }\href {https://doi.org/10.1103/PhysRevLett.130.081501}
  {\bibfield  {journal} {\bibinfo  {journal} {Phys. Rev. Lett.}\ }\textbf
  {\bibinfo {volume} {130}},\ \bibinfo {pages} {081501} (\bibinfo {year}
  {2023})},\ \Eprint {https://arxiv.org/abs/2111.13232} {arXiv:2111.13232
  [hep-th]} \BibitemShut {NoStop}%
\bibitem [{\citenamefont {Gasenzer}\ and\ \citenamefont
  {Pawlowski}(2008)}]{Gasenzer:2007za}%
  \BibitemOpen
  \bibfield  {author} {\bibinfo {author} {\bibfnamefont {T.}~\bibnamefont
  {Gasenzer}}\ and\ \bibinfo {author} {\bibfnamefont {J.~M.}\ \bibnamefont
  {Pawlowski}},\ }\bibfield  {title} {\bibinfo {title} {{Towards
  far-from-equilibrium quantum field dynamics: A functional
  renormalisation-group approach}},\ }\href
  {https://doi.org/10.1016/j.physletb.2008.10.049} {\bibfield  {journal}
  {\bibinfo  {journal} {Phys. Lett. B}\ }\textbf {\bibinfo {volume} {670}},\
  \bibinfo {pages} {135} (\bibinfo {year} {2008})},\ \Eprint
  {https://arxiv.org/abs/0710.4627} {arXiv:0710.4627 [cond-mat.other]}
  \BibitemShut {NoStop}%
\bibitem [{\citenamefont {Gasenzer}\ \emph {et~al.}(2010)\citenamefont
  {Gasenzer}, \citenamefont {Kessler},\ and\ \citenamefont
  {Pawlowski}}]{Gasenzer:2010rq}%
  \BibitemOpen
  \bibfield  {author} {\bibinfo {author} {\bibfnamefont {T.}~\bibnamefont
  {Gasenzer}}, \bibinfo {author} {\bibfnamefont {S.}~\bibnamefont {Kessler}},\
  and\ \bibinfo {author} {\bibfnamefont {J.~M.}\ \bibnamefont {Pawlowski}},\
  }\bibfield  {title} {\bibinfo {title} {{Far-from-equilibrium quantum
  many-body dynamics}},\ }\href
  {https://doi.org/10.1140/epjc/s10052-010-1430-3} {\bibfield  {journal}
  {\bibinfo  {journal} {Eur. Phys. J. C}\ }\textbf {\bibinfo {volume} {70}},\
  \bibinfo {pages} {423} (\bibinfo {year} {2010})},\ \Eprint
  {https://arxiv.org/abs/1003.4163} {arXiv:1003.4163 [cond-mat.quant-gas]}
  \BibitemShut {NoStop}%
\bibitem [{\citenamefont {Floerchinger}(2012)}]{Floerchinger:2011sc}%
  \BibitemOpen
  \bibfield  {author} {\bibinfo {author} {\bibfnamefont {S.}~\bibnamefont
  {Floerchinger}},\ }\bibfield  {title} {\bibinfo {title} {{Analytic
  Continuation of Functional Renormalization Group Equations}},\ }\href
  {https://doi.org/10.1007/JHEP05(2012)021} {\bibfield  {journal} {\bibinfo
  {journal} {JHEP}\ }\textbf {\bibinfo {volume} {05}},\ \bibinfo {pages}
  {021}},\ \Eprint {https://arxiv.org/abs/1112.4374} {arXiv:1112.4374 [hep-th]}
  \BibitemShut {NoStop}%
\bibitem [{\citenamefont {Strodthoff}\ \emph {et~al.}(2012)\citenamefont
  {Strodthoff}, \citenamefont {Schaefer},\ and\ \citenamefont {von
  Smekal}}]{Strodthoff:2011tz}%
  \BibitemOpen
  \bibfield  {author} {\bibinfo {author} {\bibfnamefont {N.}~\bibnamefont
  {Strodthoff}}, \bibinfo {author} {\bibfnamefont {B.-J.}\ \bibnamefont
  {Schaefer}},\ and\ \bibinfo {author} {\bibfnamefont {L.}~\bibnamefont {von
  Smekal}},\ }\bibfield  {title} {\bibinfo {title} {{Quark-meson-diquark model
  for two-color QCD}},\ }\href {https://doi.org/10.1103/PhysRevD.85.074007}
  {\bibfield  {journal} {\bibinfo  {journal} {Phys. Rev. D}\ }\textbf {\bibinfo
  {volume} {85}},\ \bibinfo {pages} {074007} (\bibinfo {year} {2012})},\
  \Eprint {https://arxiv.org/abs/1112.5401} {arXiv:1112.5401 [hep-ph]}
  \BibitemShut {NoStop}%
\bibitem [{\citenamefont {Kamikado}\ \emph {et~al.}(2014)\citenamefont
  {Kamikado}, \citenamefont {Strodthoff}, \citenamefont {von Smekal},\ and\
  \citenamefont {Wambach}}]{Kamikado:2013sia}%
  \BibitemOpen
  \bibfield  {author} {\bibinfo {author} {\bibfnamefont {K.}~\bibnamefont
  {Kamikado}}, \bibinfo {author} {\bibfnamefont {N.}~\bibnamefont
  {Strodthoff}}, \bibinfo {author} {\bibfnamefont {L.}~\bibnamefont {von
  Smekal}},\ and\ \bibinfo {author} {\bibfnamefont {J.}~\bibnamefont
  {Wambach}},\ }\bibfield  {title} {\bibinfo {title} {{Real-time correlation
  functions in the $O(N)$ model from the functional renormalization group}},\
  }\href {https://doi.org/10.1140/epjc/s10052-014-2806-6} {\bibfield  {journal}
  {\bibinfo  {journal} {Eur. Phys. J. C}\ }\textbf {\bibinfo {volume} {74}},\
  \bibinfo {pages} {2806} (\bibinfo {year} {2014})},\ \Eprint
  {https://arxiv.org/abs/1302.6199} {arXiv:1302.6199 [hep-ph]} \BibitemShut
  {NoStop}%
\bibitem [{\citenamefont {Tripolt}\ \emph {et~al.}(2014)\citenamefont
  {Tripolt}, \citenamefont {Strodthoff}, \citenamefont {von Smekal},\ and\
  \citenamefont {Wambach}}]{Tripolt:2013jra}%
  \BibitemOpen
  \bibfield  {author} {\bibinfo {author} {\bibfnamefont {R.-A.}\ \bibnamefont
  {Tripolt}}, \bibinfo {author} {\bibfnamefont {N.}~\bibnamefont {Strodthoff}},
  \bibinfo {author} {\bibfnamefont {L.}~\bibnamefont {von Smekal}},\ and\
  \bibinfo {author} {\bibfnamefont {J.}~\bibnamefont {Wambach}},\ }\bibfield
  {title} {\bibinfo {title} {{Spectral Functions for the Quark-Meson Model
  Phase Diagram from the Functional Renormalization Group}},\ }\href
  {https://doi.org/10.1103/PhysRevD.89.034010} {\bibfield  {journal} {\bibinfo
  {journal} {Phys. Rev. D}\ }\textbf {\bibinfo {volume} {89}},\ \bibinfo
  {pages} {034010} (\bibinfo {year} {2014})},\ \Eprint
  {https://arxiv.org/abs/1311.0630} {arXiv:1311.0630 [hep-ph]} \BibitemShut
  {NoStop}%
\bibitem [{\citenamefont {Pawlowski}\ and\ \citenamefont
  {Strodthoff}(2015)}]{Pawlowski:2015mia}%
  \BibitemOpen
  \bibfield  {author} {\bibinfo {author} {\bibfnamefont {J.~M.}\ \bibnamefont
  {Pawlowski}}\ and\ \bibinfo {author} {\bibfnamefont {N.}~\bibnamefont
  {Strodthoff}},\ }\bibfield  {title} {\bibinfo {title} {{Real time correlation
  functions and the functional renormalization group}},\ }\href
  {https://doi.org/10.1103/PhysRevD.92.094009} {\bibfield  {journal} {\bibinfo
  {journal} {Phys. Rev.}\ }\textbf {\bibinfo {volume} {D92}},\ \bibinfo {pages}
  {094009} (\bibinfo {year} {2015})},\ \Eprint
  {https://arxiv.org/abs/1508.01160} {arXiv:1508.01160 [hep-ph]} \BibitemShut
  {NoStop}%
\bibitem [{\citenamefont {Yokota}\ \emph {et~al.}(2016)\citenamefont {Yokota},
  \citenamefont {Kunihiro},\ and\ \citenamefont {Morita}}]{Yokota:2016tip}%
  \BibitemOpen
  \bibfield  {author} {\bibinfo {author} {\bibfnamefont {T.}~\bibnamefont
  {Yokota}}, \bibinfo {author} {\bibfnamefont {T.}~\bibnamefont {Kunihiro}},\
  and\ \bibinfo {author} {\bibfnamefont {K.}~\bibnamefont {Morita}},\
  }\bibfield  {title} {\bibinfo {title} {{Functional renormalization group
  analysis of the soft mode at the QCD critical point}},\ }\href
  {https://doi.org/10.1093/ptep/ptw062} {\bibfield  {journal} {\bibinfo
  {journal} {PTEP}\ }\textbf {\bibinfo {volume} {2016}},\ \bibinfo {pages}
  {073D01} (\bibinfo {year} {2016})},\ \Eprint
  {https://arxiv.org/abs/1603.02147} {arXiv:1603.02147 [hep-ph]} \BibitemShut
  {NoStop}%
\bibitem [{\citenamefont {Kamikado}\ \emph {et~al.}(2017)\citenamefont
  {Kamikado}, \citenamefont {Kanazawa},\ and\ \citenamefont
  {Uchino}}]{Kamikado:2016chk}%
  \BibitemOpen
  \bibfield  {author} {\bibinfo {author} {\bibfnamefont {K.}~\bibnamefont
  {Kamikado}}, \bibinfo {author} {\bibfnamefont {T.}~\bibnamefont {Kanazawa}},\
  and\ \bibinfo {author} {\bibfnamefont {S.}~\bibnamefont {Uchino}},\
  }\bibfield  {title} {\bibinfo {title} {{Mobile impurity in a Fermi sea from
  the functional renormalization group analytically continued to real time}},\
  }\href {https://doi.org/10.1103/PhysRevA.95.013612} {\bibfield  {journal}
  {\bibinfo  {journal} {Phys. Rev. A}\ }\textbf {\bibinfo {volume} {95}},\
  \bibinfo {pages} {013612} (\bibinfo {year} {2017})},\ \Eprint
  {https://arxiv.org/abs/1606.03721} {arXiv:1606.03721 [cond-mat.quant-gas]}
  \BibitemShut {NoStop}%
\bibitem [{\citenamefont {Jung}\ \emph {et~al.}(2017)\citenamefont {Jung},
  \citenamefont {Rennecke}, \citenamefont {Tripolt}, \citenamefont {von
  Smekal},\ and\ \citenamefont {Wambach}}]{Jung:2016yxl}%
  \BibitemOpen
  \bibfield  {author} {\bibinfo {author} {\bibfnamefont {C.}~\bibnamefont
  {Jung}}, \bibinfo {author} {\bibfnamefont {F.}~\bibnamefont {Rennecke}},
  \bibinfo {author} {\bibfnamefont {R.-A.}\ \bibnamefont {Tripolt}}, \bibinfo
  {author} {\bibfnamefont {L.}~\bibnamefont {von Smekal}},\ and\ \bibinfo
  {author} {\bibfnamefont {J.}~\bibnamefont {Wambach}},\ }\bibfield  {title}
  {\bibinfo {title} {{In-Medium Spectral Functions of Vector- and Axial-Vector
  Mesons from the Functional Renormalization Group}},\ }\href
  {https://doi.org/10.1103/PhysRevD.95.036020} {\bibfield  {journal} {\bibinfo
  {journal} {Phys. Rev. D}\ }\textbf {\bibinfo {volume} {95}},\ \bibinfo
  {pages} {036020} (\bibinfo {year} {2017})},\ \Eprint
  {https://arxiv.org/abs/1610.08754} {arXiv:1610.08754 [hep-ph]} \BibitemShut
  {NoStop}%
\bibitem [{\citenamefont {Pawlowski}\ \emph {et~al.}(2018)\citenamefont
  {Pawlowski}, \citenamefont {Strodthoff},\ and\ \citenamefont
  {Wink}}]{Pawlowski:2017gxj}%
  \BibitemOpen
  \bibfield  {author} {\bibinfo {author} {\bibfnamefont {J.~M.}\ \bibnamefont
  {Pawlowski}}, \bibinfo {author} {\bibfnamefont {N.}~\bibnamefont
  {Strodthoff}},\ and\ \bibinfo {author} {\bibfnamefont {N.}~\bibnamefont
  {Wink}},\ }\bibfield  {title} {\bibinfo {title} {{Finite temperature spectral
  functions in the O(N)-model}},\ }\href
  {https://doi.org/10.1103/PhysRevD.98.074008} {\bibfield  {journal} {\bibinfo
  {journal} {Phys. Rev. D}\ }\textbf {\bibinfo {volume} {98}},\ \bibinfo
  {pages} {074008} (\bibinfo {year} {2018})},\ \Eprint
  {https://arxiv.org/abs/1711.07444} {arXiv:1711.07444 [hep-th]} \BibitemShut
  {NoStop}%
\bibitem [{\citenamefont {Yokota}\ \emph {et~al.}(2017)\citenamefont {Yokota},
  \citenamefont {Kunihiro},\ and\ \citenamefont {Morita}}]{Yokota:2017uzu}%
  \BibitemOpen
  \bibfield  {author} {\bibinfo {author} {\bibfnamefont {T.}~\bibnamefont
  {Yokota}}, \bibinfo {author} {\bibfnamefont {T.}~\bibnamefont {Kunihiro}},\
  and\ \bibinfo {author} {\bibfnamefont {K.}~\bibnamefont {Morita}},\
  }\bibfield  {title} {\bibinfo {title} {{Tachyonic instability of the scalar
  mode prior to the QCD critical point based on the functional
  renormalization-group method in the two-flavor case}},\ }\href
  {https://doi.org/10.1103/PhysRevD.96.074028} {\bibfield  {journal} {\bibinfo
  {journal} {Phys. Rev. D}\ }\textbf {\bibinfo {volume} {96}},\ \bibinfo
  {pages} {074028} (\bibinfo {year} {2017})},\ \Eprint
  {https://arxiv.org/abs/1707.05520} {arXiv:1707.05520 [hep-ph]} \BibitemShut
  {NoStop}%
\bibitem [{\citenamefont {Wang}\ and\ \citenamefont
  {Zhuang}(2017)}]{Wang:2017vis}%
  \BibitemOpen
  \bibfield  {author} {\bibinfo {author} {\bibfnamefont {Z.}~\bibnamefont
  {Wang}}\ and\ \bibinfo {author} {\bibfnamefont {P.}~\bibnamefont {Zhuang}},\
  }\bibfield  {title} {\bibinfo {title} {{Meson spectral functions at finite
  temperature and isospin density with the functional renormalization group}},\
  }\href {https://doi.org/10.1103/PhysRevD.96.014006} {\bibfield  {journal}
  {\bibinfo  {journal} {Phys. Rev. D}\ }\textbf {\bibinfo {volume} {96}},\
  \bibinfo {pages} {014006} (\bibinfo {year} {2017})},\ \Eprint
  {https://arxiv.org/abs/1703.01035} {arXiv:1703.01035 [hep-ph]} \BibitemShut
  {NoStop}%
\bibitem [{\citenamefont {Tripolt}\ \emph {et~al.}(2019)\citenamefont
  {Tripolt}, \citenamefont {Jung}, \citenamefont {Tanji}, \citenamefont {von
  Smekal},\ and\ \citenamefont {Wambach}}]{Tripolt:2018jre}%
  \BibitemOpen
  \bibfield  {author} {\bibinfo {author} {\bibfnamefont {R.-A.}\ \bibnamefont
  {Tripolt}}, \bibinfo {author} {\bibfnamefont {C.}~\bibnamefont {Jung}},
  \bibinfo {author} {\bibfnamefont {N.}~\bibnamefont {Tanji}}, \bibinfo
  {author} {\bibfnamefont {L.}~\bibnamefont {von Smekal}},\ and\ \bibinfo
  {author} {\bibfnamefont {J.}~\bibnamefont {Wambach}},\ }\bibfield  {title}
  {\bibinfo {title} {{In-medium spectral functions and dilepton rates with the
  Functional Renormalization Group}},\ }\href
  {https://doi.org/10.1016/j.nuclphysa.2018.08.017} {\bibfield  {journal}
  {\bibinfo  {journal} {Nucl. Phys. A}\ }\textbf {\bibinfo {volume} {982}},\
  \bibinfo {pages} {775} (\bibinfo {year} {2019})},\ \Eprint
  {https://arxiv.org/abs/1807.04952} {arXiv:1807.04952 [hep-ph]} \BibitemShut
  {NoStop}%
\bibitem [{\citenamefont {Tripolt}\ \emph
  {et~al.}(2018{\natexlab{a}})\citenamefont {Tripolt}, \citenamefont {Weyrich},
  \citenamefont {von Smekal},\ and\ \citenamefont {Wambach}}]{Tripolt:2018qvi}%
  \BibitemOpen
  \bibfield  {author} {\bibinfo {author} {\bibfnamefont {R.-A.}\ \bibnamefont
  {Tripolt}}, \bibinfo {author} {\bibfnamefont {J.}~\bibnamefont {Weyrich}},
  \bibinfo {author} {\bibfnamefont {L.}~\bibnamefont {von Smekal}},\ and\
  \bibinfo {author} {\bibfnamefont {J.}~\bibnamefont {Wambach}},\ }\bibfield
  {title} {\bibinfo {title} {{Fermionic spectral functions with the Functional
  Renormalization Group}},\ }\href {https://doi.org/10.1103/PhysRevD.98.094002}
  {\bibfield  {journal} {\bibinfo  {journal} {Phys. Rev. D}\ }\textbf {\bibinfo
  {volume} {98}},\ \bibinfo {pages} {094002} (\bibinfo {year}
  {2018}{\natexlab{a}})},\ \Eprint {https://arxiv.org/abs/1807.11708}
  {arXiv:1807.11708 [hep-ph]} \BibitemShut {NoStop}%
\bibitem [{\citenamefont {Corell}\ \emph {et~al.}(2021)\citenamefont {Corell},
  \citenamefont {Cyrol}, \citenamefont {Heller},\ and\ \citenamefont
  {Pawlowski}}]{Corell:2019jxh}%
  \BibitemOpen
  \bibfield  {author} {\bibinfo {author} {\bibfnamefont {L.}~\bibnamefont
  {Corell}}, \bibinfo {author} {\bibfnamefont {A.~K.}\ \bibnamefont {Cyrol}},
  \bibinfo {author} {\bibfnamefont {M.}~\bibnamefont {Heller}},\ and\ \bibinfo
  {author} {\bibfnamefont {J.~M.}\ \bibnamefont {Pawlowski}},\ }\bibfield
  {title} {\bibinfo {title} {{Flowing with the temporal renormalization
  group}},\ }\href {https://doi.org/10.1103/PhysRevD.104.025005} {\bibfield
  {journal} {\bibinfo  {journal} {Phys. Rev. D}\ }\textbf {\bibinfo {volume}
  {104}},\ \bibinfo {pages} {025005} (\bibinfo {year} {2021})},\ \Eprint
  {https://arxiv.org/abs/1910.09369} {arXiv:1910.09369 [hep-th]} \BibitemShut
  {NoStop}%
\bibitem [{\citenamefont {Huelsmann}\ \emph {et~al.}(2020)\citenamefont
  {Huelsmann}, \citenamefont {Schlichting},\ and\ \citenamefont
  {Scior}}]{Huelsmann:2020xcy}%
  \BibitemOpen
  \bibfield  {author} {\bibinfo {author} {\bibfnamefont {S.}~\bibnamefont
  {Huelsmann}}, \bibinfo {author} {\bibfnamefont {S.}~\bibnamefont
  {Schlichting}},\ and\ \bibinfo {author} {\bibfnamefont {P.}~\bibnamefont
  {Scior}},\ }\bibfield  {title} {\bibinfo {title} {{Spectral functions from
  the real-time functional renormalization group}},\ }\href
  {https://doi.org/10.1103/PhysRevD.102.096004} {\bibfield  {journal} {\bibinfo
   {journal} {Phys. Rev. D}\ }\textbf {\bibinfo {volume} {102}},\ \bibinfo
  {pages} {096004} (\bibinfo {year} {2020})},\ \Eprint
  {https://arxiv.org/abs/2009.04194} {arXiv:2009.04194 [hep-ph]} \BibitemShut
  {NoStop}%
\bibitem [{\citenamefont {Jung}\ \emph {et~al.}(2021)\citenamefont {Jung},
  \citenamefont {Otto}, \citenamefont {Tripolt},\ and\ \citenamefont {von
  Smekal}}]{Jung:2021ipc}%
  \BibitemOpen
  \bibfield  {author} {\bibinfo {author} {\bibfnamefont {C.}~\bibnamefont
  {Jung}}, \bibinfo {author} {\bibfnamefont {J.-H.}\ \bibnamefont {Otto}},
  \bibinfo {author} {\bibfnamefont {R.-A.}\ \bibnamefont {Tripolt}},\ and\
  \bibinfo {author} {\bibfnamefont {L.}~\bibnamefont {von Smekal}},\ }\bibfield
   {title} {\bibinfo {title} {{Self-consistent O(4) model spectral functions
  from analytically continued functional renormalization group flows}},\ }\href
  {https://doi.org/10.1103/PhysRevD.104.094011} {\bibfield  {journal} {\bibinfo
   {journal} {Phys. Rev. D}\ }\textbf {\bibinfo {volume} {104}},\ \bibinfo
  {pages} {094011} (\bibinfo {year} {2021})},\ \Eprint
  {https://arxiv.org/abs/2107.10748} {arXiv:2107.10748 [hep-ph]} \BibitemShut
  {NoStop}%
\bibitem [{\citenamefont {Tan}\ \emph {et~al.}(2022)\citenamefont {Tan},
  \citenamefont {Chen},\ and\ \citenamefont {Fu}}]{Tan:2021zid}%
  \BibitemOpen
  \bibfield  {author} {\bibinfo {author} {\bibfnamefont {Y.-y.}\ \bibnamefont
  {Tan}}, \bibinfo {author} {\bibfnamefont {Y.-r.}\ \bibnamefont {Chen}},\ and\
  \bibinfo {author} {\bibfnamefont {W.-j.}\ \bibnamefont {Fu}},\ }\bibfield
  {title} {\bibinfo {title} {{Real-time dynamics of the $O(4)$ scalar theory
  within the fRG approach}},\ }\href
  {https://doi.org/10.21468/SciPostPhys.12.1.026} {\bibfield  {journal}
  {\bibinfo  {journal} {SciPost Phys.}\ }\textbf {\bibinfo {volume} {12}},\
  \bibinfo {pages} {026} (\bibinfo {year} {2022})},\ \Eprint
  {https://arxiv.org/abs/2107.06482} {arXiv:2107.06482 [hep-ph]} \BibitemShut
  {NoStop}%
\bibitem [{\citenamefont {Heller}\ and\ \citenamefont
  {Pawlowski}(2021)}]{Heller:2021wan}%
  \BibitemOpen
  \bibfield  {author} {\bibinfo {author} {\bibfnamefont {M.}~\bibnamefont
  {Heller}}\ and\ \bibinfo {author} {\bibfnamefont {J.~M.}\ \bibnamefont
  {Pawlowski}},\ }\bibfield  {title} {\bibinfo {title} {{Causal Temporal
  Renormalisation Group Flow of the Energy-Momentum Tensor}},\ }\Eprint
  {https://arxiv.org/abs/2112.12652} {arXiv:2112.12652 [hep-th]}  (\bibinfo
  {year} {2021})\BibitemShut {NoStop}%
\bibitem [{\citenamefont {Roth}\ \emph {et~al.}(2022)\citenamefont {Roth},
  \citenamefont {Schweitzer}, \citenamefont {Sieke},\ and\ \citenamefont {von
  Smekal}}]{Roth:2021nrd}%
  \BibitemOpen
  \bibfield  {author} {\bibinfo {author} {\bibfnamefont {J.~V.}\ \bibnamefont
  {Roth}}, \bibinfo {author} {\bibfnamefont {D.}~\bibnamefont {Schweitzer}},
  \bibinfo {author} {\bibfnamefont {L.~J.}\ \bibnamefont {Sieke}},\ and\
  \bibinfo {author} {\bibfnamefont {L.}~\bibnamefont {von Smekal}},\ }\bibfield
   {title} {\bibinfo {title} {{Real-time methods for spectral functions}},\
  }\href {https://doi.org/10.1103/PhysRevD.105.116017} {\bibfield  {journal}
  {\bibinfo  {journal} {Phys. Rev. D}\ }\textbf {\bibinfo {volume} {105}},\
  \bibinfo {pages} {116017} (\bibinfo {year} {2022})},\ \Eprint
  {https://arxiv.org/abs/2112.12568} {arXiv:2112.12568 [hep-ph]} \BibitemShut
  {NoStop}%
\bibitem [{\citenamefont {Roth}\ and\ \citenamefont {von
  Smekal}(2023)}]{Roth:2023wbp}%
  \BibitemOpen
  \bibfield  {author} {\bibinfo {author} {\bibfnamefont {J.~V.}\ \bibnamefont
  {Roth}}\ and\ \bibinfo {author} {\bibfnamefont {L.}~\bibnamefont {von
  Smekal}},\ }\bibfield  {title} {\bibinfo {title} {{Critical dynamics in a
  real-time formulation of the functional renormalization group}},\ }\href
  {https://doi.org/10.1007/JHEP10(2023)065} {\bibfield  {journal} {\bibinfo
  {journal} {JHEP}\ }\textbf {\bibinfo {volume} {10}},\ \bibinfo {pages}
  {065}},\ \Eprint {https://arxiv.org/abs/2303.11817} {arXiv:2303.11817
  [hep-ph]} \BibitemShut {NoStop}%
\bibitem [{\citenamefont {Horak}\ \emph {et~al.}(2024)\citenamefont {Horak},
  \citenamefont {Ihssen}, \citenamefont {Pawlowski}, \citenamefont {Wessely},\
  and\ \citenamefont {Wink}}]{Horak:2023hkp}%
  \BibitemOpen
  \bibfield  {author} {\bibinfo {author} {\bibfnamefont {J.}~\bibnamefont
  {Horak}}, \bibinfo {author} {\bibfnamefont {F.}~\bibnamefont {Ihssen}},
  \bibinfo {author} {\bibfnamefont {J.~M.}\ \bibnamefont {Pawlowski}}, \bibinfo
  {author} {\bibfnamefont {J.}~\bibnamefont {Wessely}},\ and\ \bibinfo {author}
  {\bibfnamefont {N.}~\bibnamefont {Wink}},\ }\bibfield  {title} {\bibinfo
  {title} {{Scalar spectral functions from the spectral functional
  renormalization group}},\ }\href
  {https://doi.org/10.1103/PhysRevD.110.056009} {\bibfield  {journal} {\bibinfo
   {journal} {Phys. Rev. D}\ }\textbf {\bibinfo {volume} {110}},\ \bibinfo
  {pages} {056009} (\bibinfo {year} {2024})},\ \Eprint
  {https://arxiv.org/abs/2303.16719} {arXiv:2303.16719 [hep-th]} \BibitemShut
  {NoStop}%
\bibitem [{\citenamefont {Dupuis}\ \emph {et~al.}(2021)\citenamefont {Dupuis},
  \citenamefont {Canet}, \citenamefont {Eichhorn}, \citenamefont {Metzner},
  \citenamefont {Pawlowski}, \citenamefont {Tissier},\ and\ \citenamefont
  {Wschebor}}]{Dupuis:2020fhh}%
  \BibitemOpen
  \bibfield  {author} {\bibinfo {author} {\bibfnamefont {N.}~\bibnamefont
  {Dupuis}}, \bibinfo {author} {\bibfnamefont {L.}~\bibnamefont {Canet}},
  \bibinfo {author} {\bibfnamefont {A.}~\bibnamefont {Eichhorn}}, \bibinfo
  {author} {\bibfnamefont {W.}~\bibnamefont {Metzner}}, \bibinfo {author}
  {\bibfnamefont {J.~M.}\ \bibnamefont {Pawlowski}}, \bibinfo {author}
  {\bibfnamefont {M.}~\bibnamefont {Tissier}},\ and\ \bibinfo {author}
  {\bibfnamefont {N.}~\bibnamefont {Wschebor}},\ }\bibfield  {title} {\bibinfo
  {title} {{The nonperturbative functional renormalization group and its
  applications}},\ }\href {https://doi.org/10.1016/j.physrep.2021.01.001}
  {\bibfield  {journal} {\bibinfo  {journal} {Phys. Rept.}\ }\textbf {\bibinfo
  {volume} {910}},\ \bibinfo {pages} {1} (\bibinfo {year} {2021})},\ \Eprint
  {https://arxiv.org/abs/2006.04853} {arXiv:2006.04853 [cond-mat.stat-mech]}
  \BibitemShut {NoStop}%
\bibitem [{\citenamefont {Mark\'o}\ \emph {et~al.}(2014)\citenamefont
  {Mark\'o}, \citenamefont {Reinosa},\ and\ \citenamefont
  {Sz\'ep}}]{Marko:2014hea}%
  \BibitemOpen
  \bibfield  {author} {\bibinfo {author} {\bibfnamefont {G.}~\bibnamefont
  {Mark\'o}}, \bibinfo {author} {\bibfnamefont {U.}~\bibnamefont {Reinosa}},\
  and\ \bibinfo {author} {\bibfnamefont {Z.}~\bibnamefont {Sz\'ep}},\
  }\bibfield  {title} {\bibinfo {title} {{Bose-Einstein condensation and Silver
  Blaze property from the two-loop $\Phi$-derivable approximation}},\ }\href
  {https://doi.org/10.1103/PhysRevD.90.125021} {\bibfield  {journal} {\bibinfo
  {journal} {Phys. Rev. D}\ }\textbf {\bibinfo {volume} {90}},\ \bibinfo
  {pages} {125021} (\bibinfo {year} {2014})},\ \Eprint
  {https://arxiv.org/abs/1410.6998} {arXiv:1410.6998 [hep-ph]} \BibitemShut
  {NoStop}%
\bibitem [{\citenamefont {Khan}\ \emph {et~al.}(2015)\citenamefont {Khan},
  \citenamefont {Pawlowski}, \citenamefont {Rennecke},\ and\ \citenamefont
  {Scherer}}]{Khan:2015puu}%
  \BibitemOpen
  \bibfield  {author} {\bibinfo {author} {\bibfnamefont {N.}~\bibnamefont
  {Khan}}, \bibinfo {author} {\bibfnamefont {J.~M.}\ \bibnamefont {Pawlowski}},
  \bibinfo {author} {\bibfnamefont {F.}~\bibnamefont {Rennecke}},\ and\
  \bibinfo {author} {\bibfnamefont {M.~M.}\ \bibnamefont {Scherer}},\
  }\bibfield  {title} {\bibinfo {title} {{The Phase Diagram of QC2D from
  Functional Methods}},\ }\href@noop {} {\  (\bibinfo {year} {2015})},\ \Eprint
  {https://arxiv.org/abs/1512.03673} {arXiv:1512.03673 [hep-ph]} \BibitemShut
  {NoStop}%
\bibitem [{\citenamefont {Braun}\ \emph {et~al.}(2021)\citenamefont {Braun},
  \citenamefont {D\"ornfeld}, \citenamefont {Schallmo},\ and\ \citenamefont
  {T\"opfel}}]{Braun:2020bhy}%
  \BibitemOpen
  \bibfield  {author} {\bibinfo {author} {\bibfnamefont {J.}~\bibnamefont
  {Braun}}, \bibinfo {author} {\bibfnamefont {T.}~\bibnamefont {D\"ornfeld}},
  \bibinfo {author} {\bibfnamefont {B.}~\bibnamefont {Schallmo}},\ and\
  \bibinfo {author} {\bibfnamefont {S.}~\bibnamefont {T\"opfel}},\ }\bibfield
  {title} {\bibinfo {title} {{Renormalization group studies of dense
  relativistic systems}},\ }\href {https://doi.org/10.1103/PhysRevD.104.096002}
  {\bibfield  {journal} {\bibinfo  {journal} {Phys. Rev. D}\ }\textbf {\bibinfo
  {volume} {104}},\ \bibinfo {pages} {096002} (\bibinfo {year} {2021})},\
  \Eprint {https://arxiv.org/abs/2008.05978} {arXiv:2008.05978 [hep-ph]}
  \BibitemShut {NoStop}%
\bibitem [{\citenamefont {Rennecke}\ and\ \citenamefont
  {Pisarski}(2022)}]{Rennecke:2021ovl}%
  \BibitemOpen
  \bibfield  {author} {\bibinfo {author} {\bibfnamefont {F.}~\bibnamefont
  {Rennecke}}\ and\ \bibinfo {author} {\bibfnamefont {R.~D.}\ \bibnamefont
  {Pisarski}},\ }\bibfield  {title} {\bibinfo {title} {{Moat Regimes in QCD and
  their Signatures in Heavy-Ion Collisions}},\ }\href
  {https://doi.org/10.22323/1.400.0016} {\bibfield  {journal} {\bibinfo
  {journal} {PoS}\ }\textbf {\bibinfo {volume} {CPOD2021}},\ \bibinfo {pages}
  {016} (\bibinfo {year} {2022})},\ \Eprint {https://arxiv.org/abs/2110.02625}
  {arXiv:2110.02625 [hep-ph]} \BibitemShut {NoStop}%
\bibitem [{\citenamefont {Rennecke}\ \emph {et~al.}(2023)\citenamefont
  {Rennecke}, \citenamefont {Pisarski},\ and\ \citenamefont
  {Rischke}}]{Rennecke:2023xhc}%
  \BibitemOpen
  \bibfield  {author} {\bibinfo {author} {\bibfnamefont {F.}~\bibnamefont
  {Rennecke}}, \bibinfo {author} {\bibfnamefont {R.~D.}\ \bibnamefont
  {Pisarski}},\ and\ \bibinfo {author} {\bibfnamefont {D.~H.}\ \bibnamefont
  {Rischke}},\ }\bibfield  {title} {\bibinfo {title} {{Particle interferometry
  in a moat regime}},\ }\href {https://doi.org/10.1103/PhysRevD.107.116011}
  {\bibfield  {journal} {\bibinfo  {journal} {Phys. Rev. D}\ }\textbf {\bibinfo
  {volume} {107}},\ \bibinfo {pages} {116011} (\bibinfo {year} {2023})},\
  \Eprint {https://arxiv.org/abs/2301.11484} {arXiv:2301.11484 [hep-ph]}
  \BibitemShut {NoStop}%
\bibitem [{\citenamefont {Schon}\ and\ \citenamefont
  {Thies}(2000)}]{Schon:2000qy}%
  \BibitemOpen
  \bibfield  {author} {\bibinfo {author} {\bibfnamefont {V.}~\bibnamefont
  {Schon}}\ and\ \bibinfo {author} {\bibfnamefont {M.}~\bibnamefont {Thies}},\
  }\bibfield  {title} {\bibinfo {title} {{2-D model field theories at finite
  temperature and density}},\ }\Eprint {https://arxiv.org/abs/hep-th/0008175}
  {arXiv:hep-th/0008175}  (\bibinfo {year} {2000})\BibitemShut {NoStop}%
\bibitem [{\citenamefont {Buballa}\ and\ \citenamefont
  {Carignano}(2015)}]{Buballa:2014tba}%
  \BibitemOpen
  \bibfield  {author} {\bibinfo {author} {\bibfnamefont {M.}~\bibnamefont
  {Buballa}}\ and\ \bibinfo {author} {\bibfnamefont {S.}~\bibnamefont
  {Carignano}},\ }\bibfield  {title} {\bibinfo {title} {{Inhomogeneous chiral
  condensates}},\ }\href {https://doi.org/10.1016/j.ppnp.2014.11.001}
  {\bibfield  {journal} {\bibinfo  {journal} {Prog. Part. Nucl. Phys.}\
  }\textbf {\bibinfo {volume} {81}},\ \bibinfo {pages} {39} (\bibinfo {year}
  {2015})},\ \Eprint {https://arxiv.org/abs/1406.1367} {arXiv:1406.1367
  [hep-ph]} \BibitemShut {NoStop}%
\bibitem [{\citenamefont {Litim}\ and\ \citenamefont
  {Pawlowski}(2002{\natexlab{a}})}]{Litim:2001ky}%
  \BibitemOpen
  \bibfield  {author} {\bibinfo {author} {\bibfnamefont {D.~F.}\ \bibnamefont
  {Litim}}\ and\ \bibinfo {author} {\bibfnamefont {J.~M.}\ \bibnamefont
  {Pawlowski}},\ }\bibfield  {title} {\bibinfo {title} {{Perturbation theory
  and renormalization group equations}},\ }\href
  {https://doi.org/10.1103/PhysRevD.65.081701} {\bibfield  {journal} {\bibinfo
  {journal} {Phys. Rev. D}\ }\textbf {\bibinfo {volume} {65}},\ \bibinfo
  {pages} {081701} (\bibinfo {year} {2002}{\natexlab{a}})},\ \Eprint
  {https://arxiv.org/abs/hep-th/0111191} {arXiv:hep-th/0111191} \BibitemShut
  {NoStop}%
\bibitem [{\citenamefont {Litim}\ and\ \citenamefont
  {Pawlowski}(2002{\natexlab{b}})}]{Litim:2002xm}%
  \BibitemOpen
  \bibfield  {author} {\bibinfo {author} {\bibfnamefont {D.~F.}\ \bibnamefont
  {Litim}}\ and\ \bibinfo {author} {\bibfnamefont {J.~M.}\ \bibnamefont
  {Pawlowski}},\ }\bibfield  {title} {\bibinfo {title} {Completeness and
  consistency of renormalisation group flows},\ }\href@noop {} {\bibfield
  {journal} {\bibinfo  {journal} {Phys. Rev.}\ }\textbf {\bibinfo {volume}
  {D66}},\ \bibinfo {pages} {025030} (\bibinfo {year} {2002}{\natexlab{b}})},\
  \Eprint {https://arxiv.org/abs/hep-th/0202188} {hep-th/0202188} \BibitemShut
  {NoStop}%
\bibitem [{\citenamefont {Gei\ss{}el}\ \emph {et~al.}(2024)\citenamefont
  {Gei\ss{}el}, \citenamefont {Gorda},\ and\ \citenamefont
  {Braun}}]{Geissel:2024nmx}%
  \BibitemOpen
  \bibfield  {author} {\bibinfo {author} {\bibfnamefont {A.}~\bibnamefont
  {Gei\ss{}el}}, \bibinfo {author} {\bibfnamefont {T.}~\bibnamefont {Gorda}},\
  and\ \bibinfo {author} {\bibfnamefont {J.}~\bibnamefont {Braun}},\ }\bibfield
   {title} {\bibinfo {title} {{Pressure and speed of sound in two-flavor
  color-superconducting quark matter at next-to-leading order}},\ }\Eprint
  {https://arxiv.org/abs/2403.18010} {arXiv:2403.18010 [hep-ph]}  (\bibinfo
  {year} {2024})\BibitemShut {NoStop}%
\bibitem [{\citenamefont {Wetterich}(1993)}]{Wetterich:1992yh}%
  \BibitemOpen
  \bibfield  {author} {\bibinfo {author} {\bibfnamefont {C.}~\bibnamefont
  {Wetterich}},\ }\bibfield  {title} {\bibinfo {title} {Exact evolution
  equation for the effective potential},\ }\href@noop {} {\bibfield  {journal}
  {\bibinfo  {journal} {Phys. Lett.}\ }\textbf {\bibinfo {volume} {B301}},\
  \bibinfo {pages} {90} (\bibinfo {year} {1993})}\BibitemShut {NoStop}%
\bibitem [{\citenamefont {Pawlowski}(2007)}]{Pawlowski:2005xe}%
  \BibitemOpen
  \bibfield  {author} {\bibinfo {author} {\bibfnamefont {J.~M.}\ \bibnamefont
  {Pawlowski}},\ }\bibfield  {title} {\bibinfo {title} {{Aspects of the
  functional renormalisation group}},\ }\href
  {https://doi.org/10.1016/j.aop.2007.01.007} {\bibfield  {journal} {\bibinfo
  {journal} {Annals Phys.}\ }\textbf {\bibinfo {volume} {322}},\ \bibinfo
  {pages} {2831} (\bibinfo {year} {2007})},\ \Eprint
  {https://arxiv.org/abs/hep-th/0512261} {arXiv:hep-th/0512261} \BibitemShut
  {NoStop}%
\bibitem [{\citenamefont {Gies}(2012)}]{Gies:2006wv}%
  \BibitemOpen
  \bibfield  {author} {\bibinfo {author} {\bibfnamefont {H.}~\bibnamefont
  {Gies}},\ }\bibfield  {title} {\bibinfo {title} {{Introduction to the
  functional RG and applications to gauge theories}},\ }\href
  {https://doi.org/10.1007/978-3-642-27320-9_6} {\bibfield  {journal} {\bibinfo
   {journal} {Lect. Notes Phys.}\ }\textbf {\bibinfo {volume} {852}},\ \bibinfo
  {pages} {287} (\bibinfo {year} {2012})},\ \Eprint
  {https://arxiv.org/abs/hep-ph/0611146} {arXiv:hep-ph/0611146 [hep-ph]}
  \BibitemShut {NoStop}%
\bibitem [{\citenamefont {Ihssen}\ \emph {et~al.}(2023)\citenamefont {Ihssen},
  \citenamefont {Pawlowski}, \citenamefont {Sattler},\ and\ \citenamefont
  {Wink}}]{Ihssen:2023xlp}%
  \BibitemOpen
  \bibfield  {author} {\bibinfo {author} {\bibfnamefont {F.}~\bibnamefont
  {Ihssen}}, \bibinfo {author} {\bibfnamefont {J.~M.}\ \bibnamefont
  {Pawlowski}}, \bibinfo {author} {\bibfnamefont {F.~R.}\ \bibnamefont
  {Sattler}},\ and\ \bibinfo {author} {\bibfnamefont {N.}~\bibnamefont
  {Wink}},\ }\bibfield  {title} {\bibinfo {title} {{Towards quantitative
  precision for QCD at large densities}},\ }\Eprint
  {https://arxiv.org/abs/2309.07335} {arXiv:2309.07335 [hep-th]}  (\bibinfo
  {year} {2023})\BibitemShut {NoStop}%
\bibitem [{\citenamefont {Otto}\ \emph {et~al.}(2022)\citenamefont {Otto},
  \citenamefont {Busch},\ and\ \citenamefont {Schaefer}}]{Otto:2022jzl}%
  \BibitemOpen
  \bibfield  {author} {\bibinfo {author} {\bibfnamefont {K.}~\bibnamefont
  {Otto}}, \bibinfo {author} {\bibfnamefont {C.}~\bibnamefont {Busch}},\ and\
  \bibinfo {author} {\bibfnamefont {B.-J.}\ \bibnamefont {Schaefer}},\
  }\bibfield  {title} {\bibinfo {title} {{Regulator scheme dependence of the
  chiral phase transition at high densities}},\ }\href
  {https://doi.org/10.1103/PhysRevD.106.094018} {\bibfield  {journal} {\bibinfo
   {journal} {Phys. Rev. D}\ }\textbf {\bibinfo {volume} {106}},\ \bibinfo
  {pages} {094018} (\bibinfo {year} {2022})},\ \Eprint
  {https://arxiv.org/abs/2206.13067} {arXiv:2206.13067 [hep-ph]} \BibitemShut
  {NoStop}%
\bibitem [{\citenamefont {Gies}\ and\ \citenamefont
  {Wetterich}(2004)}]{Gies:2002hq}%
  \BibitemOpen
  \bibfield  {author} {\bibinfo {author} {\bibfnamefont {H.}~\bibnamefont
  {Gies}}\ and\ \bibinfo {author} {\bibfnamefont {C.}~\bibnamefont
  {Wetterich}},\ }\bibfield  {title} {\bibinfo {title} {Universality of
  spontaneous chiral symmetry breaking in gauge theories},\ }\href@noop {}
  {\bibfield  {journal} {\bibinfo  {journal} {Phys. Rev.}\ }\textbf {\bibinfo
  {volume} {D69}},\ \bibinfo {pages} {025001} (\bibinfo {year} {2004})},\
  \Eprint {https://arxiv.org/abs/hep-th/0209183} {hep-th/0209183} \BibitemShut
  {NoStop}%
\bibitem [{\citenamefont {Braun}(2009)}]{Braun:2008pi}%
  \BibitemOpen
  \bibfield  {author} {\bibinfo {author} {\bibfnamefont {J.}~\bibnamefont
  {Braun}},\ }\bibfield  {title} {\bibinfo {title} {{The QCD Phase Boundary
  from Quark-Gluon Dynamics}},\ }\href
  {https://doi.org/10.1140/epjc/s10052-009-1136-6} {\bibfield  {journal}
  {\bibinfo  {journal} {Eur. Phys. J.}\ }\textbf {\bibinfo {volume} {C64}},\
  \bibinfo {pages} {459} (\bibinfo {year} {2009})},\ \Eprint
  {https://arxiv.org/abs/0810.1727} {arXiv:0810.1727 [hep-ph]} \BibitemShut
  {NoStop}%
\bibitem [{\citenamefont {Mitter}\ \emph {et~al.}(2015)\citenamefont {Mitter},
  \citenamefont {Pawlowski},\ and\ \citenamefont
  {Strodthoff}}]{Mitter:2014wpa}%
  \BibitemOpen
  \bibfield  {author} {\bibinfo {author} {\bibfnamefont {M.}~\bibnamefont
  {Mitter}}, \bibinfo {author} {\bibfnamefont {J.~M.}\ \bibnamefont
  {Pawlowski}},\ and\ \bibinfo {author} {\bibfnamefont {N.}~\bibnamefont
  {Strodthoff}},\ }\bibfield  {title} {\bibinfo {title} {{Chiral symmetry
  breaking in continuum QCD}},\ }\href
  {https://doi.org/10.1103/PhysRevD.91.054035} {\bibfield  {journal} {\bibinfo
  {journal} {Phys. Rev. D}\ }\textbf {\bibinfo {volume} {91}},\ \bibinfo
  {pages} {054035} (\bibinfo {year} {2015})},\ \Eprint
  {https://arxiv.org/abs/1411.7978} {arXiv:1411.7978 [hep-ph]} \BibitemShut
  {NoStop}%
\bibitem [{\citenamefont {Braun}\ \emph
  {et~al.}(2016{\natexlab{a}})\citenamefont {Braun}, \citenamefont {Fister},
  \citenamefont {Pawlowski},\ and\ \citenamefont {Rennecke}}]{Braun:2014ata}%
  \BibitemOpen
  \bibfield  {author} {\bibinfo {author} {\bibfnamefont {J.}~\bibnamefont
  {Braun}}, \bibinfo {author} {\bibfnamefont {L.}~\bibnamefont {Fister}},
  \bibinfo {author} {\bibfnamefont {J.~M.}\ \bibnamefont {Pawlowski}},\ and\
  \bibinfo {author} {\bibfnamefont {F.}~\bibnamefont {Rennecke}},\ }\bibfield
  {title} {\bibinfo {title} {{From Quarks and Gluons to Hadrons: Chiral
  Symmetry Breaking in Dynamical QCD}},\ }\href
  {https://doi.org/10.1103/PhysRevD.94.034016} {\bibfield  {journal} {\bibinfo
  {journal} {Phys. Rev.}\ }\textbf {\bibinfo {volume} {D94}},\ \bibinfo {pages}
  {034016} (\bibinfo {year} {2016}{\natexlab{a}})},\ \Eprint
  {https://arxiv.org/abs/1412.1045} {arXiv:1412.1045 [hep-ph]} \BibitemShut
  {NoStop}%
\bibitem [{\citenamefont {Cyrol}\ \emph {et~al.}(2018)\citenamefont {Cyrol},
  \citenamefont {Mitter}, \citenamefont {Pawlowski},\ and\ \citenamefont
  {Strodthoff}}]{Cyrol:2017ewj}%
  \BibitemOpen
  \bibfield  {author} {\bibinfo {author} {\bibfnamefont {A.~K.}\ \bibnamefont
  {Cyrol}}, \bibinfo {author} {\bibfnamefont {M.}~\bibnamefont {Mitter}},
  \bibinfo {author} {\bibfnamefont {J.~M.}\ \bibnamefont {Pawlowski}},\ and\
  \bibinfo {author} {\bibfnamefont {N.}~\bibnamefont {Strodthoff}},\ }\bibfield
   {title} {\bibinfo {title} {{Nonperturbative quark, gluon, and meson
  correlators of unquenched QCD}},\ }\href
  {https://doi.org/10.1103/PhysRevD.97.054006} {\bibfield  {journal} {\bibinfo
  {journal} {Phys. Rev.}\ }\textbf {\bibinfo {volume} {D97}},\ \bibinfo {pages}
  {054006} (\bibinfo {year} {2018})},\ \Eprint
  {https://arxiv.org/abs/1706.06326} {arXiv:1706.06326 [hep-ph]} \BibitemShut
  {NoStop}%
\bibitem [{\citenamefont {Fu}\ \emph {et~al.}(2020)\citenamefont {Fu},
  \citenamefont {Pawlowski},\ and\ \citenamefont {Rennecke}}]{Fu:2019hdw}%
  \BibitemOpen
  \bibfield  {author} {\bibinfo {author} {\bibfnamefont {W.-j.}\ \bibnamefont
  {Fu}}, \bibinfo {author} {\bibfnamefont {J.~M.}\ \bibnamefont {Pawlowski}},\
  and\ \bibinfo {author} {\bibfnamefont {F.}~\bibnamefont {Rennecke}},\
  }\bibfield  {title} {\bibinfo {title} {{QCD phase structure at finite
  temperature and density}},\ }\href
  {https://doi.org/10.1103/PhysRevD.101.054032} {\bibfield  {journal} {\bibinfo
   {journal} {Phys. Rev. D}\ }\textbf {\bibinfo {volume} {101}},\ \bibinfo
  {pages} {054032} (\bibinfo {year} {2020})},\ \Eprint
  {https://arxiv.org/abs/1909.02991} {arXiv:1909.02991 [hep-ph]} \BibitemShut
  {NoStop}%
\bibitem [{\citenamefont {Ihssen}\ \emph {et~al.}(2024)\citenamefont {Ihssen},
  \citenamefont {Pawlowski}, \citenamefont {Sattler},\ and\ \citenamefont
  {Wink}}]{Ihssen:2024miv}%
  \BibitemOpen
  \bibfield  {author} {\bibinfo {author} {\bibfnamefont {F.}~\bibnamefont
  {Ihssen}}, \bibinfo {author} {\bibfnamefont {J.~M.}\ \bibnamefont
  {Pawlowski}}, \bibinfo {author} {\bibfnamefont {F.~R.}\ \bibnamefont
  {Sattler}},\ and\ \bibinfo {author} {\bibfnamefont {N.}~\bibnamefont
  {Wink}},\ }\bibfield  {title} {\bibinfo {title} {{Towards quantitative
  precision in functional QCD I}},\ }\Eprint {https://arxiv.org/abs/2408.08413}
  {arXiv:2408.08413 [hep-ph]}  (\bibinfo {year} {2024})\BibitemShut {NoStop}%
\bibitem [{\citenamefont {Gao}\ and\ \citenamefont
  {Pawlowski}(2022)}]{Gao:2021vsf}%
  \BibitemOpen
  \bibfield  {author} {\bibinfo {author} {\bibfnamefont {F.}~\bibnamefont
  {Gao}}\ and\ \bibinfo {author} {\bibfnamefont {J.~M.}\ \bibnamefont
  {Pawlowski}},\ }\bibfield  {title} {\bibinfo {title} {{Phase structure of
  (2+1)-flavor QCD and the magnetic equation of state}},\ }\href
  {https://doi.org/10.1103/PhysRevD.105.094020} {\bibfield  {journal} {\bibinfo
   {journal} {Phys. Rev. D}\ }\textbf {\bibinfo {volume} {105}},\ \bibinfo
  {pages} {094020} (\bibinfo {year} {2022})},\ \Eprint
  {https://arxiv.org/abs/2112.01395} {arXiv:2112.01395 [hep-ph]} \BibitemShut
  {NoStop}%
\bibitem [{\citenamefont {Braun}\ \emph
  {et~al.}(2023{\natexlab{b}})\citenamefont {Braun} \emph
  {et~al.}}]{Braun:2023qak}%
  \BibitemOpen
  \bibfield  {author} {\bibinfo {author} {\bibfnamefont {J.}~\bibnamefont
  {Braun}} \emph {et~al.},\ }\bibfield  {title} {\bibinfo {title} {Soft modes
  in hot qcd matter},\ }\Eprint {https://arxiv.org/abs/2310.19853}
  {arXiv:2310.19853 [hep-ph]}  (\bibinfo {year}
  {2023}{\natexlab{b}})\BibitemShut {NoStop}%
\bibitem [{\citenamefont {Isserstedt}\ \emph {et~al.}(2021)\citenamefont
  {Isserstedt}, \citenamefont {Fischer},\ and\ \citenamefont
  {Steinert}}]{Isserstedt:2020qll}%
  \BibitemOpen
  \bibfield  {author} {\bibinfo {author} {\bibfnamefont {P.}~\bibnamefont
  {Isserstedt}}, \bibinfo {author} {\bibfnamefont {C.~S.}\ \bibnamefont
  {Fischer}},\ and\ \bibinfo {author} {\bibfnamefont {T.}~\bibnamefont
  {Steinert}},\ }\bibfield  {title} {\bibinfo {title} {{Thermodynamics from the
  quark condensate}},\ }\href {https://doi.org/10.1103/PhysRevD.103.054012}
  {\bibfield  {journal} {\bibinfo  {journal} {Phys. Rev. D}\ }\textbf {\bibinfo
  {volume} {103}},\ \bibinfo {pages} {054012} (\bibinfo {year} {2021})},\
  \Eprint {https://arxiv.org/abs/2012.04991} {arXiv:2012.04991 [hep-ph]}
  \BibitemShut {NoStop}%
\bibitem [{\citenamefont {Braun}\ \emph {et~al.}(2019)\citenamefont {Braun},
  \citenamefont {Leonhardt},\ and\ \citenamefont {Pawlowski}}]{Braun:2018svj}%
  \BibitemOpen
  \bibfield  {author} {\bibinfo {author} {\bibfnamefont {J.}~\bibnamefont
  {Braun}}, \bibinfo {author} {\bibfnamefont {M.}~\bibnamefont {Leonhardt}},\
  and\ \bibinfo {author} {\bibfnamefont {J.~M.}\ \bibnamefont {Pawlowski}},\
  }\bibfield  {title} {\bibinfo {title} {{Renormalization group consistency and
  low-energy effective theories}},\ }\href
  {https://doi.org/10.21468/SciPostPhys.6.5.056} {\bibfield  {journal}
  {\bibinfo  {journal} {SciPost Phys.}\ }\textbf {\bibinfo {volume} {6}},\
  \bibinfo {pages} {056} (\bibinfo {year} {2019})},\ \Eprint
  {https://arxiv.org/abs/1806.04432} {arXiv:1806.04432 [hep-ph]} \BibitemShut
  {NoStop}%
\bibitem [{\citenamefont {Helmboldt}\ \emph {et~al.}(2015)\citenamefont
  {Helmboldt}, \citenamefont {Pawlowski},\ and\ \citenamefont
  {Strodthoff}}]{Helmboldt:2014iya}%
  \BibitemOpen
  \bibfield  {author} {\bibinfo {author} {\bibfnamefont {A.~J.}\ \bibnamefont
  {Helmboldt}}, \bibinfo {author} {\bibfnamefont {J.~M.}\ \bibnamefont
  {Pawlowski}},\ and\ \bibinfo {author} {\bibfnamefont {N.}~\bibnamefont
  {Strodthoff}},\ }\bibfield  {title} {\bibinfo {title} {{Towards quantitative
  precision in the chiral crossover: masses and fluctuation scales}},\ }\href
  {https://doi.org/10.1103/PhysRevD.91.054010} {\bibfield  {journal} {\bibinfo
  {journal} {Phys. Rev. D}\ }\textbf {\bibinfo {volume} {91}},\ \bibinfo
  {pages} {054010} (\bibinfo {year} {2015})},\ \Eprint
  {https://arxiv.org/abs/1409.8414} {arXiv:1409.8414 [hep-ph]} \BibitemShut
  {NoStop}%
\bibitem [{\citenamefont {Cohen}(2003)}]{Cohen:2003kd}%
  \BibitemOpen
  \bibfield  {author} {\bibinfo {author} {\bibfnamefont {T.~D.}\ \bibnamefont
  {Cohen}},\ }\bibfield  {title} {\bibinfo {title} {{Functional integrals for
  QCD at nonzero chemical potential and zero density}},\ }\href
  {https://doi.org/10.1103/PhysRevLett.91.222001} {\bibfield  {journal}
  {\bibinfo  {journal} {Phys. Rev. Lett.}\ }\textbf {\bibinfo {volume} {91}},\
  \bibinfo {pages} {222001} (\bibinfo {year} {2003})},\ \Eprint
  {https://arxiv.org/abs/hep-ph/0307089} {arXiv:hep-ph/0307089 [hep-ph]}
  \BibitemShut {NoStop}%
\bibitem [{\citenamefont {Braun}\ \emph {et~al.}(2017)\citenamefont {Braun},
  \citenamefont {Leonhardt},\ and\ \citenamefont {Pospiech}}]{Braun:2017srn}%
  \BibitemOpen
  \bibfield  {author} {\bibinfo {author} {\bibfnamefont {J.}~\bibnamefont
  {Braun}}, \bibinfo {author} {\bibfnamefont {M.}~\bibnamefont {Leonhardt}},\
  and\ \bibinfo {author} {\bibfnamefont {M.}~\bibnamefont {Pospiech}},\
  }\bibfield  {title} {\bibinfo {title} {{Fierz-complete NJL model study: Fixed
  points and phase structure at finite temperature and density}},\ }\href
  {https://doi.org/10.1103/PhysRevD.96.076003} {\bibfield  {journal} {\bibinfo
  {journal} {Phys. Rev.}\ }\textbf {\bibinfo {volume} {D96}},\ \bibinfo {pages}
  {076003} (\bibinfo {year} {2017})},\ \Eprint
  {https://arxiv.org/abs/1705.00074} {arXiv:1705.00074 [hep-ph]} \BibitemShut
  {NoStop}%
\bibitem [{\citenamefont {Braun}\ \emph {et~al.}(2020)\citenamefont {Braun},
  \citenamefont {Leonhardt},\ and\ \citenamefont {Pospiech}}]{Braun:2019aow}%
  \BibitemOpen
  \bibfield  {author} {\bibinfo {author} {\bibfnamefont {J.}~\bibnamefont
  {Braun}}, \bibinfo {author} {\bibfnamefont {M.}~\bibnamefont {Leonhardt}},\
  and\ \bibinfo {author} {\bibfnamefont {M.}~\bibnamefont {Pospiech}},\
  }\bibfield  {title} {\bibinfo {title} {{Fierz-complete NJL model study III:
  Emergence from quark-gluon dynamics}},\ }\href
  {https://doi.org/10.1103/PhysRevD.101.036004} {\bibfield  {journal} {\bibinfo
   {journal} {Phys. Rev.}\ }\textbf {\bibinfo {volume} {D101}},\ \bibinfo
  {pages} {036004} (\bibinfo {year} {2020})},\ \Eprint
  {https://arxiv.org/abs/1909.06298} {arXiv:1909.06298 [hep-ph]} \BibitemShut
  {NoStop}%
\bibitem [{\citenamefont {Leonhardt}\ \emph {et~al.}(2020)\citenamefont
  {Leonhardt}, \citenamefont {Pospiech}, \citenamefont {Schallmo},
  \citenamefont {Braun}, \citenamefont {Drischler}, \citenamefont {Hebeler},\
  and\ \citenamefont {Schwenk}}]{Leonhardt:2019fua}%
  \BibitemOpen
  \bibfield  {author} {\bibinfo {author} {\bibfnamefont {M.}~\bibnamefont
  {Leonhardt}}, \bibinfo {author} {\bibfnamefont {M.}~\bibnamefont {Pospiech}},
  \bibinfo {author} {\bibfnamefont {B.}~\bibnamefont {Schallmo}}, \bibinfo
  {author} {\bibfnamefont {J.}~\bibnamefont {Braun}}, \bibinfo {author}
  {\bibfnamefont {C.}~\bibnamefont {Drischler}}, \bibinfo {author}
  {\bibfnamefont {K.}~\bibnamefont {Hebeler}},\ and\ \bibinfo {author}
  {\bibfnamefont {A.}~\bibnamefont {Schwenk}},\ }\bibfield  {title} {\bibinfo
  {title} {{Symmetric nuclear matter from the strong interaction}},\ }\href
  {https://doi.org/10.1103/PhysRevLett.125.142502} {\bibfield  {journal}
  {\bibinfo  {journal} {Phys. Rev. Lett.}\ }\textbf {\bibinfo {volume} {125}},\
  \bibinfo {pages} {142502} (\bibinfo {year} {2020})},\ \Eprint
  {https://arxiv.org/abs/1907.05814} {arXiv:1907.05814 [nucl-th]} \BibitemShut
  {NoStop}%
\bibitem [{\citenamefont {Klevansky}(1992)}]{Klevansky:1992qe}%
  \BibitemOpen
  \bibfield  {author} {\bibinfo {author} {\bibfnamefont {S.~P.}\ \bibnamefont
  {Klevansky}},\ }\bibfield  {title} {\bibinfo {title} {{The Nambu-Jona-Lasinio
  model of quantum chromodynamics}},\ }\href
  {https://doi.org/10.1103/RevModPhys.64.649} {\bibfield  {journal} {\bibinfo
  {journal} {Rev. Mod. Phys.}\ }\textbf {\bibinfo {volume} {64}},\ \bibinfo
  {pages} {649} (\bibinfo {year} {1992})}\BibitemShut {NoStop}%
\bibitem [{\citenamefont {Fu}\ \emph {et~al.}()\citenamefont {Fu},
  \citenamefont {Pawlowski}, \citenamefont {Pisarski}, \citenamefont
  {Rennecke}, \citenamefont {Wen},\ and\ \citenamefont {Yin}}]{FPPRWY}%
  \BibitemOpen
  \bibfield  {author} {\bibinfo {author} {\bibfnamefont {W.-j.}\ \bibnamefont
  {Fu}}, \bibinfo {author} {\bibfnamefont {J.~M.}\ \bibnamefont {Pawlowski}},
  \bibinfo {author} {\bibfnamefont {R.~D.}\ \bibnamefont {Pisarski}}, \bibinfo
  {author} {\bibfnamefont {F.}~\bibnamefont {Rennecke}}, \bibinfo {author}
  {\bibfnamefont {R.}~\bibnamefont {Wen}},\ and\ \bibinfo {author}
  {\bibfnamefont {S.}~\bibnamefont {Yin}},\ }\bibinfo {title} {The {QCD} moat
  regime and its real-time properties (in preparation, 2024)}\BibitemShut
  {NoStop}%
\bibitem [{\citenamefont {Braun}\ \emph {et~al.}(2015)\citenamefont {Braun},
  \citenamefont {Finkbeiner}, \citenamefont {Karbstein},\ and\ \citenamefont
  {Roscher}}]{Braun:2014fga}%
  \BibitemOpen
\bibfield  {title} {  }\bibfield  {author} {\bibinfo {author} {\bibfnamefont
  {J.}~\bibnamefont {Braun}}, \bibinfo {author} {\bibfnamefont
  {S.}~\bibnamefont {Finkbeiner}}, \bibinfo {author} {\bibfnamefont
  {F.}~\bibnamefont {Karbstein}},\ and\ \bibinfo {author} {\bibfnamefont
  {D.}~\bibnamefont {Roscher}},\ }\bibfield  {title} {\bibinfo {title} {{Search
  for inhomogeneous phases in fermionic models}},\ }\href
  {https://doi.org/10.1103/PhysRevD.91.116006} {\bibfield  {journal} {\bibinfo
  {journal} {Phys. Rev. D}\ }\textbf {\bibinfo {volume} {91}},\ \bibinfo
  {pages} {116006} (\bibinfo {year} {2015})},\ \Eprint
  {https://arxiv.org/abs/1410.8181} {arXiv:1410.8181 [hep-ph]} \BibitemShut
  {NoStop}%
\bibitem [{\citenamefont {Roscher}\ \emph {et~al.}(2015)\citenamefont
  {Roscher}, \citenamefont {Braun},\ and\ \citenamefont
  {Drut}}]{Roscher:2015xha}%
  \BibitemOpen
  \bibfield  {author} {\bibinfo {author} {\bibfnamefont {D.}~\bibnamefont
  {Roscher}}, \bibinfo {author} {\bibfnamefont {J.}~\bibnamefont {Braun}},\
  and\ \bibinfo {author} {\bibfnamefont {J.~E.}\ \bibnamefont {Drut}},\
  }\bibfield  {title} {\bibinfo {title} {{Phase structure of mass- and
  spin-imbalanced unitary Fermi gases}},\ }\href
  {https://doi.org/10.1103/PhysRevA.91.053611} {\bibfield  {journal} {\bibinfo
  {journal} {Phys. Rev.}\ }\textbf {\bibinfo {volume} {A91}},\ \bibinfo {pages}
  {053611} (\bibinfo {year} {2015})},\ \Eprint
  {https://arxiv.org/abs/1501.05544} {arXiv:1501.05544 [cond-mat.quant-gas]}
  \BibitemShut {NoStop}%
\bibitem [{\citenamefont {Braun}\ \emph
  {et~al.}(2016{\natexlab{b}})\citenamefont {Braun}, \citenamefont {Karbstein},
  \citenamefont {Rechenberger},\ and\ \citenamefont {Roscher}}]{Braun:2015fva}%
  \BibitemOpen
  \bibfield  {author} {\bibinfo {author} {\bibfnamefont {J.}~\bibnamefont
  {Braun}}, \bibinfo {author} {\bibfnamefont {F.}~\bibnamefont {Karbstein}},
  \bibinfo {author} {\bibfnamefont {S.}~\bibnamefont {Rechenberger}},\ and\
  \bibinfo {author} {\bibfnamefont {D.}~\bibnamefont {Roscher}},\ }\bibfield
  {title} {\bibinfo {title} {{Crystalline ground states in Polyakov-loop
  extended Nambu\textendash{}Jona-Lasinio models}},\ }\href
  {https://doi.org/10.1103/PhysRevD.93.014032} {\bibfield  {journal} {\bibinfo
  {journal} {Phys. Rev. D}\ }\textbf {\bibinfo {volume} {93}},\ \bibinfo
  {pages} {014032} (\bibinfo {year} {2016}{\natexlab{b}})},\ \Eprint
  {https://arxiv.org/abs/1510.04012} {arXiv:1510.04012 [hep-ph]} \BibitemShut
  {NoStop}%
\bibitem [{\citenamefont {Tripolt}\ \emph
  {et~al.}(2018{\natexlab{b}})\citenamefont {Tripolt}, \citenamefont
  {Schaefer}, \citenamefont {von Smekal},\ and\ \citenamefont
  {Wambach}}]{Tripolt:2017zgc}%
  \BibitemOpen
  \bibfield  {author} {\bibinfo {author} {\bibfnamefont {R.-A.}\ \bibnamefont
  {Tripolt}}, \bibinfo {author} {\bibfnamefont {B.-J.}\ \bibnamefont
  {Schaefer}}, \bibinfo {author} {\bibfnamefont {L.}~\bibnamefont {von
  Smekal}},\ and\ \bibinfo {author} {\bibfnamefont {J.}~\bibnamefont
  {Wambach}},\ }\bibfield  {title} {\bibinfo {title} {{Low-temperature behavior
  of the quark-meson model}},\ }\href
  {https://doi.org/10.1103/PhysRevD.97.034022} {\bibfield  {journal} {\bibinfo
  {journal} {Phys. Rev. D}\ }\textbf {\bibinfo {volume} {97}},\ \bibinfo
  {pages} {034022} (\bibinfo {year} {2018}{\natexlab{b}})},\ \Eprint
  {https://arxiv.org/abs/1709.05991} {arXiv:1709.05991 [hep-ph]} \BibitemShut
  {NoStop}%
\bibitem [{\citenamefont {Motta}\ \emph {et~al.}(2023)\citenamefont {Motta},
  \citenamefont {Bernhardt}, \citenamefont {Buballa},\ and\ \citenamefont
  {Fischer}}]{Motta:2023pks}%
  \BibitemOpen
  \bibfield  {author} {\bibinfo {author} {\bibfnamefont {T.~F.}\ \bibnamefont
  {Motta}}, \bibinfo {author} {\bibfnamefont {J.}~\bibnamefont {Bernhardt}},
  \bibinfo {author} {\bibfnamefont {M.}~\bibnamefont {Buballa}},\ and\ \bibinfo
  {author} {\bibfnamefont {C.~S.}\ \bibnamefont {Fischer}},\ }\bibfield
  {title} {\bibinfo {title} {{Toward a stability analysis of inhomogeneous
  phases in QCD}},\ }\href {https://doi.org/10.1103/PhysRevD.108.114019}
  {\bibfield  {journal} {\bibinfo  {journal} {Phys. Rev. D}\ }\textbf {\bibinfo
  {volume} {108}},\ \bibinfo {pages} {114019} (\bibinfo {year} {2023})},\
  \Eprint {https://arxiv.org/abs/2306.09749} {arXiv:2306.09749 [hep-ph]}
  \BibitemShut {NoStop}%
\bibitem [{\citenamefont {Pisarski}\ and\ \citenamefont
  {Rennecke}(2021)}]{Pisarski:2021qof}%
  \BibitemOpen
  \bibfield  {author} {\bibinfo {author} {\bibfnamefont {R.~D.}\ \bibnamefont
  {Pisarski}}\ and\ \bibinfo {author} {\bibfnamefont {F.}~\bibnamefont
  {Rennecke}},\ }\bibfield  {title} {\bibinfo {title} {{Signatures of Moat
  Regimes in Heavy-Ion Collisions}},\ }\href
  {https://doi.org/10.1103/PhysRevLett.127.152302} {\bibfield  {journal}
  {\bibinfo  {journal} {Phys. Rev. Lett.}\ }\textbf {\bibinfo {volume} {127}},\
  \bibinfo {pages} {152302} (\bibinfo {year} {2021})},\ \Eprint
  {https://arxiv.org/abs/2103.06890} {arXiv:2103.06890 [hep-ph]} \BibitemShut
  {NoStop}%
\bibitem [{\citenamefont {Haensch}\ \emph {et~al.}(2024)\citenamefont
  {Haensch}, \citenamefont {Rennecke},\ and\ \citenamefont {von
  Smekal}}]{Haensch:2023sig}%
  \BibitemOpen
  \bibfield  {author} {\bibinfo {author} {\bibfnamefont {M.}~\bibnamefont
  {Haensch}}, \bibinfo {author} {\bibfnamefont {F.}~\bibnamefont {Rennecke}},\
  and\ \bibinfo {author} {\bibfnamefont {L.}~\bibnamefont {von Smekal}},\
  }\bibfield  {title} {\bibinfo {title} {{Medium induced mixing, spatial
  modulations, and critical modes in QCD}},\ }\href
  {https://doi.org/10.1103/PhysRevD.110.036018} {\bibfield  {journal} {\bibinfo
   {journal} {Phys. Rev. D}\ }\textbf {\bibinfo {volume} {110}},\ \bibinfo
  {pages} {036018} (\bibinfo {year} {2024})},\ \Eprint
  {https://arxiv.org/abs/2308.16244} {arXiv:2308.16244 [hep-ph]} \BibitemShut
  {NoStop}%
\bibitem [{\citenamefont {Pannullo}\ \emph {et~al.}(2024)\citenamefont
  {Pannullo}, \citenamefont {Wagner},\ and\ \citenamefont
  {Winstel}}]{Pannullo:2024sov}%
  \BibitemOpen
  \bibfield  {author} {\bibinfo {author} {\bibfnamefont {L.}~\bibnamefont
  {Pannullo}}, \bibinfo {author} {\bibfnamefont {M.}~\bibnamefont {Wagner}},\
  and\ \bibinfo {author} {\bibfnamefont {M.}~\bibnamefont {Winstel}},\
  }\bibfield  {title} {\bibinfo {title} {{Regularization effects in the
  Nambu\textendash{}Jona-Lasinio model: Strong scheme dependence of
  inhomogeneous phases and persistence of the moat regime}},\ }\href
  {https://doi.org/10.1103/PhysRevD.110.076006} {\bibfield  {journal} {\bibinfo
   {journal} {Phys. Rev. D}\ }\textbf {\bibinfo {volume} {110}},\ \bibinfo
  {pages} {076006} (\bibinfo {year} {2024})},\ \Eprint
  {https://arxiv.org/abs/2406.11312} {arXiv:2406.11312 [hep-ph]} \BibitemShut
  {NoStop}%
\bibitem [{\citenamefont {Friedel}(1952)}]{doi:10.1080/14786440208561086}%
  \BibitemOpen
  \bibfield  {author} {\bibinfo {author} {\bibfnamefont {J.}~\bibnamefont
  {Friedel}},\ }\bibfield  {title} {\bibinfo {title} {Xiv. the distribution of
  electrons round impurities in monovalent metals},\ }\href
  {https://doi.org/10.1080/14786440208561086} {\bibfield  {journal} {\bibinfo
  {journal} {The London, Edinburgh, and Dublin Philosophical Magazine and
  Journal of Science}\ }\textbf {\bibinfo {volume} {43}},\ \bibinfo {pages}
  {153} (\bibinfo {year} {1952})}\BibitemShut {NoStop}%
\bibitem [{\citenamefont {Friedel}(1954)}]{doi:10.1080/00018735400101233}%
  \BibitemOpen
  \bibfield  {author} {\bibinfo {author} {\bibfnamefont {J.}~\bibnamefont
  {Friedel}},\ }\bibfield  {title} {\bibinfo {title} {Electronic structure of
  primary solid solutions in metals},\ }\href
  {https://doi.org/10.1080/00018735400101233} {\bibfield  {journal} {\bibinfo
  {journal} {Advances in Physics}\ }\textbf {\bibinfo {volume} {3}},\ \bibinfo
  {pages} {446} (\bibinfo {year} {1954})}\BibitemShut {NoStop}%
\bibitem [{\citenamefont {Friedel}(1958)}]{Friedel:1958}%
  \BibitemOpen
  \bibfield  {author} {\bibinfo {author} {\bibfnamefont {J.}~\bibnamefont
  {Friedel}},\ }\bibfield  {title} {\bibinfo {title} {Metallic alloys},\
  }\href@noop {} {\bibfield  {journal} {\bibinfo  {journal} {Nuovo Cimento}\
  }\textbf {\bibinfo {volume} {7}},\ \bibinfo {pages} {287} (\bibinfo {year}
  {1958})}\BibitemShut {NoStop}%
\bibitem [{\citenamefont {Fetter}\ and\ \citenamefont
  {Walecka}(1971)}]{FetterAndWalecka}%
  \BibitemOpen
  \bibfield  {author} {\bibinfo {author} {\bibfnamefont {A.~L.}\ \bibnamefont
  {Fetter}}\ and\ \bibinfo {author} {\bibfnamefont {J.~D.}\ \bibnamefont
  {Walecka}},\ }\href@noop {} {\emph {\bibinfo {title} {Quantum Theory of
  Many-Particle Systems}}}\ (\bibinfo  {publisher} {McGraw-Hill},\ \bibinfo
  {address} {Boston},\ \bibinfo {year} {1971})\BibitemShut {NoStop}%
\bibitem [{\citenamefont {Rennecke}\ and\ \citenamefont
  {Yin}()}]{Rennecke:2024}%
  \BibitemOpen
  \bibfield  {author} {\bibinfo {author} {\bibfnamefont {F.}~\bibnamefont
  {Rennecke}}\ and\ \bibinfo {author} {\bibfnamefont {S.}~\bibnamefont {Yin}},\
  }\bibinfo {title} {(in preparation)}\BibitemShut {NoStop}%
\bibitem [{\citenamefont {Koenigstein}\ \emph {et~al.}(2022)\citenamefont
  {Koenigstein}, \citenamefont {Pannullo}, \citenamefont {Rechenberger},
  \citenamefont {Steil},\ and\ \citenamefont {Winstel}}]{Koenigstein:2021llr}%
  \BibitemOpen
\bibfield  {title} {  }\bibfield  {author} {\bibinfo {author} {\bibfnamefont
  {A.}~\bibnamefont {Koenigstein}}, \bibinfo {author} {\bibfnamefont
  {L.}~\bibnamefont {Pannullo}}, \bibinfo {author} {\bibfnamefont
  {S.}~\bibnamefont {Rechenberger}}, \bibinfo {author} {\bibfnamefont {M.~J.}\
  \bibnamefont {Steil}},\ and\ \bibinfo {author} {\bibfnamefont
  {M.}~\bibnamefont {Winstel}},\ }\bibfield  {title} {\bibinfo {title}
  {{Detecting inhomogeneous chiral condensation from the bosonic two-point
  function in the (1 + 1)-dimensional Gross\textendash{}Neveu model in the
  mean-field approximation*}},\ }\href
  {https://doi.org/10.1088/1751-8121/ac820a} {\bibfield  {journal} {\bibinfo
  {journal} {J. Phys. A}\ }\textbf {\bibinfo {volume} {55}},\ \bibinfo {pages}
  {375402} (\bibinfo {year} {2022})},\ \Eprint
  {https://arxiv.org/abs/2112.07024} {arXiv:2112.07024 [hep-ph]} \BibitemShut
  {NoStop}%
\bibitem [{\citenamefont {Koenigstein}\ and\ \citenamefont
  {Pannullo}(2024)}]{Koenigstein:2023yzv}%
  \BibitemOpen
  \bibfield  {author} {\bibinfo {author} {\bibfnamefont {A.}~\bibnamefont
  {Koenigstein}}\ and\ \bibinfo {author} {\bibfnamefont {L.}~\bibnamefont
  {Pannullo}},\ }\bibfield  {title} {\bibinfo {title} {{Inhomogeneous
  condensation in the Gross-Neveu model in noninteger spatial dimensions
  1\ensuremath{\leq}d\ensuremath{<}3. II. Nonzero temperature and chemical
  potential}},\ }\href {https://doi.org/10.1103/PhysRevD.109.056015} {\bibfield
   {journal} {\bibinfo  {journal} {Phys. Rev. D}\ }\textbf {\bibinfo {volume}
  {109}},\ \bibinfo {pages} {056015} (\bibinfo {year} {2024})},\ \Eprint
  {https://arxiv.org/abs/2312.04904} {arXiv:2312.04904 [hep-ph]} \BibitemShut
  {NoStop}%
\bibitem [{\citenamefont {Koenigstein}\ and\ \citenamefont
  {Winstel}(2024)}]{Koenigstein:2024cyq}%
  \BibitemOpen
  \bibfield  {author} {\bibinfo {author} {\bibfnamefont {A.}~\bibnamefont
  {Koenigstein}}\ and\ \bibinfo {author} {\bibfnamefont {M.}~\bibnamefont
  {Winstel}},\ }\bibfield  {title} {\bibinfo {title} {{Revisiting the spatially
  inhomogeneous condensates in the $(1 + 1)$-dimensional chiral
  Gross\textendash{}Neveu model via the bosonic two-point function in the
  infinite-N limit}},\ }\href {https://doi.org/10.1088/1751-8121/ad6721}
  {\bibfield  {journal} {\bibinfo  {journal} {J. Phys. A}\ }\textbf {\bibinfo
  {volume} {57}},\ \bibinfo {pages} {335401} (\bibinfo {year} {2024})},\
  \Eprint {https://arxiv.org/abs/2405.03459} {arXiv:2405.03459 [hep-th]}
  \BibitemShut {NoStop}%
\bibitem [{\citenamefont {Braun}(2010)}]{Braun:2009si}%
  \BibitemOpen
  \bibfield  {author} {\bibinfo {author} {\bibfnamefont {J.}~\bibnamefont
  {Braun}},\ }\bibfield  {title} {\bibinfo {title} {{Thermodynamics of QCD
  low-energy models and the derivative expansion of the effective action}},\
  }\href {https://doi.org/10.1103/PhysRevD.81.016008} {\bibfield  {journal}
  {\bibinfo  {journal} {Phys. Rev.}\ }\textbf {\bibinfo {volume} {D81}},\
  \bibinfo {pages} {016008} (\bibinfo {year} {2010})},\ \Eprint
  {https://arxiv.org/abs/0908.1543} {arXiv:0908.1543 [hep-ph]} \BibitemShut
  {NoStop}%
\bibitem [{\citenamefont {Braun}\ and\ \citenamefont
  {Schallmo}(2022)}]{Braun:2021uua}%
  \BibitemOpen
  \bibfield  {author} {\bibinfo {author} {\bibfnamefont {J.}~\bibnamefont
  {Braun}}\ and\ \bibinfo {author} {\bibfnamefont {B.}~\bibnamefont
  {Schallmo}},\ }\bibfield  {title} {\bibinfo {title} {{From quarks and gluons
  to color superconductivity at supranuclear densities}},\ }\href
  {https://doi.org/10.1103/PhysRevD.105.036003} {\bibfield  {journal} {\bibinfo
   {journal} {Phys. Rev. D}\ }\textbf {\bibinfo {volume} {105}},\ \bibinfo
  {pages} {036003} (\bibinfo {year} {2022})},\ \Eprint
  {https://arxiv.org/abs/2106.04198} {arXiv:2106.04198 [hep-ph]} \BibitemShut
  {NoStop}%
\bibitem [{\citenamefont {Schaefer}\ and\ \citenamefont
  {Wagner}(2009)}]{Schaefer:2008hk}%
  \BibitemOpen
  \bibfield  {author} {\bibinfo {author} {\bibfnamefont {B.-J.}\ \bibnamefont
  {Schaefer}}\ and\ \bibinfo {author} {\bibfnamefont {M.}~\bibnamefont
  {Wagner}},\ }\bibfield  {title} {\bibinfo {title} {{The Three-flavor chiral
  phase structure in hot and dense QCD matter}},\ }\href
  {https://doi.org/10.1103/PhysRevD.79.014018} {\bibfield  {journal} {\bibinfo
  {journal} {Phys. Rev.}\ }\textbf {\bibinfo {volume} {D79}},\ \bibinfo {pages}
  {014018} (\bibinfo {year} {2009})},\ \Eprint
  {https://arxiv.org/abs/0808.1491} {arXiv:0808.1491 [hep-ph]} \BibitemShut
  {NoStop}%
\bibitem [{\citenamefont {Zinn-Justin}(2002)}]{Zinn-Justin:2002ecy}%
  \BibitemOpen
  \bibfield  {author} {\bibinfo {author} {\bibfnamefont {J.}~\bibnamefont
  {Zinn-Justin}},\ }\bibfield  {title} {\bibinfo {title} {{Quantum field theory
  and critical phenomena}},\ }\href@noop {} {\bibfield  {journal} {\bibinfo
  {journal} {Int. Ser. Monogr. Phys.}\ }\textbf {\bibinfo {volume} {113}},\
  \bibinfo {pages} {1} (\bibinfo {year} {2002})}\BibitemShut {NoStop}%
\bibitem [{fQC()}]{fQCD}%
  \BibitemOpen
  \href@noop {} {}\bibinfo {note} {{\it fQCD Collaboration:} Braun, Jens and
  Chen, Yong-rui and Fu, Wei-jie and Gao, Fei and Ihssen, Friederike and
  Gei{\ss}el, Andreas and Huang, Chuang and Pawlowski, Jan M. and Rennecke,
  Fabian and Sattler, Franz R. and Schallmo, Benedikt and Stoll, Jonas and Tan,
  Yang-yang and T{\"o}pfel, Sebastian and Turnwald, Jonas and Wen, Rui and
  Wessely, Jonas and Yin, Shi and Zorbach, Niklas (October 2024)}\BibitemShut
  {NoStop}%
\bibitem [{\citenamefont {T\"opfel}\ \emph {et~al.}(2024)\citenamefont
  {T\"opfel}, \citenamefont {Gei\ss{}el},\ and\ \citenamefont
  {Braun}}]{Topfel:2024cll}%
  \BibitemOpen
  \bibfield  {author} {\bibinfo {author} {\bibfnamefont {S.}~\bibnamefont
  {T\"opfel}}, \bibinfo {author} {\bibfnamefont {A.}~\bibnamefont
  {Gei\ss{}el}},\ and\ \bibinfo {author} {\bibfnamefont {J.}~\bibnamefont
  {Braun}},\ }\bibfield  {title} {\bibinfo {title} {{Subtleties in the
  calculation of correlation functions for hot and dense systems}},\ }\Eprint
  {https://arxiv.org/abs/2410.06674} {arXiv:2410.06674 [nucl-th]}  (\bibinfo
  {year} {2024})\BibitemShut {NoStop}%
\end{thebibliography}%
\end{document}